\documentclass[10pt]{article}
\usepackage[margin=20mm]{geometry}
\usepackage[numbers,sort&compress]{natbib}
\usepackage[T1]{fontenc}
\usepackage{lmodern}
\usepackage{microtype}
\usepackage{amsmath,amssymb,booktabs,graphicx}
\usepackage{siunitx}
\DeclareSIUnit\angstrom{\text{\AA}}
\DeclareSIUnit\hartree{Ha}
\DeclareSIUnit\millihartree{mHa}
\usepackage{flafter}
\usepackage[colorlinks=true,allcolors=blue]{hyperref}
\usepackage{placeins}
\usepackage{xurl}
\usepackage{framed}
\usepackage{needspace}

\providecommand{\keywords}[1]{\par\noindent\textbf{Keywords:} #1\par}
\newif\ifprotocolfloat

\newcommand{\EnergyViolationCount}{13}
\newcommand{\IndicativeMonetaryCost}{2700}
\newcommand{\AachenBerylliumWithin}{50.4}
\newcommand{\AachenOxygenZNEDispersion}{0.055}
\newcommand{\PittsburghBerylliumWithin}{17.5}
\newcommand{\PittsburghBerylliumEpoch}{133}
\newcommand{\PittsburghOxygenZNEDispersion}{0.036}
\newcommand{\LargestDispersion}{13}
\newcommand{\RepeatabilityIIDLow}{3.5}
\newcommand{\RepeatabilityIIDHigh}{14}
\newcommand{\BiasRatioIIDLow}{26}
\newcommand{\BiasRatioIIDHigh}{38}
\newcommand{\SixFlagFalsePositive}{0.125}
\newcommand{\ThreeFlagFalsePositive}{1.832}
\newcommand{\LithiumCoverageLow}{82.48}
\newcommand{\LithiumCoverageHigh}{93.37}
\newcommand{\WaterSinglesGap}{0.0003}
\newcommand{\OxygenAssignmentRemovalMin}{190}
\newcommand{\OxygenAssignmentRemovalMax}{226}
\newcommand{\OxygenSignP}{0.016}
\newcommand{\BerylliumSignP}{1.0}
\newcommand{\OxygenFixedCost}{2.33}
\newcommand{\OxygenShotCost}{3.58}
\newcommand{\RoundThreeControlCost}{368}
\newcommand{\RoundThreeControlCeiling}{400}
\newcommand{\RoundThreeControlJobCost}{18.4}
\newcommand{\RoundThreeZNEAachenMin}{1.57}
\newcommand{\RoundThreeZNEAachenMax}{3.15}
\newcommand{\RoundThreeZNEPittsburghMin}{2.26}
\newcommand{\RoundThreeZNEPittsburghMax}{4.47}
\newcommand{\RoundThreeShotRatioMin}{0.47}
\newcommand{\RoundThreeShotRatioMax}{0.53}
\newcommand{\RoundThreeMeanReturnedErrorThree}{8.7}
\newcommand{\RoundThreeDeviceTtwo}{313.91}
\newcommand{\RoundThreeWireTtwo}{426.14}
\newcommand{\RoundThreeWireInitialTtwo}{475.55}
\newcommand{\RoundThreeLiHGap}{0.0156}
\newcommand{\RoundThreeWaterGap}{0.1133}
\newcommand{\RoundThreeWaterSinglesGap}{0.0003}
\newcommand{\RoundThreeTREXLevel}{1}
\newcommand{\RoundThreeZNELevel}{2}
\newcommand{\RoundThreeKnownMixed}{1.720}
\newcommand{\RoundThreeInverseMixed}{0.581}
\newcommand{\RoundThreeKnownReversed}{1.738}
\newcommand{\RoundThreeInverseReversed}{0.575}
\newcommand{\RoundThreeKnownIsolated}{0.9992}
\newcommand{\RoundThreeInverseIsolated}{1.0008}
\newcommand{\RoundThreeKnownHealthy}{0.9967}
\newcommand{\RoundThreeInverseHealthy}{1.0033}
\newcommand{\RoundThreeEnergyErrorAnomalous}{17}
\newcommand{\RoundThreeEnergyErrorHealthy}{5.7}
\newcommand{\RoundThreePittsburghWireOriginal}{77}
\newcommand{\RoundThreePittsburghWireReplacement}{114}
\newcommand{\RoundThreeThermalIdleMin}{300}
\newcommand{\RoundThreeThermalIdleMax}{358}
\newcommand{\CellpilotibmaachenBeHTworawOneZeroTwoFourmeanmha}{424.71}
\newcommand{\CellpilotibmaachenHTworawOneZeroTwoFourmeanmha}{493.3}
\newcommand{\CellpilotibmaachenHTworawOneZeroTwoFourmeanreturnedsmha}{23.1}
\newcommand{\CellpilotibmaachenOTworawOneZeroTwoFourmeanmha}{142.3}
\newcommand{\CellpilotibmaachenOTwotrexOneZeroTwoFourmeanmha}{-349.5}
\newcommand{\CellpilotibmaachenOTwotrexOneZeroTwoFourmeanreturnedsmha}{16.3}
\newcommand{\CellpilotibmaachenOTwozneOneZeroTwoFourmeanmha}{50.41}
\newcommand{\CellpilotibmpittsburghBeHTworawOneZeroTwoFourmeanmha}{798.11}
\newcommand{\CellpilotibmpittsburghBeHTwotrexOneZeroTwoFourmeanmha}{881.41}
\newcommand{\CellpilotibmpittsburghHTworawOneZeroTwoFourmeanmha}{11.61}
\newcommand{\CellpilotibmpittsburghHTwotrexOneZeroTwoFourmeanmha}{-186.68}
\newcommand{\CellpilotibmpittsburghOTworawOneZeroTwoFourmeanmha}{238.59}
\newcommand{\CellpilotibmpittsburghOTwotrexOneZeroTwoFourmeanmha}{175.46}
\newcommand{\CellpilotibmpittsburghOTwozneOneZeroTwoFourmeanmha}{21.93}
\newcommand{\CellisolatedibmaachenBeHTworawOneZeroTwoFourmeanmha}{452}
\newcommand{\CellisolatedibmaachenBeHTwotrexOneZeroTwoFourmeanmha}{457}
\newcommand{\CellisolatedibmaachenOTworawOneZeroTwoFourmeanmha}{119}
\newcommand{\CellisolatedibmaachenOTworawOneZeroTwoFourwindowmeanminmha}{114}
\newcommand{\CellisolatedibmaachenOTworawOneZeroTwoFourwindowmeanmaxmha}{124}
\newcommand{\CellisolatedibmaachenOTworawFourZeroNineSixmeanmha}{119}
\newcommand{\CellisolatedibmaachenOTworawFourZeroNineSixminmha}{108}
\newcommand{\CellisolatedibmaachenOTworawFourZeroNineSixmaxmha}{129}
\newcommand{\CellisolatedibmaachenOTwozneOneZeroTwoFourmeanmha}{60.0}
\newcommand{\CellisolatedibmaachenOTwozneOneZeroTwoFourminmha}{58.6}
\newcommand{\CellisolatedibmaachenOTwozneOneZeroTwoFourmaxmha}{61.0}
\newcommand{\CellisolatedibmpittsburghBeHTworawOneZeroTwoFourmeanmha}{685}
\newcommand{\CellisolatedibmpittsburghBeHTworawOneZeroTwoFoursamplesdmha}{120.4}
\newcommand{\CellisolatedibmpittsburghBeHTworawOneZeroTwoFourmeanreturnedsmha}{21.64}
\newcommand{\CellisolatedibmpittsburghBeHTworawOneZeroTwoFourwindowmeanminmha}{588}
\newcommand{\CellisolatedibmpittsburghBeHTworawOneZeroTwoFourwindowmeanmaxmha}{838}
\newcommand{\CellisolatedibmpittsburghBeHTwotrexOneZeroTwoFourmeanmha}{692}
\newcommand{\CellisolatedibmpittsburghOTworawOneZeroTwoFourmeanmha}{250}
\newcommand{\CellisolatedibmpittsburghOTworawOneZeroTwoFourwindowmeanminmha}{228}
\newcommand{\CellisolatedibmpittsburghOTworawOneZeroTwoFourwindowmeanmaxmha}{271}
\newcommand{\CellisolatedibmpittsburghOTworawFourZeroNineSixmeanmha}{246}
\newcommand{\CellisolatedibmpittsburghOTworawFourZeroNineSixminmha}{236}
\newcommand{\CellisolatedibmpittsburghOTworawFourZeroNineSixmaxmha}{266}
\newcommand{\CellisolatedibmpittsburghOTwozneOneZeroTwoFourmeanmha}{41.3}
\newcommand{\CellisolatedibmpittsburghOTwozneOneZeroTwoFourminmha}{40.3}
\newcommand{\CellisolatedibmpittsburghOTwozneOneZeroTwoFourmaxmha}{43.2}
\newcommand{\CellisolatedibmpittsburghOTwozneOneZeroTwoFourmeanreturnedsmha}{44.75}
\newcommand{\CellpilotibmaachenOTwotrexOneZeroTwoFourAbsoluteMean}{349.50}
\newcommand{\VaribmpittsburghBeHTwowindowmeansdmha}{133.76}
\newcommand{\VaribmpittsburghBeHTwoF}{116}
\newcommand{\ModelibmaachenBeHTwopilotdensitymatrixlocalqwcresidualmha}{164.53}
\newcommand{\ModelibmaachenHTwopilotnaivedepolarizingresidualmha}{32.8}
\newcommand{\ModelibmaachenHTwopilotdensitymatrixlocalqwcresidualmha}{59.5}
\newcommand{\ModelibmaachenHTwopilotobservedovernaiveprediction}{15.02}
\newcommand{\ModelibmaachenHTwopilotthermalgatereadoutresidualmha}{59.4}
\newcommand{\ModelibmpittsburghBeHTwovalidationTwoZeroTwoSixOneZeroZeroThreeusdensitymatrixlocalqwcresidualmha}{219.30}
\newcommand{\ModelibmpittsburghBeHTwovalidationTwoZeroTwoSixOneZeroZeroThreeusthermalgatereadoutresidualmha}{220.35}
\newcommand{\ModelibmpittsburghBeHTwovalidationwindowTwoTwoZeroTwoSixOneZeroZeroThreeusdensitymatrixlocalqwcresidualmha}{228.12}
\newcommand{\ModelibmpittsburghBeHTwovalidationwindowTwoTwoZeroTwoSixOneZeroZeroThreeusobservedovernaiveprediction}{1.77}
\newcommand{\ModelibmpittsburghBeHTwovalidationwindowTwoTwoZeroTwoSixOneZeroZeroThreeusthermalgatereadoutresidualmha}{229.17}
\newcommand{\ModelibmpittsburghBeHTwovalidationwindowThreeTwoZeroTwoSixOneZeroZeroFiveusobservedrawmeanresidualmha}{629.67}
\newcommand{\ModelibmpittsburghBeHTwovalidationwindowThreeTwoZeroTwoSixOneZeroZeroFiveusdensitymatrixlocalqwcresidualmha}{232.08}
\newcommand{\ModelibmpittsburghBeHTwovalidationwindowThreeTwoZeroTwoSixOneZeroZeroFiveusthermalgatereadoutresidualmha}{232.87}
\newcommand{\RefLiHactivetruncationmha}{20.53}
\newcommand{\RefOTwoactivetruncationmha}{113.32}
\newcommand{\ChannelibmaachenmixedOneTwoFourraweffectivea}{0.131}
\newcommand{\ChannelibmaachenmixedOneTwoFourraweffectiveb}{0.324}
\newcommand{\ChannelibmaachenmixedOneTwoFourraweffectiveassignmentfidelity}{0.772}
\newcommand{\ChannelibmaachenmixedOneTwoFourrawxcontrast}{0.564}
\newcommand{\ChannelibmaachenmixedOneTwoFourrawxoffset}{-0.168}
\newcommand{\ChannelibmaachenmixedOneTwoFourrawpredictedhTworesidualmha}{549.0}
\newcommand{\ChannelibmaachenmixedOneTwoFourrawobservedhTworesidualmha}{528.5}
\newcommand{\ChannelibmaachenmixedTwoSevenraweffectiveassignmentfidelity}{0.997}
\newcommand{\ChannelibmaachenmixedTwoSevenrawpredictedhTworesidualmha}{3.4}
\newcommand{\ChannelibmaachenmixedTwoSevenrawobservedhTworesidualmha}{0.9968}
\newcommand{\IdealHTwoOneZeroTwoFourfixedtotalequalsigmamha}{1.61}
\newcommand{\IdealHTwoOOneZeroTwoFourfixedtotalequalsigmamha}{5.62}
\newcommand{\ControlibmaachenmixedOneTwoFourzerozrawValue}{0.352}
\newcommand{\ControlibmaachenmixedOneTwoFourzerozrawError}{0.021}
\newcommand{\ControlibmaachenmixedTwoSevenzerozrawValue}{0.9961}
\newcommand{\ControlibmaachenmixedTwoSevenzerozrawError}{0.0020}
\newcommand{\ControlibmaachenmixedTwoSevenzeroztrexValue}{1.720}
\newcommand{\ControlibmaachenmixedTwoSevenzeroztrexError}{0.015}
\newcommand{\ControlibmaachenhealthypairTwoSevenzeroztrexValue}{0.9967}
\newcommand{\ControlibmaachenhealthypairTwoSevenzeroztrexError}{0.0027}
\newcommand{\ControlibmaachenisolatedTwoSevenzeroztrexValue}{0.9992}
\newcommand{\ControlibmaachenisolatedTwoSevenzeroztrexError}{0.0037}
\newcommand{\ControlibmaachenreversedTwoSevenzeroztrexValue}{1.738}
\newcommand{\ControlibmaachenreversedTwoSevenzeroztrexError}{0.015}
\newcommand{\HeadlineSDLow}{75.2}
\newcommand{\HeadlineSDHigh}{295}
\newcommand{\RawRatioMin}{12}
\newcommand{\RawRatioMax}{32}
\newcommand{\CrossoverMin}{1.02}
\newcommand{\CrossoverMax}{7.20}
\newcommand{\MSELeverage}{0.70}
\newcommand{\IsolatedSpanHours}{30.19}
\newcommand{\WindowStartSpanHours}{30.03}
\newcommand{\WindowGapOne}{6.22}
\newcommand{\WindowGapTwo}{23.81}
\newcommand{\BasisRatio}{24.0}
\newcommand{\BasisGap}{35.08}
\newcommand{\BasisTZGap}{1.46}
\newcommand{\EnergyThreeSCount}{4}
\newcommand{\BracketSensitivityMax}{27.6}
\newcommand{\OxygenSectorDimension}{2}
\newcommand{\Scalarbilledminutes}{28.23}
\newcommand{\ScalarAachenpilotOTworawmeanYF}{255.65}
\newcommand{\ScalarroundTworatiomin}{2.4}
\newcommand{\ScalarroundTworatiomax}{5.1}
\newcommand{\ScalarroundTwocampaignhours}{31.69}
\newcommand{\ScalarroundTwodurationBeHTwo}{18.75}
\newcommand{\WindowDispersionibmpittsburghBeHTwo}{6.2}
\newcommand{\DispersionibmpittsburghBeHTwoOneZeroTwoFour}{5.6}
\newcommand{\PilotOxygenibmaachenTREXFull}{-236.17}
\newcommand{\PilotOxygenibmaachenCancellation}{19.48}
\newcommand{\OxygenTREXibmpittsburghMin}{187}
\newcommand{\OxygenTREXibmpittsburghMax}{207}
\newcommand{\OxygenTREXCostibmpittsburgh}{2.50}
\newcommand{\DurationLiH}{0.304}
\newcommand{\DurationOTwo}{1.416}
\newcommand{\Contrastfprod}{0.719}
\newcommand{\Contrastfprodidle}{0.573}
\newcommand{\ContrastObserved}{0.422}
\newcommand{\BudgetRatioAfterMin}{0.95}
\newcommand{\BudgetRatioBeforeMax}{2.59}
\newcommand{\BudgetRatioAfterMax}{1.78}
\newcommand{\BerylliumWindowRange}{249.59}
\newcommand{\TREXSmixed}{0.014759}
\newcommand{\TREXSreversed}{0.014625}
\newcommand{\TREXSisolated}{0.0037199}
\newcommand{\TREXShealthypair}{0.0027285}
\newcommand{\TREXSInflationisolated}{4}
\newcommand{\TREXSInflationhealthypair}{5.4}
\newcommand{\JobCount}{110}
\newcommand{\BenchmarkCount}{122}
\newcommand{\EnergyControlCount}{28}
\newcommand{\PauliControlCount}{168}
\newcommand{\ChargeSeconds}{1694}
\newcommand{\ChargeMinutes}{28.23}
\newcommand{\ChargeMinutesAbstract}{28.2}
\newcommand{\DiagnosticCount}{196}
\newcommand{\PilotJobs}{10}
\newcommand{\PilotEnergies}{38}
\newcommand{\IsolatedJobs}{84}
\newcommand{\IsolatedEnergies}{84}
\newcommand{\KnownJobs}{16}
\newcommand{\SpectrumibmaachenOTwotrexoutsidespectrummha}{349.50}
\newcommand{\SpectrumibmaachenOTwotrexoutsidespectrumoverreportedsigma}{21}
\newcommand{\SpectrumibmpittsburghHTwotrexoutsidespectrummha}{186.68}
\newcommand{\SpectrumibmpittsburghHTwotrexoutsidespectrumoverreportedsigma}{17}
\newcommand{\SpectrumibmpittsburghLiHtrexoutsidespectrummha}{0.58}
\newcommand{\SpectrumibmpittsburghLiHtrexoutsidespectrumoverreportedsigma}{0.24}
\newcommand{\Assignmentibmaachenidealx}{0.224771}
\newcommand{\Assignmentibmaachenidealz}{0.974412}
\newcommand{\Assignmentibmaachenassignmentonlypredictedresidualmha}{59.26}
\newcommand{\Assignmentibmpittsburghassignmentonlypredictedresidualmha}{-0.42}
\newcommand{\AnomalousReadoutPercent}{4.05}
\newcommand{\BondHTwo}{0.7428}
\newcommand{\BondLiH}{1.6204}
\newcommand{\BondOTwo}{1.2160}
\newcommand{\ReplayMantissa}{2.385}
\newcommand{\LinearZNEAmplification}{1.21}
\newcommand{\ShotsLow}{1024}
\newcommand{\ShotsHigh}{4096}
\newcommand{\ShotsControl}{2048}
\newcommand{\PilotCells}{22}
\newcommand{\IsolatedCells}{22}
\newcommand{\CrossoverCells}{44}
\newcommand{\IdealRepeats}{128}
\newcommand{\LiHExactHits}{123}
\newcommand{\LiHEstimatedHits}{114}
\newcommand{\IdealTotalBudget}{48128}
\newcommand{\ShotContrasts}{18}
\newcommand{\InspectedContrasts}{48}
\newcommand{\LocalSamplingCells}{10}
\newcommand{\OxygenChargeRawMin}{6}
\newcommand{\OxygenChargeMoreShotsMin}{17}
\newcommand{\OxygenChargeTREXMin}{15}
\newcommand{\OxygenChargeZNEMin}{24}
\newcommand{\OxygenChargeZNEMax}{25}
\newcommand{\QuotedRate}{96}
\newcommand{\AnsatzLiHUCCDlogicalcxdepth}{11}
\newcommand{\AnsatzLiHUCCSDlogicalcxdepth}{12}
\newcommand{\AnsatzHTwoOUCCDlogicalcxdepth}{274}
\newcommand{\AnsatzHTwoOUCCSDlogicalcxdepth}{290}
\newcommand{\HardwareInclusionLow}{6}
\newcommand{\HardwareCountLow}{6}
\newcommand{\HardwareInclusionHigh}{1}
\newcommand{\HardwareCountHigh}{3}
\newcommand{\PairedibmaachenMean}{2.40}
\newcommand{\HReconstructionOneTwoFourCI}{-0.330355}
\newcommand{\HReconstructionOneTwoFourCZ}{-0.786251}
\newcommand{\HReconstructionOneTwoFourCX}{-0.181367}
\newcommand{\HReconstructionOneTwoFourX}{-0.0859}
\newcommand{\HReconstructionOneTwoFourZ}{0.3613}
\newcommand{\HReconstructionOneTwoFourPredicted}{538.4}
\newcommand{\HReconstructionOneTwoFourObserved}{528.5}
\newcommand{\HReconstructionOneTwoFourS}{16.6}
\newcommand{\HReconstructionOneTwoFourDifference}{9.9}
\newcommand{\HReconstructionTwoSevenObserved}{1.0}
\newcommand{\AssignmentOneTwoFourZB}{0.3242}
\newcommand{\AssignmentOneTwoFourZA}{0.1313}
\newcommand{\AssignmentOneTwoFourZFactor}{5.6}
\newcommand{\AssignmentOneTwoFourXB}{0.3022}
\newcommand{\AssignmentOneTwoFourXA}{0.1338}
\newcommand{\AssignmentOneTwoFourXFactor}{5.4}
\newcommand{\TreatmentibmaachenOTwozneOneZeroTwoFourPercent}{49.7}
\newcommand{\TreatmentibmpittsburghOTwozneOneZeroTwoFourPercent}{83.5}
\newcommand{\ZNEGainReturnedMin}{3.2}
\newcommand{\ZNEGainReturnedMax}{6.6}
\newcommand{\PilotrawJobs}{6}
\newcommand{\PilotrawEnergies}{26}
\newcommand{\PilottrexJobs}{2}
\newcommand{\PilottrexEnergies}{8}
\newcommand{\PilotzneJobs}{2}
\newcommand{\PilotzneEnergies}{4}
\newcommand{\ThirdAccuracyLow}{1.9}
\newcommand{\ThirdAccuracyHigh}{430}
\newcommand{\ThirdModelLiHMin}{0.15}
\newcommand{\ThirdModelLiHMax}{0.31}
\newcommand{\ThirdModelOTwoMin}{1.1}
\newcommand{\ThirdModelOTwoMax}{2.2}
\newcommand{\ThirdModelBeHTwoMin}{16}
\newcommand{\ThirdModelBeHTwoMax}{24}
\newcommand{\ThirdGainATREX}{13}
\newcommand{\ThirdGainAZNE}{15}
\newcommand{\ThirdGainPTREX}{80}
\newcommand{\ThirdGainPZNE}{50}
\newcommand{\ThirdPilotOxygenFullRaw}{256}
\newcommand{\ThirdPilotOxygenFullTREX}{236}
\newcommand{\ThirdInflationTREXMin}{0.99}
\newcommand{\ThirdInflationTREXMax}{1.3}
\newcommand{\ThirdInflationZNEMin}{1.7}
\newcommand{\ThirdInflationZNEMax}{3.4}
\newcommand{\ThirdPilotPredMin}{0.81}
\newcommand{\ThirdPilotPredMax}{3.4}
\newcommand{\ThirdPilotUnder}{8}
\newcommand{\ThirdPilotCount}{9}
\newcommand{\ThirdFaultFactor}{15.0}
\newcommand{\ThirdEnergyDenominator}{150}
\newcommand{\ThirdPauliDenominator}{168}
\newcommand{\ThirdReadoutMedian}{0.0059}
\newcommand{\ThirdReadoutPercentile}{95}
\newcommand{\FourthCrossLiHMean}{3.3}
\newcommand{\FourthCrossLiHSE}{0.2}
\newcommand{\FourthCrossOxygenMean}{-130.7}
\newcommand{\FourthCrossOxygenSE}{11.3}
\newcommand{\FourthCrossBerylliumMean}{-233.4}
\newcommand{\FourthCrossBerylliumSE}{79.8}
\newcommand{\FourthStartOffsetMax}{1.0}
\newcommand{\FourthDurationMin}{6.5}
\newcommand{\FourthDurationMax}{9.8}
\newcommand{\FourthBracketMin}{-118.2}
\newcommand{\FourthBracketMax}{73.5}
\newcommand{\FourthBracketCount}{18}
\newcommand{\FourthNonBaselineCells}{16}
\newcommand{\FourthWindowCount}{3}
\newcommand{\FourthLocalSystems}{5}
\newcommand{\FourthLocalShotSettings}{2}
\newcommand{\FourthFidelityLossMin}{0.75}
\newcommand{\FourthFidelityLossMax}{2.22}
\newcommand{\FourthPilotCalPittsburghBeH}{0.703}
\newcommand{\FourthLithiumTREX}{1.57}
\newcommand{\FourthFullLithiumTREX}{22.1}
\newcommand{\FourthMaxViolation}{581.1}
\newcommand{\FourthMaxViolationS}{36.7}
\newcommand{\FourthBenchmarkTREXOutside}{3}
\newcommand{\FourthBenchmarkZNEOutside}{2}
\newcommand{\FourthControlTREXOutside}{8}
\newcommand{\FourthControlRawJobsEach}{3}
\newcommand{\FourthControlTREXJobsEach}{5}
\newcommand{\FourthReadoutTwoSeven}{0.001587}
\newcommand{\FourthSXOneTwoFour}{8.68e-05}
\newcommand{\FourthSXTwoSeven}{1.62e-04}
\newcommand{\FourthResetOneTwoFour}{1.30}
\newcommand{\FourthResetTwoSeven}{0.87}
\newcommand{\FourthBasisAverageDifference}{0.010}
\newcommand{\FifthBenchmarkJobs}{94}
\newcommand{\FifthRawNonFaultMaxAbs}{0.05}
\newcommand{\FifthRawNonFaultMaxS}{2.4}
\newcommand{\FifthTREXMaxEstimate}{1.7402}
\newcommand{\FifthTREXMaxS}{0.0146}
\newcommand{\FifthTREXPauliDenominator}{96}
\newcommand{\FifthMaxAbsPauli}{1.74}
\newcommand{\FifthTREXPauliOutside}{31}
\newcommand{\FifthTREXPauliOutsideAachen}{18}
\newcommand{\FifthTREXPauliOutsidePittsburgh}{13}
\newcommand{\FifthIdealHTwoZ}{0.9744}
\newcommand{\FifthIdealHTwoX}{0.2248}
\newcommand{\FifthAffinePrediction}{549.0}
\newcommand{\FifthInclusionRawLowJobs}{36}
\newcommand{\FifthInclusionRawLowTwoS}{7}
\newcommand{\FifthInclusionRawHighJobs}{18}
\newcommand{\FifthInclusionRawHighTwoS}{1}
\newcommand{\FifthInclusionTREXJobs}{18}
\newcommand{\FifthInclusionTREXTwoS}{4}
\newcommand{\FifthInclusionZNEJobs}{12}
\newcommand{\FifthInclusionZNETwoS}{10}
\newcommand{\FifthBracketPositives}{7}
\newcommand{\FifthBracketSignP}{0.48}
\newcommand{\FifthProductRatioPercent}{6.2}
\newcommand{\FifthBelowUnityLiH}{0.75}
\newcommand{\FifthBelowUnityOTwo}{0.99}
\newcommand{\FifthPittAnovaP}{0.001436}
\newcommand{\FifthPittAnovaAdjustedP}{0.008617}
\newcommand{\FifthZNEBarScatterMin}{18.3}
\newcommand{\FifthZNEBarScatterMax}{27.4}
\newcommand{\FifthZNEInflationAachenOTwo}{2.3}
\newcommand{\FifthZNEInflationPittsburghOTwo}{3.4}
\newcommand{\FifthReadoutRatio}{25.5}
\newcommand{\FifthReadoutOneTwoFour}{0.040527}
\newcommand{\FifthRawNonFaultZeroZ}{0.01}

\newcommand{\ZenodoAccession}{\href{https://doi.org/10.5281/zenodo.23739731}{10.5281/zenodo.23739731}}

\newcommand{\papertitle}{Execution bias, repeatability and mitigation cost in parameter-frozen molecular VQE energy evaluation}
\newcommand{\authoraddress}{Fraunhofer Institute for Manufacturing Technology and Advanced Materials IFAM, Wiener Strasse 12, 28359 Bremen, Germany}
\newcommand{\authorcontact}{\href{mailto:julen.larrucea@ifam.fraunhofer.de}{julen.larrucea@ifam.fraunhofer.de}; ORCID: \href{https://orcid.org/0000-0003-0582-9259}{0000-0003-0582-9259}}
\hypersetup{pdftitle={Execution bias, repeatability and mitigation cost in parameter-frozen molecular VQE energy evaluation},pdfauthor={Julen Larrucea}}

\title{\papertitle}
\author{Julen Larrucea\\{\small\parbox{.85\textwidth}{\centering\authoraddress}}\\{\small\authorcontact}}
\date{}
\begin{document}
\maketitle
\begin{abstract}A smaller statistical error bar need not mean a more accurate molecular energy. We evaluate noiselessly optimized molecular variational quantum eigensolver states on IBM Aachen and Pittsburgh, two 156-qubit Heron processors (provider-reported revision 3). Only energy evaluation runs on hardware, with parameters, layouts and compiled circuits frozen and dynamical decoupling disabled. A five-molecule pilot, three-molecule isolated cohort in three windows and known-state controls comprise \JobCount{} jobs and \ChargeMinutesAbstract{} billed quantum processing unit (QPU) minutes; the isolated cohort spans 30.2 h within a 31.7 h campaign. Raw \ShotsLow{} O$_2$ and BeH$_2$ mean residuals exceed returned uncertainty by factors \RawRatioMin--\RawRatioMax. Pittsburgh BeH$_2$ cross-job standard deviation is \DispersionibmpittsburghBeHTwoOneZeroTwoFour{} times mean returned uncertainty. The zero-noise extrapolation (ZNE) package reduces O$_2$ absolute residuals by \TreatmentibmaachenOTwozneOneZeroTwoFourPercent\% on Aachen and \TreatmentibmpittsburghOTwozneOneZeroTwoFourPercent\% on Pittsburgh at about fourfold charge. On Pittsburgh, twirled readout error extinction (TREX) removes \ThirdGainPTREX{} millihartree (mHa) per Raw \ShotsLow{} job charge versus \ThirdGainPZNE{}~mHa for ZNE. Controls reveal qubit-124 preparation or readout errors \AssignmentOneTwoFourXFactor{}--\AssignmentOneTwoFourZFactor{} times calibration; \EnergyViolationCount{} mitigated energy point estimates fall outside the mapped spectrum, with no Raw counterpart. These results support workload checks and repeated measurements before additional sampling.
\end{abstract}
\section{Introduction}
The variational quantum eigensolver (VQE) couples circuit preparation and energy measurement to classical optimization~\cite{peruzzo2014}. An inaccurate energy leaves a practical allocation problem: should additional quantum processing unit (QPU) time buy more shots, mitigation or independent repetitions? A per-job error bar cannot distinguish electronic-model approximations, ansatz restrictions and execution error. One millihartree (mHa) is $10^{-3}E_h$, where $E_h$ is the hartree. We optimize parameters noiselessly, freeze the parameters, physical layouts and instruction-set architecture (ISA) circuits, and run only the energy-evaluation step on hardware. This is not a hardware VQE: no optimization iterates on the device. The fixed-state reference lets us measure execution error at that step.

A Primitive Unified Bloc (PUB) pairs a circuit with its observables and parameters. Our five-molecule pilot shares jobs among molecular PUBs. The isolated cohort assigns one molecular PUB to each job and repeats LiH, O$_2$ and BeH$_2$ in three temporal windows. An arm is a treatment condition: Raw, twirled readout error extinction (TREX), or the zero-noise extrapolation (ZNE) package. A cell identifies a backend, system, arm and nominal shot setting. We use \emph{repeatability} for measurements of the same frozen workload on the same device at different times; cross-device comparisons test a different condition. The statistical analyses are exploratory.

Molecular ground-state preparation and mitigation benchmarks are established~\cite{mccaskey2019,kandala2017,kandala2019}. Otten and Gray separate algorithm errors by reducing individual error sources~\cite{otten2019}, while Huggins et al. reduce chemistry measurement work through low-rank factorization and particle-number post-selection~\cite{huggins2021}. H$_2$ precision studies discuss temporal variation~\cite{mihalikova2022}. Weaving et al. benchmark mitigated HCl energies from noiselessly optimized states and validate uncertainty using repeated diagonal terms~\cite{weaving2023}. Drift-detection and device-stability studies instead characterize changing processor noise~\cite{proctor2020,dasgupta2022,kim2025}. Here full molecular energies, frozen compiled circuits, cross-window dispersion and billed cost jointly support an allocation diagnostic and a physicality audit.

Variational reviews and noise-induced optimization limitations motivate separating energy evaluation from optimization~\cite{tilly2022,cerezo2021,mcclean2016,wang2021}. Primary extrapolation methods include continuous and digital noise amplification~\cite{temme2017,libenjamin2017,giurgica2020}. Learned cancellation and larger mitigated workloads~\cite{vandenberg2023,kim2023}, together with their classical simulation studies~\cite{tindall2024,begusic2024}, make both accuracy and cost relevant benchmarking endpoints~\cite{proctor2025}. Industrial chemistry requires substantially larger and more accurate electronic models than these diagnostic workloads~\cite{elfving2020}.

The author and co-workers' H$_2$ accuracy--cost benchmark varied workflow settings and compilation~\cite{larrucea2026h2}. The six-model optimizer-selection study remeasured selected points~\cite{larrucea2026selection}. Neither supplies a measurement here, and archived job identities distinguish the datasets. The present design adds two different Heron revision-3 processors, a frozen-ISA repeated-window cohort, known-state controls on the selected physical qubits, and a mapped-spectrum physicality audit, none of which appear in those studies.

\subsection{Contributions}
The following claims concern the measured circuits and acquisition design.
\begin{enumerate}
\item[(C1)] Isolated Raw \ShotsLow{} O$_2$ and BeH$_2$ mean residuals exceed mean returned uncertainty by factors \RawRatioMin--\RawRatioMax; window-based standard errors and the allocation calculation appear in Secs.~\ref{sec:hardware} and \ref{sec:allocation}.
\item[(C2)] Pittsburgh BeH$_2$ has a six-job sample standard deviation (SD) of \qty{\CellisolatedibmpittsburghBeHTworawOneZeroTwoFoursamplesdmha}{\millihartree} versus mean returned uncertainty \qty{\CellisolatedibmpittsburghBeHTworawOneZeroTwoFourmeanreturnedsmha}{\millihartree}, a factor of \DispersionibmpittsburghBeHTwoOneZeroTwoFour{} (Sec.~\ref{sec:hardware}).
\item[(C3)] Mitigation benefit depends on system and workload. Isolated O$_2$ TREX improves every window, whereas BeH$_2$ shows no mean benefit and the Aachen pilot O$_2$ response reverses sign (Secs.~\ref{sec:references} and \ref{sec:hardware}; Fig.~\ref{fig:references}(a)).
\item[(C4)] Known-state controls expose qubit-124 response errors, while \EnergyViolationCount{} of \ThirdEnergyDenominator{} energies and 31 of the 96 TREX Pauli control estimates fall outside mapped physical bounds as mitigated point estimates, with no Raw estimate outside those bounds (Sec.~\ref{sec:controls}).
\end{enumerate}

\section{Methods}
\subsection{Reference hierarchy and calibration model}
Let $F$ denote full-space full configuration interaction (FCI) energy in the stated basis and spin sector, $A$ the exact active-space energy, and $V$ the ideal energy of the frozen variational state. The execution estimate $Y$ is the energy returned by finite ideal sampling, raw hardware or mitigated hardware. With identical energy offsets,
\begin{equation}
Y-F=(A-F)+(V-A)+(Y-V).
\label{eq:hierarchy}
\end{equation}
The left-hand side is the signed difference between the reported energy and full-space FCI. On the right, $A-F$ measures the effect of restricting the orbital space, $V-A$ the error of the frozen variational state within that space, and $Y-V$ the difference introduced when its energy is estimated with finite sampling and the chosen hardware treatment. We denote this last difference by $r=Y-V$; for a cell of repeated jobs, the execution bias is its mean $\bar r$. Define the maximally mixed-state energy $E_{\rm mix}=\operatorname{Tr}(H)/2^{n_q}$ for the $n_q$-qubit tapered Hamiltonian with its offsets. For the calibration-product model, let $r_j$ be a native gate infidelity and $r_q^{\rm ro}$ a calibrated readout error. Then
\begin{equation}
f_{\rm gate}=\prod_{j\in{\rm gates}}(1-r_j),\quad
f_{\rm ro}=\prod_{q\in{\rm layout}}(1-r_q^{\rm ro}),\quad
P_{\rm prod}=(1-f_{\rm gate}f_{\rm ro})(E_{\rm mix}-V).
\label{eq:product}
\end{equation}
The product over two-qubit gate occurrences is $f_{2q}$. These quantities use the calibration snapshot matching each acquisition window.

\paragraph{Chemistry and reference states.}
All systems are neutral in STO-3G. Restricted Hartree--Fock (HF) singlets initialize H$_2$, LiH, BeH$_2$ and H$_2$O. Restricted open-shell HF (ROHF) initializes triplet O$_2$. Full FCI is spin checked, and active references diagonalize within fixed alpha and beta particle numbers and target $\langle S^2\rangle$. Geometries come from archived NIST Chemistry WebBook B3LYP/6-31G(d) structure files~\cite{nistwebbook}. The original rounded Cartesian coordinates are used without symmetry restoration; provenance details are given in the supplement. Their small bond-length asymmetries are compatible with coordinate rounding (Table~\ref{S-tab:geometries}).

HF states and one-repetition unitary coupled cluster doubles (UCCD) circuits use parity mapping with two-qubit reduction and additional $\mathbb Z_2$ tapering. The latter removes one qubit for H$_2$, none for LiH, two for O$_2$, and one for BeH$_2$ and H$_2$O in the selected HF sectors. Three seeded sequential least squares programming (SLSQP) starts minimize exact statevector expectations, followed by central-gradient stationarity checks and scans for single-parameter states (Sec.~\ref{S-sec:reference-methods}).

LiH uses the validated shallow 2e2o circuit; its \RefLiHactivetruncationmha{}~mHa active-space truncation dominates its error budget (Sec.~\ref{sec:references}). O$_2$, BeH$_2$ and H$_2$O use the 4e4o frontier active spaces of Refs.~\cite{larrucea2026h2,larrucea2026selection}. Matching a reconstructed ROHF orbital gauge to both archived integral tensors identifies O$_2$ indices 6--9 as one bonding $\pi_u$ component, two antibonding $\pi_g^*$ orbitals and an antibonding $\sigma_u^*$ orbital. The other degenerate $\pi_u$ component is frozen, so correlation treats the pair asymmetrically and the active space is not symmetry complete. Within the tapered space, the selected triplet sector and HF-connected block both have dimension \OxygenSectorDimension{}, allowing one doubles parameter to reach active-space FCI. This does not describe general O$_2$ multireference chemistry (Sec.~\ref{S-sec:orbital-audit}).

UCCD preserves the validated workload with small stationary gaps. A frozen-integral singles-and-doubles (UCCSD) comparison lowers LiH's gap below $10^{-5}$~mHa and H$_2$O's to \WaterSinglesGap{}~mHa with little extra logical depth (Table~\ref{S-tab:ansatz}).

These simplified workloads still permit an execution diagnostic: the residual $r=Y-V$ compares each measured energy with the ideal energy of the same frozen state.

\paragraph{Processors and execution options.}
The archived provider probe identifies both processors as 156-qubit Heron revision 3, with basis \texttt{cz}, \texttt{id}, \texttt{rz}, \texttt{sx}, \texttt{x}, \texttt{xslow}.\footnote{Family, revision and basis strings are quoted verbatim from the acquisition probe, including \texttt{xslow}.}
Parameters are bound before level-3 transpilation, with seed 271828 and fractional gates disabled for a common discrete ISA. Table~\ref{tab:circuits} gives frozen resources and calibration regimes.

\begin{table*}[!htbp]\centering\small
\textbf{(a) Frozen resources and pilot residuals}\par\smallskip
\resizebox{\linewidth}{!}{\begin{tabular}{llllllll}
\toprule
Backend & System & Layout & CZ & Depth & $f_{2q}$ & Predicted & Observed\\
\midrule
ibm\_aachen & H2 & \num{124} & \num{0} & \num{4} & \num{1.000} & \num{32.8} & \num{493.3}\\
ibm\_aachen & LiH & 135, 139 & \num{2} & \num{11} & \num{0.999} & \num{5.5} & \num{8.8}\\
ibm\_aachen & O2 & 134, 135, 139, 155 & \num{11} & \num{39} & \num{0.992} & \num{70.7} & \num{142.3}\\
ibm\_aachen & BeH2 & 134, 139, 135, 155, 154 & \num{156} & \num{465} & \num{0.896} & \num{194.6} & \num{424.7}\\
ibm\_aachen & H2O & 134, 135, 139, 155, 154 & \num{199} & \num{506} & \num{0.869} & \num{405.8} & \num{1390.4}\\
ibm\_pittsburgh & H2 & \num{61} & \num{0} & \num{4} & \num{1.000} & \num{8.3} & \num{11.6}\\
ibm\_pittsburgh & LiH & 154, 153 & \num{2} & \num{11} & \num{0.998} & \num{3.4} & \num{3.7}\\
ibm\_pittsburgh & O2 & 153, 154, 155, 139 & \num{11} & \num{39} & \num{0.990} & \num{292.8} & \num{238.6}\\
ibm\_pittsburgh & BeH2 & 153, 155, 154, 139, 135 & \num{156} & \num{458} & \num{0.872} & \num{352.7} & \num{798.1}\\
ibm\_pittsburgh & H2O & 153, 154, 155, 139, 135 & \num{199} & \num{507} & \num{0.842} & \num{672.4} & \num{1503.6}\\
\bottomrule
\end{tabular}
}
\par\medskip\textbf{(b) Raw eigenstate controls on qubit 124}\par\smallskip
\begin{tabular}{lrrrr}
\toprule
Control & Expected & Returned & $s$ & Departure/$s$\\
\midrule
$|0\rangle:Z$ & \num{1} & \num{0.3516} & \num{0.0207} & \num{-31}\\
$|1\rangle:Z$ & \num{-1} & \num{-0.7373} & \num{0.0149} & \num{18}\\
$|+\rangle:X$ & \num{1} & \num{0.3955} & \num{0.0203} & \num{-30}\\
$|-\rangle:X$ & \num{-1} & \num{-0.7324} & \num{0.0150} & \num{18}\\
\bottomrule
\end{tabular}

\par\medskip\textbf{(c) Aachen H$_2$ calibration}\par\smallskip
\begin{tabular}{lllll}
\toprule
Qubit & $T_1$ (\unit{\micro\second}) & $T_2$ (\unit{\micro\second}) & Readout & SX error\\
\midrule
\num{124} & \num{192.51} & \num{307.38} & \num{0.040527} & \num{8.68e-05}\\
\num{27} & \num{286.58} & \num{272.66} & \num{0.001587} & \num{1.62e-04}\\
\bottomrule
\end{tabular}

\caption{(a) Native controlled-Z (CZ) count, depth, product $f_{2q}$ of two-qubit gate factors, predicted $P_{\rm prod}$ and observed pilot residual from $V$, in mHa. Layout lists physical qubits in logical-qubit order; Raw has three pilot jobs per system except H$_2$O (one). (b) Raw eigenstate controls on qubit 124 show returned values and $s$; the cohort design and qubits are specified in Methods. (c) The device median readout error is \ThirdReadoutMedian{}, and qubit 124 is at the \ThirdReadoutPercentile{}th percentile in the archived snapshot.}
\label{tab:circuits}\label{tab:knownstates}\end{table*}

Raw requests resilience level 0, with readout mitigation, gate twirling, measurement twirling and dynamical decoupling (DD) all disabled. Archived resolved options confirm that DD was disabled, matching the Runtime default; DD can suppress idle decoherence but adds pulses~\cite{pokharel2018,ezzell2023,ibm_ddoptions}. TREX requests level \RoundThreeTREXLevel{}, measurement twirling and readout mitigation~\cite{vandenberg2022}, with gate twirling off. The ZNE package requests level \RoundThreeZNELevel{}, both twirling modes, readout mitigation, factors $(1,3,5)$ and exponential extrapolation with linear fallback. Runtime chooses the first heuristically successful extrapolator and can differ by Pauli term~\cite{ibm_zne}. Mitigated nominal-\ShotsLow{} jobs resolve 32 randomizations with 32 shots each. Initialization uses a \qty{250}{\micro\second} repetition delay, \FourthResetOneTwoFour{} times qubit 124's $T_1$ and \FourthResetTwoSeven{} times qubit 27's. The common delay should reset qubit 124 more effectively, so it cannot by itself explain that qubit's larger $0\!\to\!1$ error. Qubit-specific excited-state initialization or a readout-discriminator offset remains an untested hypothesis.

The known-state cohort has \FourthControlRawJobsEach{} Raw and \FourthControlTREXJobsEach{} TREX jobs on each processor. Its $|0\rangle$, $|1\rangle$, $|+\rangle$ and $|-\rangle$ Pauli tests use the recorded single-qubit layouts: qubits 27, 38 and 124 on Aachen, and 61, 77, 112 and 114 on Pittsburgh. The cohort also includes H$_2$ state and energy controls. The \ThirdEnergyDenominator{} energy denominator combines benchmark and molecular control estimates; the \ThirdPauliDenominator{} Pauli denominator contains control estimates only. Pauli coverage is non-uniform across qubits and arms: the 168 control estimates comprise 96 TREX and 72 Raw estimates, 84 on each processor, and not every tested qubit contributes all four eigenstate tests in both arms. Table~\ref{S-tab:fifth-pauli} gives the per-qubit, per-arm counts.

Requested and returned options are archived separately: resolved resilience levels, learned calibration shots, complete executed-circuit counts and total shots were not returned (Sec.~\ref{S-sec:resolved-options}). ZNE was limited to LiH and O$_2$ to bound acquisition cost before amplifying deeper circuits.

\paragraph{Sampling and statistical analysis.}
For ideal qubit-wise commuting (QWC) groups $H_g$, $N$ independent shots per group give
\begin{equation}
\sigma^2=\sum_g v_g/N,\qquad v_g=\langle H_g^2\rangle-\langle H_g\rangle^2.
\label{eq:variance}
\end{equation}
This retains within-group covariance. At total budget $B$ and $G$ groups, equal allocation gives variance $G\sum_gv_g/B$ and continuous variance-optimal allocation gives variance $(\sum_g\sqrt{v_g})^2/B$~\cite{verteletskyi2020,rubin2018}. The square roots appear with the ideal results in Sec.~\ref{sec:references}.  Runtime resolves a different measurement and randomization workload, so nominal shots do not equate total work.

\paragraph{Interpretive conventions.}
Hardware bars show returned Estimator standard errors. Three windows describe repeatability without fitting a drift law.

The Estimator \texttt{stds} field, denoted $s$, is a returned standard-error estimate. With gate twirling it reflects randomization variation, and with ZNE it describes an extrapolator at zero noise~\cite{ibm_estimator}. We reserve $\sigma$ for exact ideal SDs. SE$_w$ is the SD of three window means divided by $\sqrt3$; SE$_R={\rm SE}_w/\bar s$ is the window-mean standard error in units of mean returned uncertainty. Statistics use full-precision records; the displayed table rounds each column. We group by window before forming SE$_w$ so that two closely spaced Raw jobs do not count as independent observations of the between-window variation. For cells with one job per window this reduces to the ordinary three-observation standard error. The exploratory F test assumes independent Gaussian within-window errors with a common variance. Hardware inclusion counts summarize $|Y-V|\leq ks$ for $k=1$ and $2$.

The calibration comparisons use separately defined gate-depolarizing-plus-assignment, on-gate thermal and idle-gap thermal models (Sec.~\ref{S-sec:attribution}). Complete provider rotations and bitstrings are unavailable; individual-bitstring symmetry post-selection~\cite{bonet2018} cannot be reconstructed from these expectation-value payloads.

Experiment control adapts routines from the preceding chemistry studies~\cite{larrucea2026h2,larrucea2026selection}. Frozen inputs, Qiskit's QPY binary circuit files, resolved metadata and SHA-256 identities support offline replay. The pinned environment includes Python 3.12, Qiskit 2.5.2, Qiskit Nature 0.7.2, \texttt{qiskit-ibm-runtime} 0.49.0, PySCF 2.10 and SciPy 1.17.1. All numerical outputs are programmatically checked against archived inputs. AI assistance is disclosed in the Acknowledgements.

\section{Reference hierarchy and ideal sampling}\label{sec:references}
Eq.~(\ref{eq:hierarchy}) separates active-space truncation, the stationary ansatz gap and the execution residual. Basis incompleteness remains an additional error relative to the physical molecule. Table~\ref{tab:references} gives the corresponding reference energies and $E_{\rm mix}$.

\begin{table*}[!htbp]\centering\small
\resizebox{\linewidth}{!}{\begin{tabular}{lrrrrr}
\toprule
Quantity & H2 & LiH & O2 & BeH2 & H2O\\
\midrule
Bonds (\unit{\angstrom}) & \num{0.7428} & \num{1.6204} & \num{1.2160} & 1.3294, 1.3295 & 0.9685, 0.9686\\
Angle ($^\circ$) & -- & -- & -- & \num{179.944} & \num{103.601}\\
Active orbitals & 0, 1 & 1, 2 & 6, 7, 8, 9 & 1, 2, 3, 4 & 3, 4, 5, 6\\
Qubits & \num{1} & \num{2} & \num{4} & \num{5} & \num{5}\\
Parameters & \num{1} & \num{1} & \num{1} & \num{8} & \num{8}\\
Pauli terms & \num{3} & \num{9} & \num{50} & \num{119} & \num{155}\\
QWC groups & \num{2} & \num{4} & \num{14} & \num{36} & \num{47}\\
$F$ (\unit{\hartree}) & \num{-1.1372533} & \num{-7.8819358} & \num{-147.7464826} & \num{-15.5951273} & \num{-75.0156066}\\
$A$ (\unit{\hartree}) & \num{-1.1372533} & \num{-7.8614080} & \num{-147.6331577} & \num{-15.5660242} & \num{-74.9723181}\\
$V$ (\unit{\hartree}) & \num{-1.1372533} & \num{-7.8613924} & \num{-147.6331577} & \num{-15.5660235} & \num{-74.9722048}\\
$E_{mix}$ (\unit{\hartree}) & \num{-0.330355} & \num{-7.614969} & \num{-145.919457} & \num{-14.379967} & \num{-72.918692}\\
$A-F$ (\unit{\millihartree}) & $<10^{-5}$ & \num{20.53} & \num{113.32} & \num{29.10} & \num{43.29}\\
$V-A$ (\unit{\millihartree}) & $<10^{-5}$ & \num{0.0156} & $<10^{-5}$ & \num{0.0007} & \num{0.113}\\
\bottomrule
\end{tabular}
}
\caption{STO-3G model definitions and references. Energies $F$, $A$, $V$ and $E_{\rm mix}$ include common offsets. Bonds and angles are calculated from the archived NIST Chemistry WebBook Cartesian coordinates optimized at B3LYP/6-31G(d)~\cite{nistwebbook}; orbital indices are zero based, and occupied orbitals outside the active space are frozen. Nuclear and inactive-orbital offsets and full Cartesians are in Tables~\ref{S-tab:geometries} and \ref{S-tab:energy-offsets}; gaps below $10^{-5}$~mHa are numerically unresolved.}
\label{tab:references}\end{table*}

\begin{figure*}[!htbp]\centering
\includegraphics[width=\linewidth]{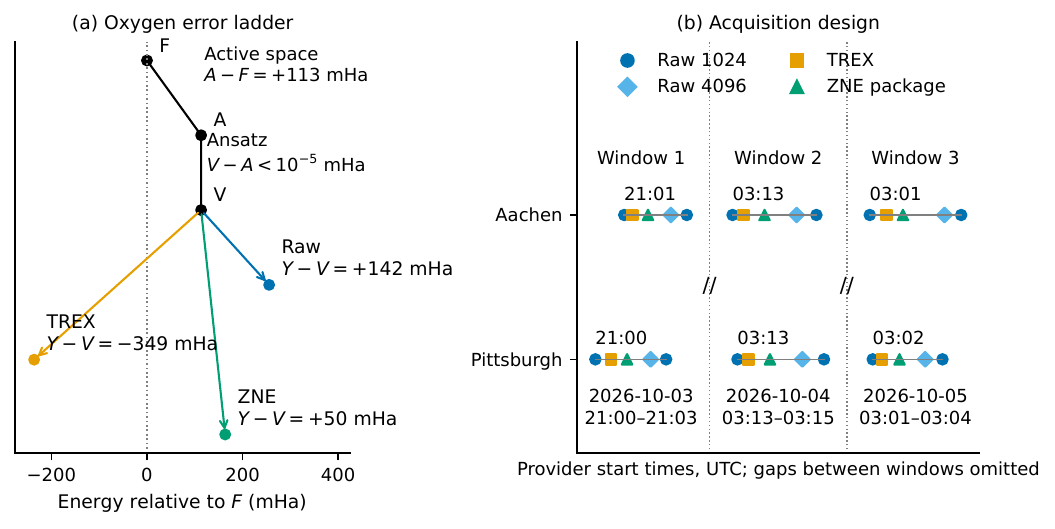}
\caption{(a) Aachen pilot O$_2$ energy ladder relative to full-space $F$: $A$ is the exact active reference, $V$ the frozen ideal energy, black links the active-space and ansatz contributions, and colored arrows the execution residual $Y-V$. Raw is a three-job mean and each mitigation estimate is one PUB; cancellation reduces $|Y-F|$ from \ThirdPilotOxygenFullRaw{} to \ThirdPilotOxygenFullTREX{}~mHa under TREX even as $|Y-V|$ grows. (b) Start times show the O$_2$ subset of each isolated window on Aachen and Pittsburgh; each complete device window has 14 jobs over \FourthDurationMin--\FourthDurationMax{}~min. Window gaps are omitted, order was fixed, and BeH$_2$ had no ZNE arm.}
\label{fig:references}\end{figure*}

The largest active-space error is O$_2$'s \qty{\RefOTwoactivetruncationmha}{\millihartree}. Figure~\ref{fig:references}(a) shows how a negative mitigation residual can cancel positive model error and produce a misleadingly closer energy to $F$. The full arithmetic is in Sec.~\ref{S-sec:oxygen-cancellation}. We use $1$~kcal/mol $=1.594$~mHa as a numerical reference scale for energy residuals; this comparison does not establish chemical accuracy for relative energies or chemical observables. At the same H$_2$ geometry, the STO-3G to cc-pVTZ FCI difference is \qty{\BasisGap}{\millihartree}, \BasisRatio{} times the \qty{\BasisTZGap}{\millihartree} cc-pVTZ--cc-pVQZ gap; no complete-basis extrapolation is attempted.

\begin{table*}[!htbp]\centering\small
\resizebox{\linewidth}{!}{\begin{tabular}{lrrrrrr}
\toprule
System & QWC groups & Measurements at $N=1024$ & $\sigma_{1024}$ & $\sigma_{4096}$ & $\sigma_{B,\mathrm{equal}}$ & $\sigma_{B,\mathrm{opt}}$\\
\midrule
H$_2$ & \num{2} & \num{2048} & \num{7.810} & \num{3.905} & \num{1.611} & \num{1.611}\\
LiH & \num{4} & \num{4096} & \num{0.588} & \num{0.294} & \num{0.172} & \num{0.131}\\
O$_2$ & \num{14} & \num{14336} & \num{4.181} & \num{2.091} & \num{2.282} & \num{1.685}\\
BeH$_2$ & \num{36} & \num{36864} & \num{4.230} & \num{2.115} & \num{3.702} & \num{2.783}\\
H$_2$O & \num{47} & \num{48128} & \num{5.621} & \num{2.811} & \num{5.621} & \num{4.363}\\
\bottomrule
\end{tabular}
}
\caption{Exact ideal standard deviations $\sigma$, in mHa, for QWC measurements at $V$. Equal per-group shots and equal or variance-optimal allocation at common total budget $B=\num{\IdealTotalBudget}$ define local ideal baselines. This equals H$_2$O's own 47 groups times \ShotsLow{} shots; its equal-allocation result therefore matches its nominal-shot result. Runtime uses a different measurement policy. H$_2$'s two group variances coincide, making its two fixed-budget columns identical.}
\label{tab:sampling}\end{table*}

In Table~\ref{tab:sampling}, at equal total budget, the ideal H$_2$O and H$_2$ sampling SDs are \IdealHTwoOOneZeroTwoFourfixedtotalequalsigmamha{} and \qty{\IdealHTwoOneZeroTwoFourfixedtotalequalsigmamha}{\millihartree}. The local repetitions use the same frozen states and ideal group probabilities throughout. Their spread validates the sampling calculation; it does not predict a hardware error bar, whose state variances and measurement policy can differ. For LiH, replacing estimated group variances with exact variances raises two-uncertainty inclusion from \LiHEstimatedHits{}/\IdealRepeats{} to \LiHExactHits{}/\IdealRepeats{} in the same simulations (Sec.~\ref{S-sec:additional}; Fig.~\ref{S-fig:sampling}).

\section{Hardware repeatability and intervention cost}\label{sec:hardware}
\paragraph{Baseline and design.}
The isolated cohort tests shallow LiH and O$_2$ circuits and a deeper BeH$_2$ circuit. Each window has two Raw \ShotsLow{} jobs bracketing TREX, the ZNE package (gate twirling, noise amplification and extrapolation), and Raw \ShotsHigh{} measurements; BeH$_2$ ZNE was not run. The first provider start to final finish spans \qty{\IsolatedSpanHours}{\hour}. Window starts span \qty{\WindowStartSpanHours}{\hour}, separated by \WindowGapOne{} and \qty{\WindowGapTwo}{\hour} (Table~\ref{S-tab:timing}). All arms reuse the pilot's bound ISA circuit and mapped observable. Fixed order confounds treatment with time within the window.

The repeated design holds the chemical state and compiled workload fixed while allowing the device response to change. Bracketing raw jobs estimate the local baseline around each intervention. Within-window disagreement tests short-term scatter; changes between window means test repeatability over the acquisition span. These endpoints answer different resource questions, so the analysis keeps their variation separate.

The first O$_2$ Raw jobs on the two devices start within \FourthStartOffsetMax{}~min in each matched window. Aachen minus Pittsburgh window-matched mean residuals are \FourthCrossLiHMean{}~mHa for LiH (SE$_w$ \FourthCrossLiHSE{}~mHa), \FourthCrossOxygenMean{}~mHa for O$_2$ (SE$_w$ \FourthCrossOxygenSE{}~mHa), and \FourthCrossBerylliumMean{}~mHa for BeH$_2$ (SE$_w$ \FourthCrossBerylliumSE{}~mHa). Here SE$_w$ is the SD of the three paired device differences divided by $\sqrt3$. These device contrasts share temporal windows and use different physical layouts.

\begin{figure*}[!htbp]\centering
\includegraphics[width=\linewidth]{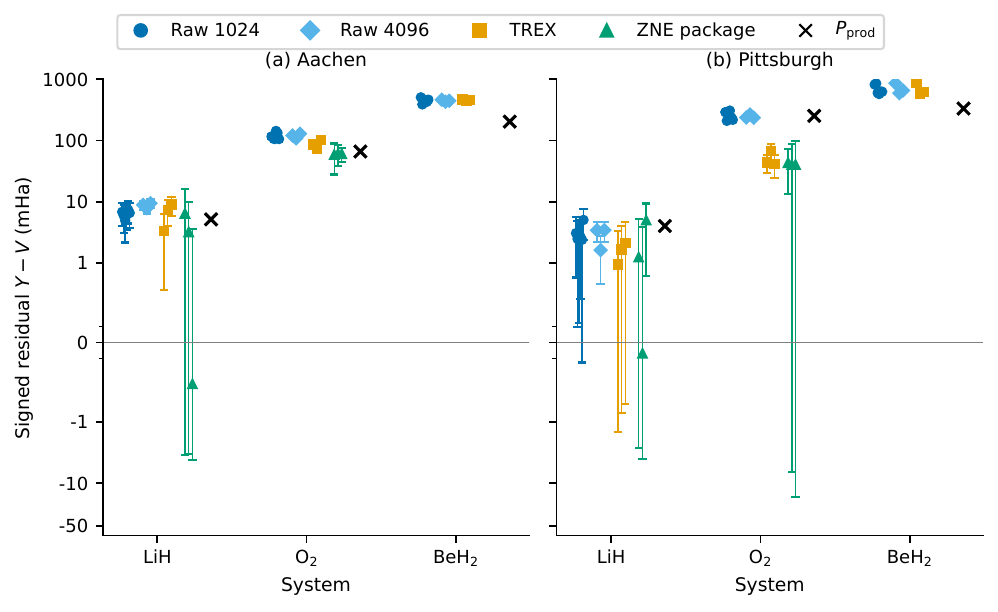}
\caption{Individual isolated residuals $Y-V$ on (a) Aachen and (b) Pittsburgh, with returned Estimator standard errors $s$. Markers identify Raw \ShotsLow{}, Raw \ShotsHigh{}, TREX and the ZNE package; black crosses show the calibration prediction $P_{\rm prod}$ from Eq.~(\ref{eq:product}) averaged over matching windows. The signed logarithmic scale is linear within $\pm0.3$~mHa. Raw \ShotsLow{} has six jobs per system and device; other supported cells have three, with BeH$_2$ ZNE unmeasured.}
\label{fig:followup}\end{figure*}

\paragraph{Bias versus returned uncertainty.}
Figure~\ref{fig:followup} compares individual residuals with the calibration model of Eq.~(\ref{eq:product}). All raw cell means are positive. Mixing the state toward $E_{\rm mix}>V$ predicts that sign, although it does not identify a depolarizing channel. Define the retained contrast
\begin{equation}
f_{\rm obs}=\frac{E_{\rm mix}-Y}{E_{\rm mix}-V}
=1-\frac{Y-V}{E_{\rm mix}-V},\qquad f_{\rm cal}=f_{\rm gate}f_{\rm ro}.
\label{eq:contrast}
\end{equation}
Pittsburgh BeH$_2$ gives $f_{\rm obs}=\ContrastObserved{}$ versus $f_{\rm cal}=\Contrastfprod{}$, an approximately twofold difference in the lost contrast. The $f_{\rm cal}$ value averages the isolated-window snapshots; the pilot Pittsburgh BeH$_2$ snapshot gives $f_{\rm cal}=\FourthPilotCalPittsburghBeH{}$, whose predicted residual $P_{\rm prod}$ appears in Table~\ref{tab:circuits}. The per-window values appear in Tables~\ref{S-tab:products} and \ref{S-tab:calibration-model}. Tapering removes symmetry sectors, so this reduced-space contrast is not a device fidelity (Fig.~\ref{S-fig:contrast-scatter}).

Across six Raw \ShotsLow{} cells, $(1-f_{\rm obs})/(1-f_{\rm cal})$ ranges from \FourthFidelityLossMin{} to \FourthFidelityLossMax{} (Table~\ref{S-tab:fourth-fidelity}). Pittsburgh LiH (\FifthBelowUnityLiH{}) and O$_2$ (\FifthBelowUnityOTwo{}) are the two cells below unity, where observed contrast loss is smaller than this calibration model predicts.

\begin{table*}[!htbp]\centering\scriptsize
\resizebox{\linewidth}{!}{\begin{tabular}{llrrrrrrrrrrrr}
\toprule
System & Arm & $n_{\rm job}$ & Mean & SE$_w$ & SD & $\bar s$ & SD/$\bar s$ & $R$ & SE$_R$ & $C$ & $\Delta r$ & $G$ & $|\bar r|/|A-F|$\\
\midrule
\multicolumn{14}{l}{\textbf{Aachen}}\\
BeH2 & Raw 1024 & \num{6} & \num{452} & \num{4.42} & \num{39.6} & \num{20.51} & \num{1.9} & \num{22} & \num{0.22} & \num{1.0} & \num{0.0} & \num{0.0} & \num{16}\\
BeH2 & Raw 4096 & \num{3} & \num{446} & \num{9.55} & \num{16.5} & \num{10.30} & \num{1.6} & \num{43} & \num{0.93} & \num{3.6} & \num{-5.6} & \num{1.6} & \num{15}\\
BeH2 & TREX & \num{3} & \num{457} & \num{8.08} & \num{14.0} & \num{20.54} & \num{0.68} & \num{22} & \num{0.39} & \num{1.9} & \num{5.5} & \num{-2.9} & \num{16}\\
LiH & Raw 1024 & \num{6} & \num{6.30} & \num{0.411} & \num{0.84} & \num{2.827} & \num{0.30} & \num{2.2} & \num{0.15} & \num{1.0} & \num{0.0} & \num{0.0} & \num{0.31}\\
LiH & Raw 4096 & \num{3} & \num{8.70} & \num{0.464} & \num{0.80} & \num{1.487} & \num{0.54} & \num{5.9} & \num{0.31} & \num{2.0} & \num{2.4} & \num{-1.2} & \num{0.42}\\
LiH & TREX & \num{3} & \num{6.51} & \num{1.68} & \num{2.91} & \num{3.104} & \num{0.94} & \num{2.1} & \num{0.54} & \num{4.0} & \num{0.2} & \num{-0.1} & \num{0.32}\\
LiH & ZNE package & \num{3} & \num{3.11} & \num{1.92} & \num{3.33} & \num{6.756} & \num{0.49} & \num{0.46} & \num{0.28} & \num{5.0} & \num{-3.2} & \num{0.6} & \num{0.15}\\
O2 & Raw 1024 & \num{6} & \num{119} & \num{3.04} & \num{15.6} & \num{9.996} & \num{1.6} & \num{12} & \num{0.30} & \num{1.0} & \num{0.0} & \num{0.0} & \num{1.1}\\
O2 & Raw 4096 & \num{3} & \num{119} & \num{6.10} & \num{10.6} & \num{4.956} & \num{2.1} & \num{24} & \num{1.2} & \num{2.8} & \num{0.0} & \num{0.0} & \num{1.1}\\
O2 & TREX & \num{3} & \num{87} & \num{7.83} & \num{13.6} & \num{10.71} & \num{1.3} & \num{8.1} & \num{0.73} & \num{2.5} & \num{-32.4} & \num{13.0} & \num{0.77}\\
O2 & ZNE package & \num{3} & \num{60.01} & \num{0.721} & \num{1.25} & \num{22.89} & \num{0.055} & \num{2.6} & \num{0.032} & \num{4.0} & \num{-59.2} & \num{14.8} & \num{0.53}\\
\multicolumn{14}{l}{\textbf{Pittsburgh}}\\
BeH2 & Raw 1024 & \num{6} & \num{685} & \num{77.2} & \num{120.4} & \num{21.64} & \num{5.6} & \num{32} & \num{3.6} & \num{1.0} & \num{0.0} & \num{0.0} & \num{24}\\
BeH2 & Raw 4096 & \num{3} & \num{704} & \num{80.9} & \num{140.2} & \num{10.79} & \num{13} & \num{65} & \num{7.5} & \num{3.4} & \num{18.9} & \num{-5.6} & \num{24}\\
BeH2 & TREX & \num{3} & \num{692} & \num{91.0} & \num{157.6} & \num{24.42} & \num{6.5} & \num{28} & \num{3.7} & \num{1.8} & \num{6.9} & \num{-3.9} & \num{24}\\
LiH & Raw 1024 & \num{6} & \num{3.02} & \num{0.348} & \num{1.01} & \num{2.464} & \num{0.41} & \num{1.2} & \num{0.14} & \num{1.0} & \num{0.0} & \num{0.0} & \num{0.15}\\
LiH & Raw 4096 & \num{3} & \num{2.82} & \num{0.601} & \num{1.04} & \num{1.192} & \num{0.87} & \num{2.4} & \num{0.50} & \num{2.0} & \num{-0.2} & \num{0.1} & \num{0.14}\\
LiH & TREX & \num{3} & \num{1.57} & \num{0.338} & \num{0.59} & \num{2.446} & \num{0.24} & \num{0.64} & \num{0.14} & \num{4.0} & \num{-1.5} & \num{0.4} & \num{0.076}\\
LiH & ZNE package & \num{3} & \num{2.05} & \num{1.51} & \num{2.62} & \num{4.070} & \num{0.64} & \num{0.50} & \num{0.37} & \num{5.0} & \num{-1.0} & \num{0.2} & \num{0.10}\\
O2 & Raw 1024 & \num{6} & \num{250} & \num{12.4} & \num{40.0} & \num{13.23} & \num{3.0} & \num{19} & \num{0.94} & \num{1.0} & \num{0.0} & \num{0.0} & \num{2.2}\\
O2 & Raw 4096 & \num{3} & \num{246} & \num{9.77} & \num{16.9} & \num{6.547} & \num{2.6} & \num{38} & \num{1.5} & \num{2.8} & \num{-3.8} & \num{1.3} & \num{2.2}\\
O2 & TREX & \num{3} & \num{51} & \num{8.53} & \num{14.8} & \num{17.14} & \num{0.86} & \num{3.0} & \num{0.50} & \num{2.5} & \num{-199.0} & \num{79.6} & \num{0.45}\\
O2 & ZNE package & \num{3} & \num{41.30} & \num{0.942} & \num{1.63} & \num{44.75} & \num{0.036} & \num{0.92} & \num{0.021} & \num{4.2} & \num{-208.6} & \num{50.1} & \num{0.36}\\
\bottomrule
\end{tabular}
}
\caption{Isolated residuals $r=Y-V$, in mHa: job count $n_{\rm job}$, mean, window-mean standard error SE$_w$, sample SD, mean returned uncertainty $\bar s$, bias ratio $R=|\bar r|/\bar s$, and SE$_R={\rm SE}_w/\bar s$, the window-mean standard error in units of mean returned uncertainty. Means are integers when cell SD exceeds 10~mHa and otherwise have two decimal places. Charge multiple $C$ is relative to one Raw 1024 job for the same system and backend; $\Delta r$ is the change in mean signed residual; $G=(|\bar r_{\rm raw}|-|\bar r_{\rm arm}|)/C$. The final column is $|\bar r|/|A-F|$. Ratios use full-precision observations and display two significant digits, including trailing zeros; $\Delta r$, $C$ and $G$ are shown to one decimal place; per-job ranges are in Table~\ref{S-tab:full-descriptive}.}
\label{tab:followup}\label{tab:efficiency}\end{table*}

Comparing mean Raw \ShotsLow{} execution residuals with $|A-F|$ separates limiting terms: $|\bar r|/|A-F|$ spans \ThirdModelLiHMin--\ThirdModelLiHMax{} for LiH, \ThirdModelOTwoMin--\ThirdModelOTwoMax{} for O$_2$, and \ThirdModelBeHTwoMin--\ThirdModelBeHTwoMax{} for BeH$_2$ (Table~\ref{tab:followup}). Active-space truncation is larger for LiH, while execution error dominates BeH$_2$. Over the ten pilot raw cells, the calibration product underpredicts \ThirdPilotUnder{} of the \ThirdPilotCount{} cells outside the anomalous Aachen H$_2$ cell of Sec.~\ref{sec:controls}, with observed-to-predicted ratios \ThirdPilotPredMin--\ThirdPilotPredMax{}; Aachen H$_2$ reaches \ThirdFaultFactor{} (Table~\ref{tab:circuits}).

Quadrupling nominal shots gives returned-uncertainty ratios \RoundThreeShotRatioMin--\RoundThreeShotRatioMax{} across \ShotContrasts{} contrasts, consistent with the returned estimator's local square-root scaling. Table~\ref{S-tab:paired-all} gives paired Raw \ShotsHigh{} minus bracketing Raw \ShotsLow{} contrasts for all six cells. Aachen LiH's mean Raw \ShotsHigh{} minus Raw \ShotsLow{} contrast is \qty{\PairedibmaachenMean}{\millihartree} and keeps the same sign in all three windows, small in absolute terms but systematic. 

The first-to-last Raw \ShotsLow{} differences within a window range from \FourthBracketMin{} to \FourthBracketMax{}~mHa over \FourthBracketCount{} matched pairs (Table~\ref{S-tab:fourth-brackets}). This measured range bounds observed baseline movement over the fixed treatment sequence; it does not isolate drift from sampling scatter. An exact two-sided sign test finds \FifthBracketPositives{} positive differences among \FourthBracketCount{} pairs ($p=\FifthBracketSignP{}$, descriptive only).

\begin{figure*}[!htbp]\centering
\includegraphics[width=\linewidth]{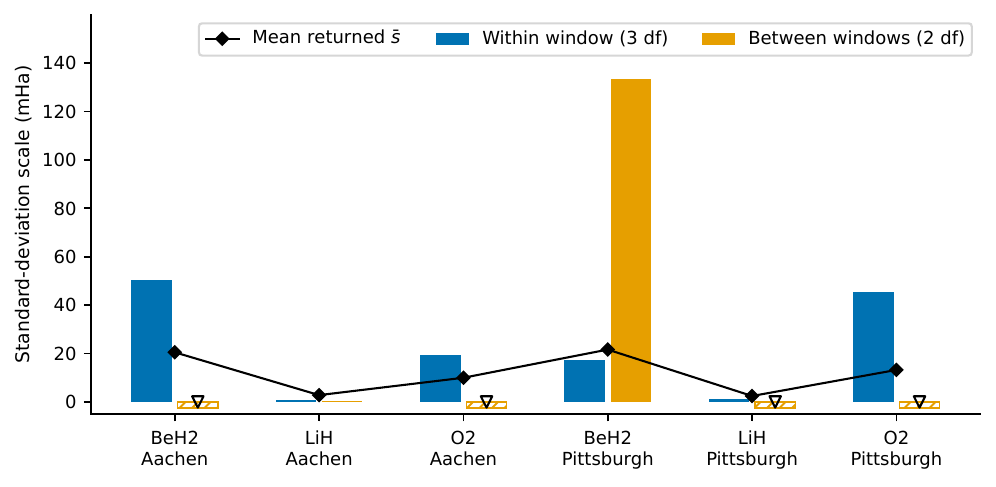}
\caption{Within-window and between-window moment SD components for six Raw \ShotsLow{} jobs in three windows, compared with mean returned uncertainty $\bar s$. Negative between-window moment variance estimates are displayed at zero with hatching and an open downward marker. The between-window component has two degrees of freedom; the within-window component has three.}
\label{fig:temporal}\end{figure*}

\paragraph{Repeatability.}
Pittsburgh BeH$_2$'s window means span \CellisolatedibmpittsburghBeHTworawOneZeroTwoFourwindowmeanminmha--\qty{\CellisolatedibmpittsburghBeHTworawOneZeroTwoFourwindowmeanmaxmha}{\millihartree}. Its within-window SD is \qty{\PittsburghBerylliumWithin}{\millihartree}, compared with a between-window component of \qty{\PittsburghBerylliumEpoch}{\millihartree}; Aachen instead has within-window SD \qty{\AachenBerylliumWithin}{\millihartree} and a negative untruncated between-window variance estimate (Fig.~\ref{fig:temporal}). An exploratory balanced F test gives Pittsburgh $F(2,3)=\num{\VaribmpittsburghBeHTwoF}$, $p=\FifthPittAnovaP{}$ and Bonferroni-adjusted $p=\FifthPittAnovaAdjustedP{}$, the only Bonferroni-adjusted rejection among six Raw \ShotsLow{} cells (Table~\ref{S-tab:fifth-anova}). The other tests have little power at these degrees of freedom.

Calibration models help locate missing contributions. For Aachen pilot H$_2$ the gate-depolarizing-plus-assignment model predicts \qty{\ModelibmaachenHTwopilotdensitymatrixlocalqwcresidualmha}{\millihartree}, and on-gate thermal relaxation gives \qty{\ModelibmaachenHTwopilotthermalgatereadoutresidualmha}{\millihartree}, against $P_{\rm prod}=\qty{\ModelibmaachenHTwopilotnaivedepolarizingresidualmha}{\millihartree}$. The observed residual is \qty{\CellpilotibmaachenHTworawOneZeroTwoFourmeanmha}{\millihartree}; all three models underpredict it by at least a factor of 8. An idle-gap thermal model predicts \RoundThreeThermalIdleMin--\qty{\RoundThreeThermalIdleMax}{\millihartree} for Pittsburgh BeH$_2$, below its observed window means (Table~\ref{S-tab:calibration-model}). Sec.~\ref{S-sec:duration-models} separates those channel models from a phenomenological total-duration attenuation that can double-count gate decoherence.

\begin{figure*}[!htbp]\centering
\includegraphics[width=\linewidth]{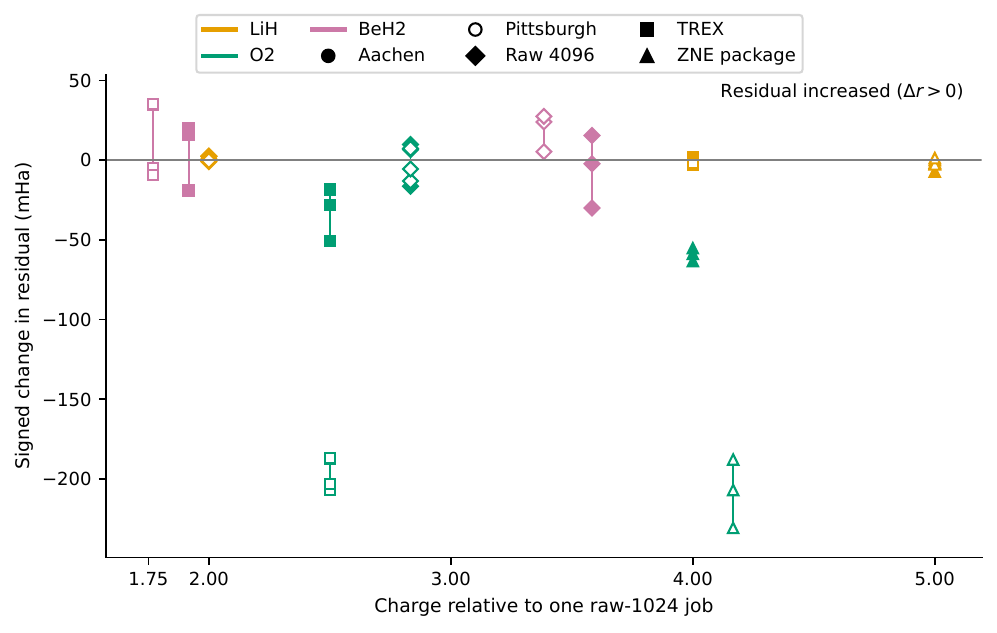}
\caption{Signed residual changes from $V$ relative to each window's mean Raw \ShotsLow{} bracket versus charge relative to one Raw \ShotsLow{} job of the same system and backend. Colors identify systems, filled markers Aachen and open markers Pittsburgh; diamonds, squares and triangles identify Raw \ShotsHigh{}, TREX and the ZNE package. Positive changes raise the signed residual; a negative change can still increase absolute error if the estimate crosses $V$. Each cell contributes three window points; per-cell and absolute-reduction views appear in Figs.~\ref{S-fig:cost-panels} and \ref{S-fig:absolute-benefit}.}
\label{fig:costbenefit}\end{figure*}

\paragraph{Mitigation and cost.}
Figure~\ref{fig:costbenefit} shows each window's signed change at its charge. TREX increases BeH$_2$ mean residuals on Aachen from \CellisolatedibmaachenBeHTworawOneZeroTwoFourmeanmha{} to \qty{\CellisolatedibmaachenBeHTwotrexOneZeroTwoFourmeanmha}{\millihartree}, and on Pittsburgh from \CellisolatedibmpittsburghBeHTworawOneZeroTwoFourmeanmha{} to \qty{\CellisolatedibmpittsburghBeHTwotrexOneZeroTwoFourmeanmha}{\millihartree}; window-level signs vary on both processors. An exact sign test over six window-level TREX changes gives $p=\BerylliumSignP{}$ (descriptive only). For O$_2$, ZNE has the lowest mean residual on both devices, reducing absolute error by \TreatmentibmaachenOTwozneOneZeroTwoFourPercent\% and \TreatmentibmpittsburghOTwozneOneZeroTwoFourPercent\% at approximately four times raw charge. Pittsburgh's ZNE residual \qty{\CellisolatedibmpittsburghOTwozneOneZeroTwoFourmeanmha}{\millihartree} is comparable to returned $\bar s=\CellisolatedibmpittsburghOTwozneOneZeroTwoFourmeanreturnedsmha{}$~mHa, but its gain over the raw bracket is \ZNEGainReturnedMin--\ZNEGainReturnedMax{} combined returned standard errors. Here the combined error is $s_{\rm comb}=\sqrt{s_{\rm ZNE}^2+(s_{\rm Raw,first}^2+s_{\rm Raw,last}^2)/4}$, assuming independent returned errors. The gain is resolvable on this conditional statistical scale.

Per unit Raw \ShotsLow{} charge, O$_2$ TREX removes \ThirdGainPTREX{}~mHa on Pittsburgh versus \ThirdGainPZNE{}~mHa for ZNE. Aachen reverses that ordering (\ThirdGainATREX{} versus \ThirdGainAZNE{}~mHa). Thus the arm with the smallest residual need not give the largest improvement per charge (Table~\ref{tab:followup}).

Removing the independent assignment channels from the on-gate thermal model predicts a Pittsburgh O$_2$ improvement of \OxygenAssignmentRemovalMin--\qty{\OxygenAssignmentRemovalMax}{\millihartree}, overlapping the observed TREX improvement of \OxygenTREXibmpittsburghMin--\qty{\OxygenTREXibmpittsburghMax}{\millihartree}. The calibration probabilities are marginal assignment errors, whereas TREX learns its response within the submitted workload. The mixed-PUB pilot and isolated cohort differ in both time and PUB multiplicity, so their contrasting TREX results cannot separate those two changes. Tables~\ref{S-tab:cohort-comparison} and \ref{S-tab:products} give the cohort means and separate gate and readout products.

At fixed nominal shots, TREX multiplies mean returned uncertainty by \ThirdInflationTREXMin--\ThirdInflationTREXMax{} across supported cells; ZNE multiplies it by \ThirdInflationZNEMin--\ThirdInflationZNEMax{}. O$_2$ ZNE cross-job SD divided by $\bar s$ is \AachenOxygenZNEDispersion{} on Aachen and \PittsburghOxygenZNEDispersion{} on Pittsburgh, whereas the largest raw ratio is \LargestDispersion{} for Pittsburgh BeH$_2$ Raw \ShotsHigh{}. Returned $s$ therefore does not put the arms on a common repeatability scale; repeated jobs supply that comparison. Linear interpolation of the raw brackets changes contrasts by up to \qty{\BracketSensitivityMax}{\millihartree} without changing any O$_2$ TREX improvement sign (Table~\ref{S-tab:brackets}).

\subsection{Qubit-124 reconstruction and physicality}\label{sec:controls}
The level-3 transpiler selected Aachen H$_2$ qubit 124 without a user-specified layout. Its calibrated readout error was \FifthReadoutOneTwoFour{}, \FifthReadoutRatio{} times qubit 27's \FourthReadoutTwoSeven{} and above the device median (Table~\ref{tab:circuits}). The calibration indicated worse readout but did not predict the observed magnitude. The offset-inclusive one-qubit Hamiltonian is
\begin{equation}
H=\HReconstructionOneTwoFourCI{}I\HReconstructionOneTwoFourCZ{}Z\HReconstructionOneTwoFourCX{}X
\quad(\mathrm{Ha}).
\label{eq:h2}
\end{equation}
Measurements of the \emph{molecular state} give $\langle Z\rangle=\HReconstructionOneTwoFourZ{}$ and $\langle X\rangle=\HReconstructionOneTwoFourX{}$, reconstructing a residual of \qty{\HReconstructionOneTwoFourPredicted}{\millihartree}. The separate control-cohort energy PUB gives \qty{\HReconstructionOneTwoFourObserved}{\millihartree} with returned $s=\HReconstructionOneTwoFourS{}$~mHa. The \qty{\HReconstructionOneTwoFourDifference}{\millihartree} difference between the \qty{\HReconstructionOneTwoFourPredicted}{\millihartree} reconstruction and the \qty{\HReconstructionOneTwoFourObserved}{\millihartree} energy PUB is smaller than that PUB's returned uncertainty. Fitting an affine response to the four eigenstate controls and applying it to the ideal frozen-state expectations $\langle Z\rangle_{\rm ideal}=\FifthIdealHTwoZ{}$ and $\langle X\rangle_{\rm ideal}=\FifthIdealHTwoX{}$ gives $\langle Z\rangle=0.338$ and $\langle X\rangle=-0.042$. It reproduces the observed sign reversal of $\langle X\rangle$ but underestimates its magnitude by about a factor of two. The predicted residual is \qty{\FifthAffinePrediction}{\millihartree}. The $|0\rangle:Z$ and $|+\rangle:X$ controls are different states and cannot be substituted into Eq.~(\ref{eq:h2}).

Under an effective assignment interpretation, the four eigenstate controls give Z-basis $p(1|0)=\AssignmentOneTwoFourZB{}$ and $p(0|1)=\AssignmentOneTwoFourZA{}$, with X-basis error probabilities \AssignmentOneTwoFourXB{} and \AssignmentOneTwoFourXA{}. Their basis-average errors exceed the calibrated readout value by factors \AssignmentOneTwoFourZFactor{} and \AssignmentOneTwoFourXFactor{} (Table~\ref{S-tab:effective-assignment}). The two basis-average errors differ by only \FourthBasisAverageDifference{}, which disfavors a basis-change-only explanation. The larger $0\!\to\!1$ than $1\!\to\!0$ error runs opposite to the expected direction of $T_1$ decay during readout. Qubit 124 has lower calibrated SX error than qubit 27 (\num{\FourthSXOneTwoFour} versus \num{\FourthSXTwoSeven}); these observations point toward preparation or readout, without identifying a channel. The pilot calibration product predicts only \qty{\ModelibmaachenHTwopilotnaivedepolarizingresidualmha}{\millihartree}, or \FifthProductRatioPercent\% on this scale; it comes from a different acquisition. A three-standard-error Pauli screen restricted to the Raw controls fires on all seven tested qubits: the returned standard error shrinks as $|\langle P\rangle|\to 1$, so small readout errors are resolvable at the control shot setting. The largest non-124 Raw departure in units of $s$ is 5.2$s$, at an absolute deviation of only 0.03. The largest non-124 Raw departure in absolute magnitude is \FifthRawNonFaultMaxAbs{}, or \FifthRawNonFaultMaxS{}$s$, and the largest non-124 Raw $|0\rangle$:Z deviation is \FifthRawNonFaultZeroZ{}, versus 0.65 for qubit 124; magnitude, rather than significance alone, separates the qubits (Table~\ref{S-tab:fifth-screen}). The largest departure in the full control set, 50.7$s$ and 0.74 in absolute units, is the Aachen TREX $|+\rangle$:X control on qubit 27 (estimate \FifthTREXMaxEstimate{}, against an expected $+1$, returned $s=\FifthTREXMaxS{}$); it is a mitigation over-correction. TREX controls can exceed unit magnitude and are therefore excluded from the Raw screen. Archived QPY and observable replay did not reveal a mapping error. A Sampler run on the same circuit and layout, retaining bitstring counts, remains necessary to separate initialization and assignment.

For the replay check, we load each frozen circuit in Qiskit's binary QPY format, apply its recorded layout to the logical observable, and compare both coefficients and noiseless energy with the archived operator and $V$. The maximum energy difference is $\ReplayMantissa\times10^{-10}$~Ha (Sec.~\ref{S-sec:attribution}). That agreement tests the saved preparation and observable together. It leaves provider-generated basis rotations and measurement circuits untested. The corresponding H$_2$ diagnostic on low-readout-error qubit 27 yielded a residual of \HReconstructionTwoSevenObserved{}~mHa, compared with \HReconstructionOneTwoFourObserved{}~mHa on qubit 124. The measurements do not isolate qubit selection from other acquisition conditions; the contrast motivates checking alternative physical layouts before increasing the molecular shot budget.

Spectral checks use the mapped Hamiltonian with matching total-energy offsets. Of \ThirdEnergyDenominator{} energy estimates, \EnergyViolationCount{} lie below the mapped minimum and none lie above the mapped maximum: \FourthBenchmarkTREXOutside{} benchmark TREX, \FourthBenchmarkZNEOutside{} benchmark ZNE and \FourthControlTREXOutside{} control TREX outputs; no Raw energy does. The largest departure, \FourthMaxViolation{}~mHa or \FourthMaxViolationS{} returned uncertainties, is an Aachen H$_2$ TREX control energy; \EnergyThreeSCount{} lie more than three uncertainties below the minimum. Of \ThirdPauliDenominator{} control Pauli estimates, \FifthTREXPauliOutside{} of the \FifthTREXPauliDenominator{} TREX Pauli controls exceed unit magnitude (\FifthTREXPauliOutsideAachen{} on Aachen, \FifthTREXPauliOutsidePittsburgh{} on Pittsburgh) and no Raw estimate does. The 168 control estimates are split 84/84 across processors; the largest $|\langle P\rangle|$ is \FifthMaxAbsPauli{} (Table~\ref{S-tab:fifth-pauli}). Readout mitigation and extrapolation can produce point estimates outside the spectrum of the measured observable because the mitigated estimator need not correspond to a physical quantum state~\cite{vandenberg2022,nation2021}. These are estimator diagnostics rather than claims that every point is a statistically established physical violation. The response-inversion amplification and shared-job learning hypotheses are analyzed in Sec.~\ref{S-sec:trex-workload}. An energy below the exact mapped minimum cannot be interpreted as a better variational state, even if it lies closer to the full-space chemistry reference.

\section{Shot crossover and diagnostic allocation}\label{sec:allocation}
For nominal per-configuration shots $N$, let $\bar r$ be the cell mean residual and $\bar s$ the mean returned uncertainty. A stationary bias-plus-sampling model with $s(N)\propto N^{-1/2}$ gives
\begin{equation}
R=|\bar r|/\bar s,\qquad L=(1+R^2)^{-1},\qquad N^*=N/R^2.
\label{eq:crossover}
\end{equation}
Here $L$ is the maximum fractional mean squared error (MSE) reduction from eliminating sampling. Raw \ShotsLow{} O$_2$ and BeH$_2$ imply $N^*=\CrossoverMin$--$\CrossoverMax$, a model extrapolation below the tested range (Table~\ref{S-tab:supp-crossover}). Opposite-sign residuals can cancel in $\bar r$, making $R$ small and $N^*$ large even when individual errors are substantial.

This allocation rule is conditional on a known fixed-state reference. Increasing shots reduces the estimator variance in the model but leaves its bias term unchanged. On a new molecule, a small returned error bar alone does not supply that bias estimate: a classically tractable reference workload, repeated controls or a separately justified noise model is needed. Before increasing the shot count, identify the dominant source of error and compare any residual reduction with its additional charge.

Pittsburgh LiH TREX has mean execution residual \FourthLithiumTREX{}~mHa, below the \qty{1.594}{\millihartree} reference scale for that component alone. Adding active-space and ansatz contributions gives $Y-F=\FourthFullLithiumTREX{}$~mHa; no isolated cell falls below the reference scale for the full-space absolute energy.

Across isolated jobs, $V$ falls within one and two returned uncertainties according to the arm-specific counts in Table~\ref{S-tab:fifth-inclusion}. For the four Raw 1024 O$_2$ and BeH$_2$ cells, eliminating sampling improves this stationary MSE estimate by less than \MSELeverage\%. Pittsburgh LiH illustrates the distinction between precision and inclusion: $V$ falls within $2s$ in \HardwareInclusionLow{}/\HardwareCountLow{} Raw \ShotsLow{} jobs but only \HardwareInclusionHigh{}/\HardwareCountHigh{} Raw \ShotsHigh{} jobs (Table~\ref{S-tab:coverage}).

\subsection{Proposed diagnostic protocol}\label{sec:protocol}
\small
\emph{The steps below have not been validated on an independent dataset.}
\begin{enumerate}
\item Replay the compiled circuit and mapped observable noiselessly. Their energy, including identical offsets, should reproduce the frozen-state reference $V$ within $10^{-9}$~Ha. Keep $F$ and $A$ as separate chemistry-model references.
\item Test known X- and Z-eigenstates with mitigation disabled on the selected qubits. Compare absolute deviations, deviations relative to returned uncertainty, and calibration expectations; large discrepancies motivate preparation and readout checks. In this archive the three-standard-error criterion alone fires on all seven tested qubits because the returned standard error shrinks as $|\langle P\rangle|\to 1$, so the rule is not independently validated.
For qubit 124, the fitted response predicts \qty{\ChannelibmaachenmixedOneTwoFourrawpredictedhTworesidualmha}{\millihartree} versus the observed \qty{\HReconstructionOneTwoFourObserved}{\millihartree}, whereas the calibration product predicts \qty{\ModelibmaachenHTwopilotnaivedepolarizingresidualmha}{\millihartree}. This contrast motivates the control step, although the fitted eigenstates differ from the molecular state.
\item Obtain three or more windows with raw brackets. Cross-job SD above $2\bar s$ flags excess scatter; under independent Gaussian jobs with known common $s$, its false-positive probability is \SixFlagFalsePositive\% for six jobs and \ThreeFlagFalsePositive\% for three.
\item On a known reference, additional shots have priority while $\bar s>0.5|\bar r|$ ($R<2$). At $R=2$, even eliminating sampling reduces the model MSE by only 20\%.
\item Compare signed residuals, root mean squared error (RMSE), absolute-residual reduction and charge across windows before choosing an intervention or spending ceiling.
\end{enumerate}

\section{Billed resource use}
The \PilotJobs{} pilot jobs return \PilotEnergies{} benchmark energies and the \IsolatedJobs{} isolated jobs return \IsolatedEnergies{}, giving \BenchmarkCount{} benchmark energies from \FifthBenchmarkJobs{} jobs. The \KnownJobs{} control jobs return a further \DiagnosticCount{} scalars: \EnergyControlCount{} molecular energies and \PauliControlCount{} Pauli estimates. The campaign therefore comprises \JobCount{} jobs. Pilot raw, TREX and ZNE arms have \PilotrawJobs{}, \PilottrexJobs{} and \PilotzneJobs{} jobs, returning \PilotrawEnergies{}, \PilottrexEnergies{} and \PilotzneEnergies{} energies, respectively. Their \PilotCells{} cells are two processors times five raw, four TREX and two ZNE molecular cases (Table~\ref{S-tab:cohorts}).

The charge is \qty{\ChargeSeconds}{\second}, or \qty{\ChargeMinutes}{\minute}. For isolated O$_2$, one Raw \ShotsLow{} job costs \qty{\OxygenChargeRawMin}{\second}; Raw \ShotsHigh{}, TREX and ZNE cost \OxygenChargeMoreShotsMin{}, \OxygenChargeTREXMin{} and \OxygenChargeZNEMin--\qty{\OxygenChargeZNEMax}{\second}, respectively. The two-point raw interpolation is \qty{\OxygenFixedCost}{\second} plus \qty{\OxygenShotCost}{\milli\second} per nominal shot. At the historical starting rate of \num{\QuotedRate}~USD per minute, checked 5 October 2026~\cite{ibm_pricing}, \qty{\ChargeMinutes}{\minute} corresponds to about USD~\IndicativeMonetaryCost{}.

Job-mode billed usage includes mitigation learning and twirled circuits~\cite{ibm_usage}. The provider does not itemize those components. Table~\ref{S-tab:system-costs} gives system-specific charges; queue time and classical computation are excluded, and provider-specific overhead limits extrapolation to larger active spaces.

\FloatBarrier
\section{Discussion}
A practitioner should verify the chemistry model and compiled state, test known states on the selected qubits, and measure independent windows before buying smaller error bars. The six Raw \ShotsLow{} cell means range from \ThirdAccuracyLow{} to \ThirdAccuracyHigh{} times the $1$~kcal/mol reference scale (1.594~mHa); none falls below it. These energies serve as execution diagnostics.

Repeated residuals and billed cost favor different interventions for different systems. In the isolated cohort, O$_2$ improves under both packages, BeH$_2$ worsens on average under TREX, and more shots sharpen the reported uncertainty without reliably reducing bias. Larger hardware chemistry demonstrations~\cite{arute2020,eddins2022} use other state-preparation strategies, reinforcing the need to separate model and execution targets.
No single device ranking holds across LiH, O$_2$ and BeH$_2$; the matched differences change sign by molecule.

\subsection{Scope and limitations}\label{sec:limits}
\begin{itemize}
\item Two 156-qubit Heron devices with provider-reported revision 3 and three windows bound the evidence. Fixed arm order confounds treatment with time, and all raw baselines disable DD, matching the Runtime default; the mitigation arms are therefore compared against a provider-default-off baseline rather than a best-effort one. Enabling DD could change residuals, temporal variability and relative mitigation benefits, particularly for deeper BeH$_2$ circuits; these effects require a controlled comparison. The next campaign enables DD and randomizes arm order within each window.
\item  The \CrossoverCells{} crossover cells comprise \PilotCells{} pilot and \IsolatedCells{} isolated cells; \InspectedContrasts{} window contrasts are \FourthNonBaselineCells{} non-baseline cells across \FourthWindowCount{} windows; \LocalSamplingCells{} local sampling cells are \FourthLocalSystems{} systems at \FourthLocalShotSettings{} shot settings. These are exploratory; only the six-cell F-test screen has the stated Bonferroni correction.
\item Three windows cannot establish a drift law or population coverage. Window SEs and iid intervals are assumption-dependent, and shared windows can correlate device contrasts and sign tests.
\item Calibration models omit correlated errors, leakage and provider measurement details. The known-state controls use only single-qubit layouts, so preparation and readout on the multi-qubit LiH, O$_2$ and BeH$_2$ circuits remain uncharacterized. Fault lists, execution-time target metadata, learned responses and counts-level controls are unavailable. No optimization ran on hardware, so optimizer--hardware feedback lies outside the design.
\end{itemize}

\section{Conclusions}
Frozen energy evaluation separates chemistry-model and ansatz errors from execution residuals. Raw \ShotsLow{} O$_2$ and BeH$_2$ mean residuals exceed returned uncertainty by factors \RawRatioMin--\RawRatioMax. Independent repetitions expose variation absent from a per-job error bar. For Pittsburgh O$_2$, TREX removes \ThirdGainPTREX{}~mHa per Raw \ShotsLow{} job charge versus \ThirdGainPZNE{}~mHa for ZNE; Aachen reverses that ordering. When measured bias dominates returned uncertainty, check the selected qubits and cross-window dispersion before allocating more shots.
The ZNE package gives the lowest O$_2$ residuals, reducing them by \TreatmentibmaachenOTwozneOneZeroTwoFourPercent\% on Aachen and \TreatmentibmpittsburghOTwozneOneZeroTwoFourPercent\% on Pittsburgh. It inflates returned uncertainty \FifthZNEInflationAachenOTwo{}-fold on Aachen and \FifthZNEInflationPittsburghOTwo{}-fold on Pittsburgh, or 1.7--3.4-fold across all four ZNE cells. Across isolated jobs, $V$ falls within two returned errors in \FifthInclusionRawLowTwoS{}/\FifthInclusionRawLowJobs{} Raw \ShotsLow{}, \FifthInclusionRawHighTwoS{}/\FifthInclusionRawHighJobs{} Raw \ShotsHigh{}, \FifthInclusionTREXTwoS{}/\FifthInclusionTREXJobs{} TREX and \FifthInclusionZNETwoS{}/\FifthInclusionZNEJobs{} ZNE jobs; one-error counts appear in Table~\ref{S-tab:fifth-inclusion}. For O$_2$ under the ZNE package, the returned bar exceeds cross-job scatter by \FifthZNEBarScatterMin--\FifthZNEBarScatterMax{} times, so arm comparisons require repeated-job dispersion.

\section*{Data availability statement}
The data supporting this study are openly available from Zenodo at \ZenodoAccession{}. Zenodo is the archival copy of record; the MolQBench web application at \url{https://quantum.larrucea.eu/molqbench/} provides an interactive convenience mirror. The deposited data are licensed CC BY 4.0.

\section*{Acknowledgements}
The author acknowledges the Fraunhofer Quantum Now program for IBM Quantum access. This work was carried out within EQeS, subproject \emph{Chemistry and analysis at the application level} (Teilvorhaben: Chemie und Analyse auf Anwendungsebene), funded by the German Federal Ministry of Research, Technology and Space (BMFTR) under contract 13N17338 in the Research Program Quantum Systems.

Generative AI tools (OpenAI Codex) assisted with language editing and analysis and plotting scripts. AI-assisted outputs were checked against archived observations, reference calculations and cited sources; figures and tables are generated programmatically. The author retains full responsibility for the content.
\section*{Author contributions}
J.L. conceived the study, secured funding, developed methodology and software, performed investigation, validation and analysis, curated data, prepared figures, and wrote and revised the manuscript.
\section*{Competing interests}
The author declares no competing interests.

\bibliographystyle{iopart-num}
\bibliography{references}
\clearpage
\begin{center}\Large\bfseries Supplementary Information\end{center}
\setcounter{section}{0}\setcounter{figure}{0}\setcounter{table}{0}
\renewcommand{\thesection}{S\arabic{section}}
\renewcommand{\thefigure}{S\arabic{figure}}
\renewcommand{\thetable}{S\arabic{table}}
\section{Conventions and abbreviations}
The public data accession is \ZenodoAccession.
\paragraph{Interpretive conventions.} Deterministic model values carry no inferential error bars. Hardware bars show returned Estimator standard errors $s$. Hardware inclusion counts summarize this dataset. They do not estimate population coverage. Three windows describe repeatability without fitting a drift model.

We use quantum processing unit (QPU), instruction-set architecture (ISA),
Twirled readout error extinction (TREX), zero-noise extrapolation (ZNE),
Primitive Unified Bloc (PUB), full configuration interaction (FCI),
qubit-wise commuting (QWC), standard deviation (SD), mean absolute error (MAE),
root mean squared error (RMSE), and as soon as possible (ASAP) scheduling.
CPU denotes central processing unit. QPY denotes Qiskit's binary circuit format. Self-consistent-field (SCF)
calculations define the molecular orbital gauge. Unitary coupled cluster (UCC)
uses doubles (UCCD) or singles and doubles (UCCSD); sequential least squares
programming (SLSQP) performs local optimization.
\section{Exploratory hardware pilot}
\begin{figure}[!htbp]
\centering
\includegraphics[width=\linewidth]{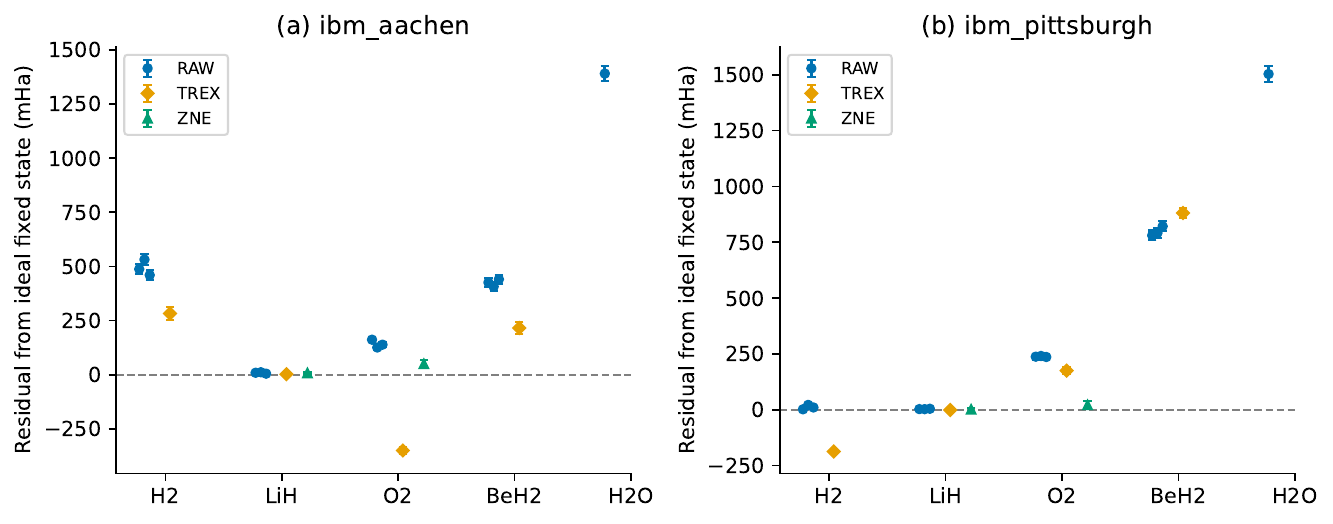}
\caption{Pilot residuals from $V$: panels (a) Aachen and (b) Pittsburgh,
for raw, TREX and the ZNE package. Bars are returned $s$. Each core raw cell has
three jobs, while H$_2$O and each mitigation cell have one in a single window.
}
\label{S-fig:pilot}
\end{figure}

Figure~\ref{S-fig:pilot} and Table~\ref{S-tab:pilot} summarize the pilot. It first evaluated the frozen states on Aachen. Three raw
jobs repeated the four core systems, with H$_2$O included only in the first job;
one TREX job evaluated the core systems and one ZNE job evaluated LiH and O$_2$.
The same saved ISA circuit and layout-applied observable were used across arms.
All local compiled-state expectation checks agreed with the frozen ideal values
within $10^{-9}$ Ha. The pilot occupies one short temporal window per processor
and is not pooled with the isolated-molecule cohorts.

\begin{table}[htbp]
\centering
\small
\resizebox{\linewidth}{!}{\begin{tabular}{llllllllll}
\toprule
Backend & System & Arm & $N$ & $n$ & Mean & SD & Min & Max & $\bar s$\\
\midrule
ibm\_aachen & BeH2 & raw & \num{1024} & \num{3} & \num{424.71} & \num{16.90} & \num{407.14} & \num{440.85} & \num{20.376}\\
ibm\_aachen & BeH2 & trex & \num{1024} & \num{1} & \num{216.02} & -- & \num{216.02} & \num{216.02} & \num{27.535}\\
ibm\_aachen & H2 & raw & \num{1024} & \num{3} & \num{493.35} & \num{36.18} & \num{460.31} & \num{532.02} & \num{23.149}\\
ibm\_aachen & H2 & trex & \num{1024} & \num{1} & \num{282.87} & -- & \num{282.87} & \num{282.87} & \num{30.869}\\
ibm\_aachen & H2O & raw & \num{1024} & \num{1} & \num{1390.41} & -- & \num{1390.41} & \num{1390.41} & \num{35.403}\\
ibm\_aachen & LiH & raw & \num{1024} & \num{3} & \num{8.79} & \num{3.33} & \num{5.22} & \num{11.81} & \num{3.04}\\
ibm\_aachen & LiH & trex & \num{1024} & \num{1} & \num{2.89} & -- & \num{2.89} & \num{2.89} & \num{2.7506}\\
ibm\_aachen & LiH & zne & \num{1024} & \num{1} & \num{7.81} & -- & \num{7.81} & \num{7.81} & \num{6.0297}\\
ibm\_aachen & O2 & raw & \num{1024} & \num{3} & \num{142.33} & \num{18.29} & \num{125.89} & \num{162.03} & \num{10.686}\\
ibm\_aachen & O2 & trex & \num{1024} & \num{1} & \num{-349.50} & -- & \num{-349.50} & \num{-349.50} & \num{16.269}\\
ibm\_aachen & O2 & zne & \num{1024} & \num{1} & \num{50.41} & -- & \num{50.41} & \num{50.41} & \num{16.013}\\
ibm\_pittsburgh & BeH2 & raw & \num{1024} & \num{3} & \num{798.11} & \num{21.08} & \num{780.99} & \num{821.66} & \num{22.225}\\
ibm\_pittsburgh & BeH2 & trex & \num{1024} & \num{1} & \num{881.41} & -- & \num{881.41} & \num{881.41} & \num{22.083}\\
ibm\_pittsburgh & H2 & raw & \num{1024} & \num{3} & \num{11.61} & \num{9.97} & \num{2.24} & \num{22.08} & \num{8.6683}\\
ibm\_pittsburgh & H2 & trex & \num{1024} & \num{1} & \num{-186.68} & -- & \num{-186.68} & \num{-186.68} & \num{11.185}\\
ibm\_pittsburgh & H2O & raw & \num{1024} & \num{1} & \num{1503.61} & -- & \num{1503.61} & \num{1503.61} & \num{36.242}\\
ibm\_pittsburgh & LiH & raw & \num{1024} & \num{3} & \num{3.67} & \num{0.94} & \num{3.07} & \num{4.75} & \num{2.4344}\\
ibm\_pittsburgh & LiH & trex & \num{1024} & \num{1} & \num{-0.59} & -- & \num{-0.59} & \num{-0.59} & \num{2.3536}\\
ibm\_pittsburgh & LiH & zne & \num{1024} & \num{1} & \num{2.03} & -- & \num{2.03} & \num{2.03} & \num{7.2491}\\
ibm\_pittsburgh & O2 & raw & \num{1024} & \num{3} & \num{238.59} & \num{2.43} & \num{236.49} & \num{241.25} & \num{12.904}\\
ibm\_pittsburgh & O2 & trex & \num{1024} & \num{1} & \num{175.46} & -- & \num{175.46} & \num{175.46} & \num{14.144}\\
ibm\_pittsburgh & O2 & zne & \num{1024} & \num{1} & \num{21.93} & -- & \num{21.93} & \num{21.93} & \num{18.63}\\
\bottomrule
\end{tabular}
}
\caption{Exploratory signed residuals from $V$, in mHa: $n$, mean, sample SD,
minimum, maximum and mean returned $s$. Raw has three jobs per core case;
H$_2$O, TREX and the ZNE package each have one per supported case, with unavailable
sample SD omitted.}
\label{S-tab:pilot}
\end{table}

Aachen's raw H$_2$ mean residual is \CellpilotibmaachenHTworawOneZeroTwoFourmeanmha{} mHa, despite its circuit having no
two-qubit gates. Its mean reported uncertainty is only \qty{\CellpilotibmaachenHTworawOneZeroTwoFourmeanreturnedsmha}{\millihartree}. Its stored
layout uses physical qubit 124. The calibration snapshot reports a readout error
of approximately \AnomalousReadoutPercent{}\% on that qubit, but this single calibration number does not
explain or identify the observed residual. Circuit-depth correlations alone would
miss this anomalous control. Layout and basis checks, returned measurement metadata
and targeted preparation and readout controls must therefore precede any attribution
to a particular physical error mechanism.

For O$_2$ on Aachen, raw residuals average +\CellpilotibmaachenOTworawOneZeroTwoFourmeanmha{} mHa, whereas the single TREX
observation is \CellpilotibmaachenOTwotrexOneZeroTwoFourmeanmha{} mHa. This lower mitigated energy is further from the ideal
state than the raw mean, while its reported uncertainty is only \qty{\CellpilotibmaachenOTwotrexOneZeroTwoFourmeanreturnedsmha}{\millihartree}. The
single ZNE result is +\CellpilotibmaachenOTwozneOneZeroTwoFourmeanmha{} mHa. The sign reversal matters: the mitigated estimate moves farther from the ideal
state despite a lower reported energy. Nuclear and inactive offsets are
identical across each comparison, so these execution residuals do not include
active-space truncation.

Pittsburgh's raw H$_2$ mean residual is +\CellpilotibmpittsburghHTworawOneZeroTwoFourmeanmha{} mHa, with mean reported
uncertainty \RoundThreeMeanReturnedErrorThree{} mHa. Its single TREX residual is \CellpilotibmpittsburghHTwotrexOneZeroTwoFourmeanmha{} mHa. For O$_2$,
the raw mean, TREX and ZNE residuals are +\CellpilotibmpittsburghOTworawOneZeroTwoFourmeanmha{}, +\CellpilotibmpittsburghOTwotrexOneZeroTwoFourmeanmha{} and +\CellpilotibmpittsburghOTwozneOneZeroTwoFourmeanmha{} mHa,
respectively. BeH$_2$ has a raw mean residual of +\CellpilotibmpittsburghBeHTworawOneZeroTwoFourmeanmha{} mHa and a TREX
residual of +\CellpilotibmpittsburghBeHTwotrexOneZeroTwoFourmeanmha{} mHa. Thus neither mitigation nor a simple processor ordering
is uniformly favorable across these cases. The processor contrast is descriptive:
layouts, calibrations and observation times differ, and this pilot sampled only
one temporal window per processor.

\subsection{Offline diagnostic checks}
QPY replay and mapped-observable checks are detailed in the attribution checklist (Sec.~\ref{S-sec:attribution}). Diagonalization of the entire tapered-qubit operator shows that the
Aachen TREX O$_2$ and Pittsburgh TREX H$_2$ estimates lie below its spectral
minimum by \SpectrumibmaachenOTwotrexoutsidespectrummha{} and \SpectrumibmpittsburghHTwotrexoutsidespectrummha{} mHa, respectively: \SpectrumibmaachenOTwotrexoutsidespectrumoverreportedsigma{} and \SpectrumibmpittsburghHTwotrexoutsidespectrumoverreportedsigma{} times their
reported uncertainties. Pittsburgh TREX LiH is only \SpectrumibmpittsburghLiHtrexoutsidespectrummha{} mHa below its minimum,
or \SpectrumibmpittsburghLiHtrexoutsidespectrumoverreportedsigma{} reported standard deviations. These checks use the entire operator spectrum. Finite-sample or mitigated
estimates need not correspond to an expectation in a physical density matrix;
these excursions diagnose physicality. We preserve them without clipping or excluding observations.

The ideal H$_2$ state has $\langle X\rangle=\Assignmentibmaachenidealx{}$ and
$\langle Z\rangle=\Assignmentibmaachenidealz{}$. For a stationary classical assignment channel,
with $a=P(0|1)$ and $b=P(1|0)$, a binary expectation $m$ becomes
$(1-a-b)m+a-b$. Applying the archived calibration probabilities to both ideal
basis expectations predicts residuals of +\Assignmentibmaachenassignmentonlypredictedresidualmha{} mHa on Aachen and \Assignmentibmpittsburghassignmentonlypredictedresidualmha{} mHa
on Pittsburgh. This deliberately restricted model assumes ideal preparation
and basis rotations. It cannot reproduce the observed raw means and does not
identify a cause; direct controls must test its assumptions.

Noise-factor traces and nonfinite fit arrays are examined in
Sec.~\ref{S-sec:zne-traces}. The archive contains the circuit audit, spectral
checks, assignment-channel calculation and noise-factor estimates.

Table~\ref{S-tab:supp-budget} lists pilot charges by processor. Each total is below its regional spending gate. Job-level charges are shared by molecular PUBs in the pilot, and the molecular measurement work differs between arms; pilot comparisons therefore have unequal costs. Individual molecular charges are available for the isolated cohort.
\FloatBarrier

\section{Targeted controls and isolated-molecule validation}
\subsection{Known-state tests diagnose a workload-dependent discrepancy}
The pilot's anomalous control motivated an explicitly exploratory follow-up.
On Aachen qubit 124, raw preparation of $|0\rangle$ returned
$\langle Z\rangle=\ControlibmaachenmixedOneTwoFourzerozrawValue{}\pm\ControlibmaachenmixedOneTwoFourzerozrawError{}$, whereas the prospectively
calibration-selected qubit 27 returned $\ControlibmaachenmixedTwoSevenzerozrawValue{}\pm\ControlibmaachenmixedTwoSevenzerozrawError{}$.
The corresponding raw H$_2$ energy residuals were +\ChannelibmaachenmixedOneTwoFourrawobservedhTworesidualmha{} and +\ChannelibmaachenmixedTwoSevenrawobservedhTworesidualmha{} mHa.
These controls demonstrate a discrepancy in the preparation and measurement
path. They cannot separate initialization, assignment and basis-rotation errors.
The archived calibration's readout error alone did not flag its magnitude.

\begin{figure}[!htbp]
\centering
\includegraphics[width=\linewidth]{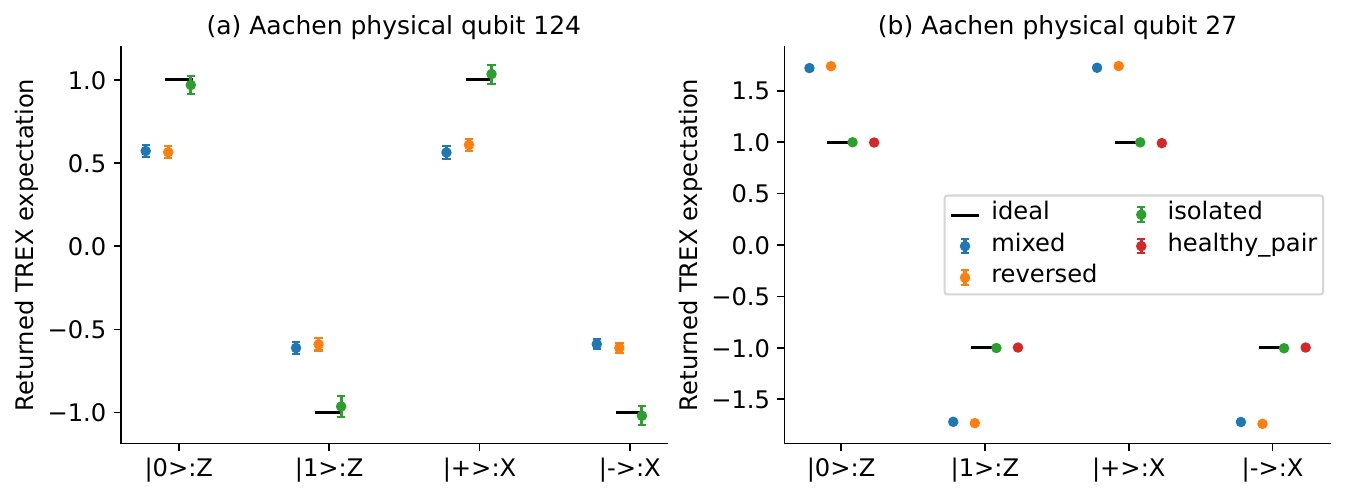}
\caption{Aachen TREX controls with known X and Z eigenstate expectations (black marks).
Mixed and reversed batches contain the same two physical qubits, 124 and 27.
Isolated execution assigns one job to each qubit. The healthy pair contains 27 and
38. Error bars are returned standard uncertainties; PUBs share a job and mitigation learning. The healthy-pair condition
does not contain qubit 124. Individual estimates and complete metadata are
archived.}
\label{S-fig:controls}
\end{figure}

In a single mixed-qubit TREX job, the healthy control's nominal $|0\rangle$
expectation became $\ControlibmaachenmixedTwoSevenzeroztrexValue{}\pm\ControlibmaachenmixedTwoSevenzeroztrexError{}$. Reversing the two PUB blocks
returned $\ControlibmaachenreversedTwoSevenzeroztrexValue{}\pm\ControlibmaachenreversedTwoSevenzeroztrexError{}$, preserving the large excursion.
In a separate job confined to qubit 27, the estimate was
$\ControlibmaachenisolatedTwoSevenzeroztrexValue{}\pm\ControlibmaachenisolatedTwoSevenzeroztrexError{}$. A mixed job with two prospectively calibration-selected
healthy qubits returned $\ControlibmaachenhealthypairTwoSevenzeroztrexValue{}\pm\ControlibmaachenhealthypairTwoSevenzeroztrexError{}$ on qubit 27.
The discrepant qubit's own isolated TREX controls also approached their known
expectations. Figure~\ref{S-fig:controls} shows the complementary X and Z tests.
The mixed and reversed jobs and the isolated and healthy-pair jobs give different corrected responses. Their sequential acquisition leaves both temporal effects and workload effects as possible explanations. TREX follows a multiplicative response correction whose sampling overhead grows when the response is weak~\cite{vandenberg2022}. The observed excursions motivate checking that response in the workload, but do not show which internal correction was applied. The variants were not randomized in time, and the learned
response matrices were not exposed in the retained result payloads.

Pittsburgh's mixed controls on qubits 61 and 77 did not show comparable severe
amplification, and their isolated controls were close to the known values.
The Pittsburgh reversed condition selected qubit 114 instead of 77 after a
calibration update and is therefore not a strict order-only comparison.
Later diagnostic variants freeze the original selection. Results are retained
under their physical-qubit identities. Each control measures a fixed state in its recorded workload and physical layout.

\section{Methods}
\subsection{Electronic structure and physical-sector references}\label{S-sec:reference-methods}
All systems are neutral. H$_2$, LiH, BeH$_2$ and H$_2$O use singlet references;
O$_2$ uses a triplet with $(N_\alpha,N_\beta)=(3,1)$ in the active space.
Diatomic bond lengths are \BondHTwo{}, \BondLiH{} and \BondOTwo{}~\AA\ for H$_2$, LiH and O$_2$,
respectively. The archived BeH$_2$ geometry is nearly linear, with the small
transverse displacement retained; H$_2$O is bent. Exact Cartesian coordinates,
geometry sources and orbital-selection matrices are included in the public-deposit package (Table~\ref{S-tab:geometries}).
In the frozen O$_2$ orbital gauge, the selected spatial orbitals are zero-based
indices 6--9. The orbital reconstruction and sector dimension are documented in Sec.~\ref{S-sec:orbital-audit}. Frozen integrals define the repeated hardware problem.

Molecule definitions record Cartesian geometry in angstroms, basis, charge and spin ($N_\alpha-N_\beta$). PySCF-backed Qiskit Nature problems use RHF for singlets and ROHF for open-shell systems. Inherited active-space methods take priority over advisory indices. Selection matrices and orbital indices are retained where exposed by the public transformer API. Full and active molecular-orbital integrals, qubit Pauli coefficients and energy constants are archived together. They suffice to replay the frozen Hamiltonian even if a new SCF run chooses a different orbital gauge.

For the active reference, form the Jordan--Wigner Hamiltonian and select computational states with the required $(N_\alpha,N_\beta)$. Diagonalize $S^2$ in that subspace, project onto its target eigenvalue $S(S+1)$, and diagonalize the projected Hamiltonian. This avoids assigning spin from arbitrary vectors in a degenerate energy eigenspace. Require the resulting Hamiltonian eigenvector residual below $10^{-8}$ Ha. Dense active references are limited to at most six spatial orbitals by the current implementation. Independent full-space FCI uses a spin-fixed solver with an $S^2$ check and SCF and FCI convergence tolerances of $10^{-12}$. Agreement of the independently constructed Hartree--Fock references is required within $10^{-7}$ Ha.

\subsection{Ansatz, optimization and replay}
The initial state is Hartree--Fock. The primary ansatz is one-repetition Qiskit Nature UCC with doubles excitations. Parity mapping with two-qubit reduction is followed by symmetry tapering. The parameter order is stored explicitly. SLSQP minimizes exact statevector expectations, using finite-difference gradients for stationarity checks. Three runs start from zero and two seeded Gaussian perturbations with standard deviation 0.2 radians. Central-difference gradients use a $10^{-5}$-radian step. The maximum absolute gradient component must be at most $10^{-7}$ Ha/radian for an accepted stationary result. Solver status, evaluation counts, initial and final vectors, gradients and the energy spread among accepted starts are retained. One-parameter cases additionally undergo a 65-point numerical scan on $[-\pi,\pi]$ with local scalar polishing. Agreement with the selected optimum is required within $10^{-8}$ Ha. The multistart and scan checks test stationarity and consistency. They do not certify global optimality.

Optimized states are checked against mapped particle-number and angular-momentum observables where retained by tapering, and must not fall below the physical active reference beyond numerical tolerance. The local research archive retains parameterized and bound QPY circuits, Pauli observables, offsets, measurement models and compressed integral arrays. SHA-256 identities guard replay. The public dataset provides the reported observations, all frozen molecular and control circuit inputs with mapped observables, integral tensors, calibration snapshots and software versions.

\subsection{Ideal finite sampling and uncertainty}
For each QWC group, rotate X bases with Hadamards and Y bases with $S^\dagger$ followed by Hadamards. Calculate the complete ideal computational-basis probabilities and draw a multinomial count vector. Weighted Pauli eigenvalues evaluated on each outcome give the group energy without treating correlated terms as independent. Draws are independently seeded by a stable hash of the master seed, molecule, shot count and repetition index. This preserves identical local samples after restart. Full count vectors and unbiased group variance estimates are stored. The analytic mean from the measurement model must agree with the direct statevector expectation within $10^{-9}$ Ha.

\subsection{Execution records and statistical analysis}
Each immutable execution record contains timestamps, inputs, circuit and observable identities, parameters, requested and observed shots, execution options, measured energy, uncertainty definition, references, provenance and usage fields. Unavailable hardware, job and calibration information is null. Local calculations consume no QPU time. Atomic publication prevents a partial execution file from being mistaken for a completed record. A run with a changed configuration or source digest cannot silently resume.

Local summaries report signed bias, absolute bias, MAE, sample standard deviation, RMSE and empirical coverage against explicitly named references. Wilson intervals apply to independent local repetitions. Hardware summaries maintain individual job and window identities and use descriptive paired window contrasts. With only three temporal blocks, cluster bootstrap intervals would suggest unsupported precision and are not reported. Balanced one-way moment estimates of the within-window and between-window variance components are reported with three within and two between degrees of freedom. They are small-sample point estimates, and negative window estimates are archived with their signs. Gaussian nominal coverage is conditional on estimator distribution and uncertainty meaning. All \JobCount{} hardware jobs completed successfully; there were no failed jobs or automatic resubmissions in the reported campaign.

\subsection{Pilot execution and resource gates}
The initial pilot is a sequential five-job workload per processor: one raw job
containing all five systems, two further raw jobs containing the four core systems,
one TREX job containing those core systems and one ZNE job containing LiH and
O$_2$. Each job requests \ShotsLow{} shots per measurement configuration; ZNE uses the
specified noise factors. Aachen is evaluated first, and the Pittsburgh pilot is
conditional on successful completion of the Aachen pilot. No confirmatory cohort
is launched by the pilot runner.

Each compiled circuit is checked locally against its frozen ideal expectation
after observable layout application, with a tolerance of $10^{-9}$ Ha. The same
saved ISA and observable are reused across pilot jobs. The runner records durable
submission intent before the API call, and then records the returned job identity.
An ambiguous submission is not automatically retried. It maintains raw PUB data,
resolved metadata, circuit identities and job-level usage before releasing the
next job. Charges are recorded per job.

The regional spending gate is 240 seconds, with sequential accounting using final
billed usage and conservative per-job estimates. Each job also has a provider
quantum-execution timeout, with a ten-second allowance reserved beyond that
limit. Quantum-execution timeout and final billing are distinct quantities, so
these gates reduce overspend risk; final billing can still exceed a timeout.
Pending usage, which the SDK may expose as zero, halts further submission instead
of being interpreted as free execution. Failures and partial results halt the
pilot and retain available charge metadata. Pilot observations are analyzed separately from the isolated follow-up measurements.

\subsection{Targeted follow-up and temporal windows}
The response controls request \ShotsControl{} shots and use four known X and Z eigenstates plus
separate X and Z and Hamiltonian expectations of frozen H$_2$. Mixed, isolated,
reversed and healthy-pair variants retain physical-qubit choices and
source snapshots. Each isolated-molecule window uses 14 separate jobs per processor with
fixed ordering: low-shot raw, TREX, optional ZNE, high-shot raw and a final low-shot
raw observation for each molecule. Circuits are loaded from the initial ISA archive
and checked against the current Target and their exact expectation before release.
Individual molecular jobs provide directly billed molecular costs. The earlier
mixed pilot still has shared job-level charges. Control Hamiltonian values are
electronic energies, whereas molecular records add the common nuclear and inactive
offsets. Residual comparisons maintain these conventions. Spectral comparisons first convert electronic-energy controls to the common total-energy scale (Sec.~\ref{S-sec:physicality}).

Cumulative follow-up charges include every diagnostic variant and all three windows.
The initial acquisition target was 600 seconds per region. The third window was
released within an explicitly authorized cumulative ceiling of 900 seconds.
A missing or pending charge halts subsequent jobs. A
submission with uncertain identity is never automatically repeated. Provider execution times define each temporal window. Per-window contrasts are descriptive and maintain
intervention ordering. No population coverage conclusion is inferred from three blocks.

\subsection{Transpilation, measurement options and software}\label{S-sec:resolved-options}
The ZNE package resolves 32 randomizations with 32 shots each at each nominal
\ShotsLow{}-shot configuration, trying exponential extrapolation with linear fallback;
Runtime selects successful fits heuristically and can select different fits
for different Pauli terms~\cite{ibm_zne}. Readout learning requests 32 randomizations
with automatic shots per randomization for TREX and ZNE; resolved calibration-shot
counts and resolved \texttt{resilience\_level} are not returned. Requested levels are \RoundThreeTREXLevel~for TREX and \RoundThreeZNELevel~for the ZNE package. Noise amplification is unspecified in the archived
returned options. The installed API documents gate folding as the default,
which leaves the execution-time method unconfirmed. Requested and returned options
are archived separately.

The frozen circuits were transpiled with optimization level 3 and seed 271828,
after parameter binding, without fractional gates. Table~\ref{S-tab:circuits}
reports state-preparation resources. Basis rotations and mitigation operations
added by Runtime are additional work. QPY circuits, physical layouts and
mapped observables are archived separately for each processor.

\begin{table}[htbp]
\centering
\begin{tabular}{lllll}
\toprule
System & Qubits & Native 2Q gates & Aachen depth & Pittsburgh depth\\
\midrule
H2 & \num{1} & \num{0} & \num{4} & \num{4}\\
LiH & \num{2} & \num{2} & \num{11} & \num{11}\\
O2 & \num{4} & \num{11} & \num{39} & \num{39}\\
BeH2 & \num{5} & \num{156} & \num{465} & \num{458}\\
H2O & \num{5} & \num{199} & \num{506} & \num{507}\\
\bottomrule
\end{tabular}

\caption{Frozen state-preparation circuit resources. The native two-qubit gate
counts coincide across the two processors; depths exclude Runtime measurement
and mitigation additions. H$_2$ and H$_2$O belong to the pilot, whereas isolated
replication uses LiH, O$_2$ and BeH$_2$.}
\label{S-tab:circuits}
\end{table}

Raw execution disables readout mitigation, gate and measurement twirling and dynamical
decoupling. TREX enables measurement twirling and readout mitigation, with 32
randomizations and 32 shots per randomization at requested \ShotsLow{} shots. The ZNE
package additionally enables gate twirling and uses noise factors $(1,3,5)$ with
exponential extrapolation with linear fallback. Readout noise learning uses 32
randomizations and automatic shots per randomization. All arms initialize qubits
and request a \qty{250}{\micro\second} repetition delay; PEC and dynamical decoupling are disabled.
Each job records requested options and resolved provider metadata. The term
``ZNE'' therefore denotes this whole package. Requested shots do not include all noise-learning and
noise-amplified work. The archived software environment uses Python 3.12,
Qiskit 2.5.2, Qiskit Nature 0.7.2, Qiskit IBM Runtime 0.49.0, PySCF 2.10 and
SciPy 1.17.1; exact versions accompany each source snapshot.

\section{Attribution checklist and calibration models}\label{S-sec:attribution}
The checklist records the checks performed and evidence absent from the archive. The successful replay did not reveal a layout or observable error; it does not test provider-generated measurement circuits.
An unavailable check is \emph{UNRESOLVED}. Calibration medians, the anomalous-qubit timeline and both control qubits are tabulated in Tables~\ref{S-tab:device-calibration}, \ref{S-tab:qubit-timeline} and \ref{S-tab:qubit-calibrations}.
\begin{enumerate}
\item[(a)] \textbf{PASS}: all ten pilot QPY and mapped-observable energy replays,
maximum discrepancy $\ReplayMantissa{}\times10^{-10}$~Ha.
\item[(b)] \textbf{PARTIAL}: saved layouts equal QPY final layouts, and every
used two-qubit edge appears in archived calibration properties. Full coupling
maps or Targets at each provider execution time were not retained.
\item[(c)] \textbf{PASS}: reapplying the saved QPY layout to the logical observable
reproduces every stored coefficient within $10^{-12}$~Ha.
\item[(d)] \textbf{UNRESOLVED}: returned metadata do not expose provider-generated
basis rotations or individual measurement circuits.
\item[(e)] \textbf{PARTIAL}: the acquisition code checks native instruction support
against the current Target before each isolated window; properties snapshots
and their \texttt{last\_update\_date} are archived. Per-job Target versions and
execution-time refreshes are unavailable, so unchanged calibration is not claimed.
\end{enumerate}

\begin{table}[htbp]\centering\scriptsize
\resizebox{\linewidth}{!}{\begin{tabular}{lllllll}
\toprule
Acquisition & Qubits & $T_1$ & $T_2$ & 1Q error & 2Q error & Readout\\
\midrule
controls-2026-10-03-eu & \num{156} & \num{217.92} & \num{235.58} & \num{0.00026} & \num{0.00157} & \num{0.0059}\\
controls-2026-10-03-us & \num{156} & \num{286.97} & \num{313.91} & \num{0.00029} & \num{0.00162} & \num{0.0048}\\
controls-healthy-2026-10-03-eu & \num{156} & \num{217.92} & \num{235.58} & \num{0.00026} & \num{0.00157} & \num{0.0059}\\
controls-healthy-2026-10-03-us & \num{156} & \num{286.97} & \num{313.91} & \num{0.00029} & \num{0.00162} & \num{0.0048}\\
controls-isolated-2026-10-03-eu & \num{156} & \num{217.92} & \num{235.58} & \num{0.00026} & \num{0.00157} & \num{0.0059}\\
controls-isolated-2026-10-03-us & \num{156} & \num{286.97} & \num{313.91} & \num{0.00029} & \num{0.00162} & \num{0.0048}\\
controls-reversed-2026-10-03-eu & \num{156} & \num{217.92} & \num{235.58} & \num{0.00026} & \num{0.00157} & \num{0.0059}\\
controls-reversed-2026-10-03-us & \num{156} & \num{286.97} & \num{313.91} & \num{0.00029} & \num{0.00162} & \num{0.0050}\\
pilot-2026-10-03-eu & \num{156} & \num{217.92} & \num{235.58} & \num{0.00026} & \num{0.00157} & \num{0.0059}\\
pilot-2026-10-03-us & \num{156} & \num{286.97} & \num{313.91} & \num{0.00029} & \num{0.00162} & \num{0.0048}\\
validation-2026-10-03-eu & \num{156} & \num{217.92} & \num{235.58} & \num{0.00026} & \num{0.00157} & \num{0.0059}\\
validation-2026-10-03-us & \num{156} & \num{286.97} & \num{313.91} & \num{0.00029} & \num{0.00162} & \num{0.0050}\\
validation-window2-2026-10-03-eu & \num{156} & \num{209.35} & \num{242.04} & \num{0.00025} & \num{0.00157} & \num{0.0052}\\
validation-window2-2026-10-03-us & \num{156} & \num{286.97} & \num{313.91} & \num{0.00029} & \num{0.00162} & \num{0.0046}\\
validation-window3-2026-10-05-eu & \num{156} & \num{220.70} & \num{242.38} & \num{0.00025} & \num{0.00151} & \num{0.0058}\\
validation-window3-2026-10-05-us & \num{156} & \num{258.02} & \num{299.27} & \num{0.00029} & \num{0.00159} & \num{0.0049}\\
\bottomrule
\end{tabular}
}
\caption{Archived calibration medians by acquisition: $T_1/T_2$ in $\mu$s,
gate and readout error probabilities; 1Q medians exclude virtual RZ gates.
The snapshots record device properties at their archived timestamps;
backend versions, basis gates and per-layout error rates are archived.}
\label{S-tab:device-calibration}
\end{table}
\begin{table}[htbp]\centering\scriptsize
\resizebox{\linewidth}{!}{\begin{tabular}{llllll}
\toprule
Acquisition & Calibration time & $T_1$ (\unit{\micro\second}) & $T_2$ (\unit{\micro\second}) & Readout & SX\\
\midrule
controls-2026-10-03-eu & 2026-10-03T22:21:07+02:00 & \num{192.51} & \num{307.38} & \num{0.0405} & \num{0.00009}\\
controls-healthy-2026-10-03-eu & 2026-10-03T22:21:07+02:00 & \num{192.51} & \num{307.38} & \num{0.0405} & \num{0.00009}\\
controls-isolated-2026-10-03-eu & 2026-10-03T22:21:07+02:00 & \num{192.51} & \num{307.38} & \num{0.0405} & \num{0.00009}\\
controls-reversed-2026-10-03-eu & 2026-10-03T22:21:07+02:00 & \num{192.51} & \num{307.38} & \num{0.0405} & \num{0.00009}\\
pilot-2026-10-03-eu & 2026-10-03T20:56:57+02:00 & \num{192.51} & \num{307.38} & \num{0.0405} & \num{0.00009}\\
validation-2026-10-03-eu & 2026-10-03T22:21:07+02:00 & \num{192.51} & \num{307.38} & \num{0.0405} & \num{0.00009}\\
validation-window2-2026-10-03-eu & 2026-10-04T04:59:44+02:00 & \num{235.51} & \num{394.48} & \num{0.0415} & \num{0.00009}\\
validation-window3-2026-10-05-eu & 2026-10-05T03:46:26+02:00 & \num{189.46} & \num{297.26} & \num{0.0356} & \num{0.00009}\\
\bottomrule
\end{tabular}
}
\caption{Physical qubit 124 calibration timeline from every Aachen acquisition
snapshot, with calibration timestamps supplied by the provider.
Coherence times use microseconds. Readout and SX entries are error probabilities.
}\label{S-tab:qubit-timeline}
\end{table}
\begin{table}[htbp]\centering\scriptsize
\resizebox{\linewidth}{!}{\begin{tabular}{llllll}
\toprule
Acquisition & Qubit & $T_1$ (\unit{\micro\second}) & $T_2$ (\unit{\micro\second}) & Readout error & SX error\\
\midrule
controls-2026-10-03-eu & \num{124} & \num{192.51} & \num{307.38} & \num{0.0405} & \num{8.68e-05}\\
controls-2026-10-03-eu & \num{27} & \num{286.58} & \num{272.66} & \num{0.0016} & \num{1.62e-04}\\
controls-healthy-2026-10-03-eu & \num{124} & \num{192.51} & \num{307.38} & \num{0.0405} & \num{8.68e-05}\\
controls-healthy-2026-10-03-eu & \num{27} & \num{286.58} & \num{272.66} & \num{0.0016} & \num{1.62e-04}\\
controls-isolated-2026-10-03-eu & \num{124} & \num{192.51} & \num{307.38} & \num{0.0405} & \num{8.68e-05}\\
controls-isolated-2026-10-03-eu & \num{27} & \num{286.58} & \num{272.66} & \num{0.0016} & \num{1.62e-04}\\
controls-reversed-2026-10-03-eu & \num{124} & \num{192.51} & \num{307.38} & \num{0.0405} & \num{8.68e-05}\\
controls-reversed-2026-10-03-eu & \num{27} & \num{286.58} & \num{272.66} & \num{0.0016} & \num{1.62e-04}\\
pilot-2026-10-03-eu & \num{124} & \num{192.51} & \num{307.38} & \num{0.0405} & \num{8.68e-05}\\
pilot-2026-10-03-eu & \num{27} & \num{286.58} & \num{272.66} & \num{0.0016} & \num{1.62e-04}\\
validation-2026-10-03-eu & \num{124} & \num{192.51} & \num{307.38} & \num{0.0405} & \num{8.68e-05}\\
validation-2026-10-03-eu & \num{27} & \num{286.58} & \num{272.66} & \num{0.0016} & \num{1.62e-04}\\
validation-window2-2026-10-03-eu & \num{124} & \num{235.51} & \num{394.48} & \num{0.0415} & \num{8.82e-05}\\
validation-window2-2026-10-03-eu & \num{27} & \num{242.74} & \num{243.32} & \num{0.0012} & \num{1.22e-04}\\
validation-window3-2026-10-05-eu & \num{124} & \num{189.46} & \num{297.26} & \num{0.0356} & \num{8.82e-05}\\
validation-window3-2026-10-05-eu & \num{27} & \num{267.15} & \num{273.45} & \num{0.0027} & \num{1.22e-04}\\
\bottomrule
\end{tabular}
}
\caption{All archived calibration snapshots for Aachen H$_2$ control qubits 124 and 27. Fault-list status was not archived.}\label{S-tab:qubit-calibrations}
\end{table}

An effective affine Z response from raw known eigenstates has contrast
$c_Z=(m_0-m_1)/2$ and offset $d_Z=(m_0+m_1)/2$.
Under an assignment-channel interpretation,
$a=(1-c_Z+d_Z)/2$ and $b=(1-c_Z-d_Z)/2$, while X controls determine a
separate contrast and offset. On mixed Aachen qubit 124, the fitted values are
$a=\ChannelibmaachenmixedOneTwoFourraweffectivea$, $b=\ChannelibmaachenmixedOneTwoFourraweffectiveb$, $c_X=\ChannelibmaachenmixedOneTwoFourrawxcontrast$ and $d_X=\ChannelibmaachenmixedOneTwoFourrawxoffset$;
effective average assignment fidelity is \ChannelibmaachenmixedOneTwoFourraweffectiveassignmentfidelity{}, compared with the archived
readout error \AnomalousReadoutPercent{}\%. Healthy qubit 27 has effective fidelity \ChannelibmaachenmixedTwoSevenraweffectiveassignmentfidelity{}.
Applying the fitted affine X and Z responses to the frozen H$_2$ state predicts residuals
\ChannelibmaachenmixedOneTwoFourrawpredictedhTworesidualmha{} and \qty{\ChannelibmaachenmixedTwoSevenrawpredictedhTworesidualmha}{\millihartree}, compared with the same controls' observed \ChannelibmaachenmixedOneTwoFourrawobservedhTworesidualmha{} and
\qty{\ChannelibmaachenmixedTwoSevenrawobservedhTworesidualmha}{\millihartree} (returned $s=\RoundThreeEnergyErrorAnomalous$ and \qty{\RoundThreeEnergyErrorHealthy}{\millihartree}).
The eigenstate controls used for fitting and the molecular controls used for comparison share the same later job. This checks consistency within that workload. Predictive accuracy requires an independent execution. The fitted response includes initialization, state preparation, basis rotation and readout, so its assignment-channel interpretation is not unique.

The calibration-product model is $f_{\rm prod}=\prod_j(1-r_j)$ over all native
gate occurrences, multiplied by one $(1-r_{\rm readout})$ per occupied qubit.
Its global-depolarization prediction is
$\Delta E=(1-f_{\rm prod})(E_{\rm identity}-V)$, where $E_{\rm identity}$ includes
the Hamiltonian identity coefficient and common offsets. This is a signed contrast. A second CPU model evolves the compact ISA density matrix through independent depolarizing channels derived
from each gate's average infidelity, then applies archived asymmetric readout
probabilities to local QWC measurements. It keeps exact means, analytic local
sampling $\sigma$ and a seeded nominal-\ShotsLow{} sample, omitting idle relaxation,
leakage, correlated errors and noisy basis rotations. Measurement allocation
differs from Runtime.
An additional archived-properties model uses Aer's standard device gate channels,
including thermal relaxation during gates from $T_1$, $T_2$ and gate durations,
with depolarization adjusted to the reported gate infidelity~\cite{aer_noise}.
It clamps $T_2$ to $2T_1$ where required and applies the same readout channel;
provider measurement rotations remain unavailable. A further simulation inserts thermal-relaxation channels only in the idle gaps of the continuous-time ASAP schedule, while gate channels include their own thermal noise. The results are listed separately from the phenomenological total-duration attenuation used in Fig.~\ref{S-fig:idle-budget}. Each model uses ideal local QWC rotations and independent assignment channels.

\begin{table}[htbp]\centering\scriptsize
\resizebox{\linewidth}{!}{\begin{tabular}{lllllllll}
\toprule
Backend & System & Window & $f_{\rm prod}$ & $P_{\rm prod}$ & Observed/$P_{\rm prod}$ & Gate depol. & Gate thermal & Gate + idle\\
\midrule
Aachen & BeH2 & Pilot & \num{0.836} & \num{194.61} & \num{2.18} & \num{164.53} & \num{166.38} & \num{290.23}\\
Aachen & BeH2 & Window 1 & \num{0.836} & \num{194.61} & \num{2.30} & \num{164.53} & \num{166.38} & \num{290.23}\\
Aachen & BeH2 & Window 2 & \num{0.840} & \num{190.00} & \num{2.42} & \num{159.31} & \num{158.02} & \num{276.77}\\
Aachen & BeH2 & Window 3 & \num{0.810} & \num{225.89} & \num{1.98} & \num{192.40} & \num{194.58} & \num{312.68}\\
Aachen & H2 & Pilot & \num{0.959} & \num{32.84} & \num{15.02} & \num{59.52} & \num{59.41} & \num{59.41}\\
Aachen & H2O & Pilot & \num{0.802} & \num{405.77} & \num{3.43} & \num{496.37} & \num{490.62} & \num{657.87}\\
Aachen & LiH & Pilot & \num{0.978} & \num{5.51} & \num{1.59} & \num{3.51} & \num{3.49} & \num{3.49}\\
Aachen & LiH & Window 1 & \num{0.978} & \num{5.51} & \num{1.16} & \num{3.51} & \num{3.49} & \num{3.49}\\
Aachen & LiH & Window 2 & \num{0.984} & \num{3.94} & \num{1.41} & \num{2.57} & \num{2.57} & \num{2.57}\\
Aachen & LiH & Window 3 & \num{0.976} & \num{5.97} & \num{1.17} & \num{5.04} & \num{5.02} & \num{5.02}\\
Aachen & O2 & Pilot & \num{0.959} & \num{70.73} & \num{2.01} & \num{45.78} & \num{43.86} & \num{48.40}\\
Aachen & O2 & Window 1 & \num{0.959} & \num{70.73} & \num{1.61} & \num{45.78} & \num{43.86} & \num{48.40}\\
Aachen & O2 & Window 2 & \num{0.970} & \num{51.23} & \num{2.43} & \num{35.81} & \num{32.99} & \num{37.81}\\
Aachen & O2 & Window 3 & \num{0.955} & \num{76.28} & \num{1.56} & \num{55.46} & \num{52.78} & \num{57.39}\\
Pittsburgh & BeH2 & Pilot & \num{0.703} & \num{352.73} & \num{2.26} & \num{231.94} & \num{232.99} & \num{309.70}\\
Pittsburgh & BeH2 & Window 1 & \num{0.727} & \num{323.57} & \num{2.59} & \num{219.30} & \num{220.35} & \num{299.77}\\
Pittsburgh & BeH2 & Window 2 & \num{0.720} & \num{331.55} & \num{1.77} & \num{228.12} & \num{229.17} & \num{307.37}\\
Pittsburgh & BeH2 & Window 3 & \num{0.709} & \num{345.02} & \num{1.83} & \num{232.08} & \num{232.87} & \num{357.67}\\
Pittsburgh & H2 & Pilot & \num{0.990} & \num{8.29} & \num{1.40} & \num{-0.20} & \num{-0.30} & \num{-0.30}\\
Pittsburgh & H2O & Pilot & \num{0.673} & \num{672.42} & \num{2.24} & \num{914.04} & \num{911.78} & \num{1002.83}\\
Pittsburgh & LiH & Pilot & \num{0.986} & \num{3.42} & \num{1.07} & \num{2.52} & \num{2.50} & \num{2.50}\\
Pittsburgh & LiH & Window 1 & \num{0.986} & \num{3.33} & \num{0.83} & \num{2.41} & \num{2.38} & \num{2.38}\\
Pittsburgh & LiH & Window 2 & \num{0.983} & \num{4.08} & \num{0.64} & \num{2.79} & \num{2.76} & \num{2.76}\\
Pittsburgh & LiH & Window 3 & \num{0.981} & \num{4.63} & \num{0.80} & \num{3.64} & \num{3.60} & \num{3.60}\\
Pittsburgh & O2 & Pilot & \num{0.829} & \num{292.78} & \num{0.81} & \num{268.23} & \num{266.67} & \num{269.65}\\
Pittsburgh & O2 & Window 1 & \num{0.862} & \num{235.99} & \num{1.06} & \num{214.67} & \num{213.03} & \num{216.10}\\
Pittsburgh & O2 & Window 2 & \num{0.851} & \num{254.70} & \num{1.06} & \num{236.46} & \num{234.86} & \num{237.90}\\
Pittsburgh & O2 & Window 3 & \num{0.844} & \num{267.56} & \num{0.85} & \num{251.32} & \num{249.74} & \num{254.38}\\
\bottomrule
\end{tabular}
}
\caption{Calibration-product, gate-depolarizing and assignment-channel, and gate-thermal predictions versus
raw residuals from $V$, in mHa, for every saved circuit and matching window. The final column adds thermal channels during scheduled idle gaps to the gate-thermal model. Observed/product is the signed residual ratio. All channel simulations use ideal local QWC rotations.}
\label{S-tab:calibration-model}
\end{table}

For Aachen pilot H$_2$ the calibration-product prediction is \qty{\ModelibmaachenHTwopilotnaivedepolarizingresidualmha}{\millihartree} and observed/product
is \ModelibmaachenHTwopilotobservedovernaiveprediction{}. The gate-depolarizing and assignment-channel model predicts \qty{\ModelibmaachenHTwopilotdensitymatrixlocalqwcresidualmha}{\millihartree}. For pilot BeH$_2$ the same density-matrix model
predicts \qty{\ModelibmaachenBeHTwopilotdensitymatrixlocalqwcresidualmha}{\millihartree} versus observed \qty{\CellpilotibmaachenBeHTworawOneZeroTwoFourmeanmha}{\millihartree}. Across the isolated Pittsburgh
BeH$_2$ windows the density predictions are \ModelibmpittsburghBeHTwovalidationTwoZeroTwoSixOneZeroZeroThreeusdensitymatrixlocalqwcresidualmha{}, \ModelibmpittsburghBeHTwovalidationwindowTwoTwoZeroTwoSixOneZeroZeroThreeusdensitymatrixlocalqwcresidualmha{} and \qty{\ModelibmpittsburghBeHTwovalidationwindowThreeTwoZeroTwoSixOneZeroZeroFiveusdensitymatrixlocalqwcresidualmha}{\millihartree},
whereas the observed window means span \CellisolatedibmpittsburghBeHTworawOneZeroTwoFourwindowmeanminmha{}--\qty{\CellisolatedibmpittsburghBeHTworawOneZeroTwoFourwindowmeanmaxmha}{\millihartree}. Thus these models
remain far from the observations. Including on-gate thermal relaxation gives
\qty{\ModelibmaachenHTwopilotthermalgatereadoutresidualmha}{\millihartree} for Aachen pilot H$_2$ and \ModelibmpittsburghBeHTwovalidationTwoZeroTwoSixOneZeroZeroThreeusthermalgatereadoutresidualmha{}, \ModelibmpittsburghBeHTwovalidationwindowTwoTwoZeroTwoSixOneZeroZeroThreeusthermalgatereadoutresidualmha{} and \qty{\ModelibmpittsburghBeHTwovalidationwindowThreeTwoZeroTwoSixOneZeroZeroFiveusthermalgatereadoutresidualmha}{\millihartree} for the
Pittsburgh BeH$_2$ windows. This additional calibration-derived model does not
close those gaps. The omitted processes leave the source of the remaining residual magnitude and variability unidentified.
No provider ticket or later fault flag is present in the archive. Counts-level controls were not performed. The archived results therefore leave preparation and assignment contributions inseparable.

\section{TREX amplification and a workload hypothesis}\label{S-sec:trex-workload}
\begin{table}[htbp]\centering\small\begin{tabular}{lllll}
\toprule
Processor & Qubit & $N$ & $>3s$ & Max $|z|$\\
\midrule
Aachen & \num{27} & \num{42} & \num{19} & \num{50.7}\\
Aachen & \num{38} & \num{12} & \num{4} & \num{4.6}\\
Aachen & \num{124} & \num{30} & \num{22} & \num{31.3}\\
Pittsburgh & \num{61} & \num{30} & \num{6} & \num{4.1}\\
Pittsburgh & \num{77} & \num{30} & \num{4} & \num{7.0}\\
Pittsburgh & \num{112} & \num{12} & \num{2} & \num{5.2}\\
Pittsburgh & \num{114} & \num{12} & \num{2} & \num{5.2}\\
\bottomrule
\end{tabular}

\caption{Pauli-control departures by physical qubit. The $3s$ screen fires on every tested qubit, so it does not uniquely identify qubit 124.}
\label{S-tab:fifth-screen}\end{table}
Twirled readout error extinction (TREX) uses an inverse learned diagonal
response~\cite{vandenberg2022,cai2023}. In a scalar illustration $m_{\rm corr}=m/\epsilon$,
the sampling variance is amplified by $1/\epsilon^2$, and the output need not
lie in the physical interval. Matrix-inverse mitigation likewise permits
quasiprobabilities~\cite{nation2021}. An expectation greater than one is therefore
not itself evidence of an implementation fault.

If the uncorrected healthy expectation were exactly one, the mixed and reversed
values \RoundThreeKnownMixed\ and \RoundThreeKnownReversed\ imply effective $\epsilon=\RoundThreeInverseMixed$ and \RoundThreeInverseReversed;
isolated \RoundThreeKnownIsolated\ and healthy-pair \RoundThreeKnownHealthy\ imply \RoundThreeInverseIsolated\ and \RoundThreeInverseHealthy.
These reciprocal-output summaries describe the measured amplification:
raw input and learning conditions need not be identical. Returned $s$
does increase: mixed \TREXSmixed{} and reversed \TREXSreversed{} versus isolated \TREXSisolated{}
and healthy-pair \TREXShealthypair{}, about \TREXSInflationisolated{} and \TREXSInflationhealthypair{} times the mixed versus isolated and
mixed versus healthy-pair comparisons. That inflation does not include the distance
of the amplified expectation from the known value.

\paragraph{H1: shared workload response.}
A job-level learning and twirling policy influenced by a strongly nonideal support
could produce amplification on a healthy support in a mixed job, preserved under
order reversal but removed on isolation or healthy pairing. The Aachen data
are consistent with that signature, but shared learning matrices are not returned,
and sequential nonrandomized conditions leave temporal or preparation alternatives.
Pittsburgh's reversed condition also substituted physical qubit \RoundThreePittsburghWireReplacement\ for \RoundThreePittsburghWireOriginal, confounding order with qubit identity there.
The large corrected outputs coincide with mixed and reversed jobs, while isolated and healthy-pair controls are near their expected values. Temporal ordering and the missing learned responses prevent assigning this contrast to shared learning.

\begin{figure}[htbp]\centering
\includegraphics[width=.95\linewidth]{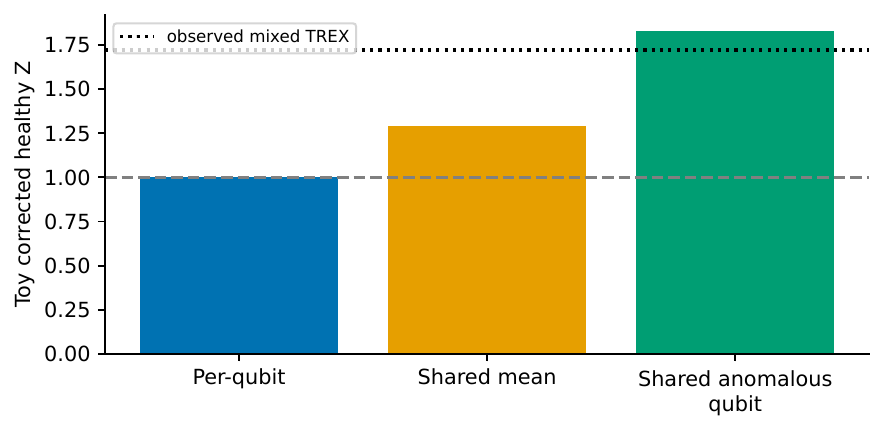}
\caption{CPU-only hypothetical corrections of healthy Z, using fitted raw
contrasts from the mixed controls: correct per-qubit, shared mean and shared
bad-qubit response. The dotted line is one observed mixed TREX output.}\label{S-fig:trex-toy}
\end{figure}

The shared-contrast and bad-qubit-only toys illustrate different assumed corrections (Fig.~\ref{S-fig:trex-toy}). Their resemblance to the measured amplification cannot identify the provider's correction because its learned matrices are unavailable. The decisive follow-up is specified in Sec.~\ref{S-sec:prospective}.

\section{ZNE diagnostics and returned measurement metadata}\label{S-sec:zne-traces}
\begin{figure}[htbp]\centering
\includegraphics[width=.95\linewidth]{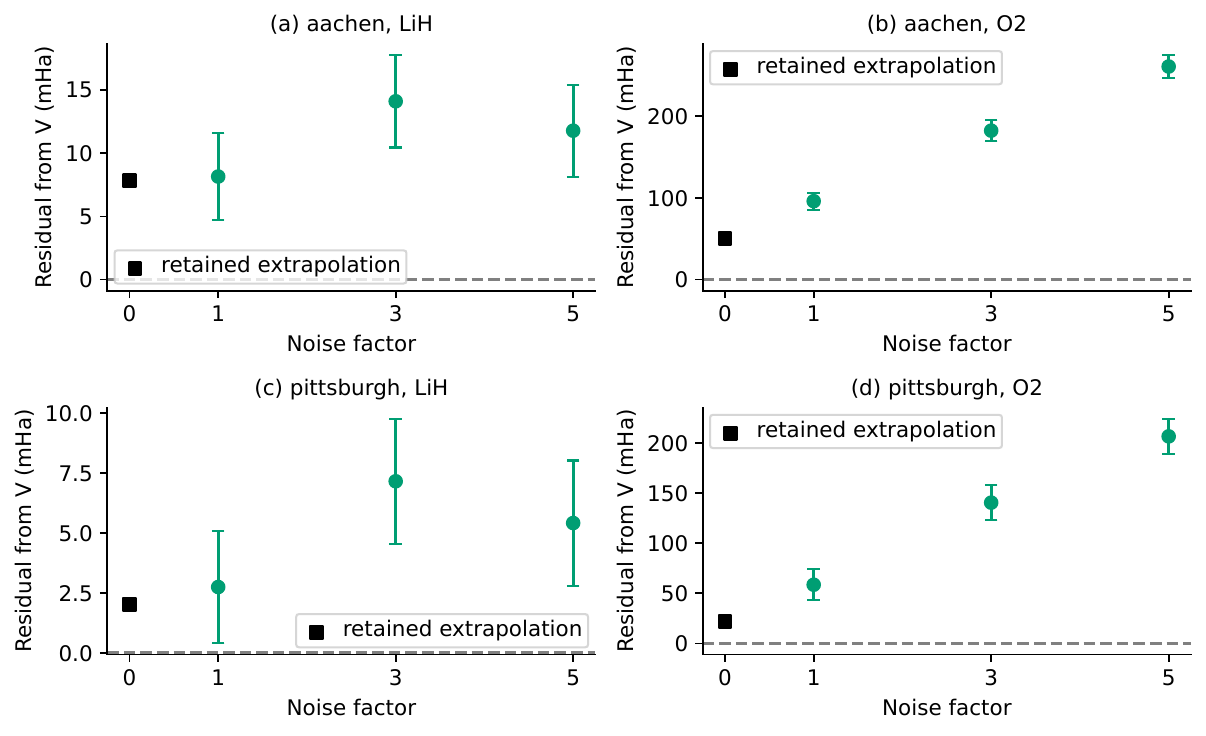}
\caption{Pilot energy residuals from $V$ at noise factors 1,3,5 and the stored
zero-noise estimate, one ZNE PUB per processor and system (panels a--d).
Bars at nonzero factors are returned per-factor $s$. The stored point is shown
without an extrapolation interval.}\label{S-fig:zne-factors}
\end{figure}

The ZNE package combines gate twirling, noise amplification and extrapolation (Fig.~\ref{S-fig:zne-factors}). The experiment compares that bundle with raw execution.
All four pilot PUBs have nonfinite homogeneous exponential arrays and finite
aggregate linear arrays. LiH has nine mapped Pauli terms and O$_2$ fifty
(Table~\ref{tab:references}). The stored \texttt{evs} is the
predeclared provider estimate; aggregate arrays do not expose whether every
individual Pauli-term fit was finite, because Runtime may combine heterogeneous
per-term extrapolators~\cite{ibm_zne}. Returned variance-amplification comparisons and the independent linear-intercept model are archived. Table~\ref{S-tab:resolved-pub} summarizes the returned measurement fields.

\begin{table}[htbp]\centering\small
\begin{tabular}{llllll}
\toprule
Arm & Nominal shots & Precision & Randomizations & Bases & Total shots\\
\midrule
raw & \num{1024} & \num{0.03125} & \num{1} & Not returned & Not returned\\
raw & \num{2048} & \num{0.0220971} & \num{1} & Not returned & Not returned\\
raw & \num{4096} & \num{0.015625} & \num{1} & Not returned & Not returned\\
trex & \num{1024} & \num{0.03125} & \num{32} & Not returned & Not returned\\
trex & \num{2048} & \num{0.0220971} & \num{32} & Not returned & Not returned\\
zne & \num{1024} & \num{0.03125} & \num{32} & Not returned & Not returned\\
\bottomrule
\end{tabular}

\caption{Resolved PUB metadata. The archive contains exact per-PUB records.
The resolved shots field excludes the total work across measurement bases, amplification factors and noise learning.
No complete executed-circuit and shot count is inferred from nominal settings.}
\label{S-tab:resolved-pub}
\end{table}

\section{Additional descriptive and ideal-model results}\label{S-sec:additional}
\begin{figure}[htbp]\centering
\includegraphics[width=.95\linewidth]{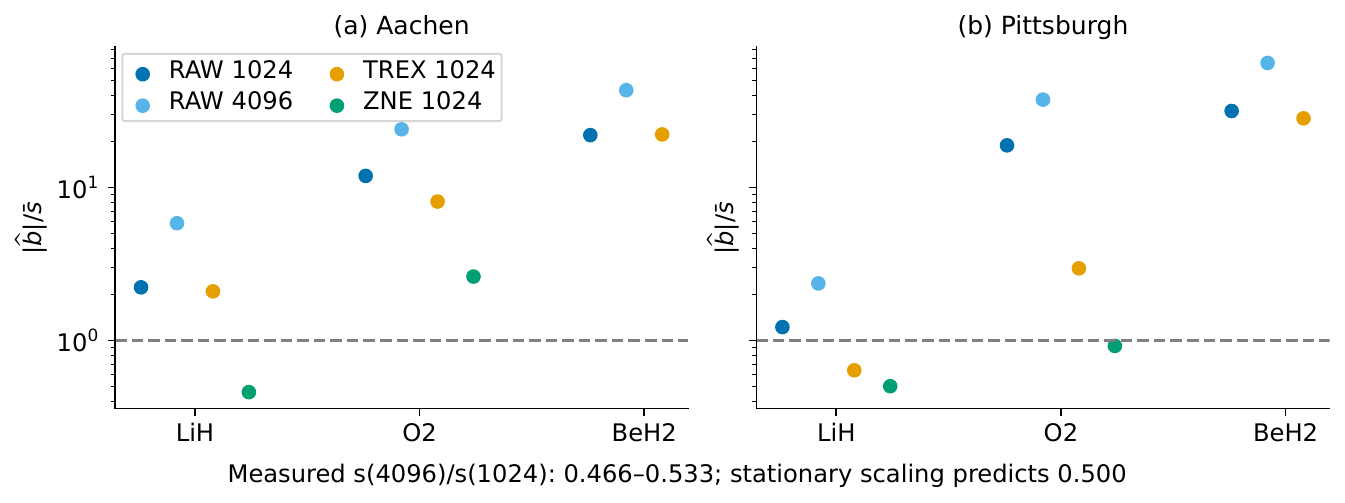}
\caption{Measured residual-to-returned-uncertainty ratios at $V$, log scale:
(a) Aachen and (b) Pittsburgh, six raw-\ShotsLow{} or three other jobs per cell.
Colors distinguish nominal shots and intervention packages.}\label{S-fig:crossover-all}
\end{figure}

Figure~\ref{S-fig:crossover-all} shows the isolated ratios. Pilot crossover ratios are archived separately: these bundled workloads have one window,
so their window SD is unknown and their single-observation extrapolations are
especially unstable. They are not pooled with isolated-job crossover cells.
\begin{table}[htbp]\centering\scriptsize
\resizebox{\linewidth}{!}{\begin{tabular}{lllllllll}
\toprule
Backend & System & Arm & $N$ & $\bar r$ & $\bar s$ & $R$ & $L$ & $N^*$\\
\midrule
ibm\_aachen & BeH2 & raw & \num{1024} & \num{452} & \num{20.512} & \num{22} & \num{0.0021} & \num{2.1}\\
ibm\_aachen & BeH2 & raw & \num{4096} & \num{446} & \num{10.305} & \num{43} & \num{0.00053} & \num{2.2}\\
ibm\_aachen & BeH2 & trex & \num{1024} & \num{457} & \num{20.544} & \num{22} & \num{0.002} & \num{2.1}\\
ibm\_aachen & LiH & raw & \num{1024} & \num{6.3} & \num{2.8266} & \num{2.2} & \num{0.17} & \num{2.1e+02}\\
ibm\_aachen & LiH & raw & \num{4096} & \num{8.7} & \num{1.4874} & \num{5.8} & \num{0.028} & \num{1.2e+02}\\
ibm\_aachen & LiH & trex & \num{1024} & \num{6.5} & \num{3.1042} & \num{2.1} & \num{0.19} & \num{2.3e+02}\\
ibm\_aachen & LiH & zne & \num{1024} & \num{3.1} & \num{6.7559} & \num{0.46} & \num{0.83} & \num{4.8e+03}\\
ibm\_aachen & O2 & raw & \num{1024} & \num{119} & \num{9.9959} & \num{12} & \num{0.007} & \num{7.2}\\
ibm\_aachen & O2 & raw & \num{4096} & \num{119} & \num{4.9564} & \num{24} & \num{0.0017} & \num{7.1}\\
ibm\_aachen & O2 & trex & \num{1024} & \num{87} & \num{10.707} & \num{8.1} & \num{0.015} & \num{16}\\
ibm\_aachen & O2 & zne & \num{1024} & \num{60.0} & \num{22.886} & \num{2.6} & \num{0.13} & \num{1.5e+02}\\
ibm\_pittsburgh & BeH2 & raw & \num{1024} & \num{685} & \num{21.637} & \num{32} & \num{0.001} & \num{1}\\
ibm\_pittsburgh & BeH2 & raw & \num{4096} & \num{704} & \num{10.793} & \num{65} & \num{0.00023} & \num{0.96}\\
ibm\_pittsburgh & BeH2 & trex & \num{1024} & \num{692} & \num{24.415} & \num{28} & \num{0.0012} & \num{1.3}\\
ibm\_pittsburgh & LiH & raw & \num{1024} & \num{3.0} & \num{2.4641} & \num{1.2} & \num{0.4} & \num{6.8e+02}\\
ibm\_pittsburgh & LiH & raw & \num{4096} & \num{2.8} & \num{1.1918} & \num{2.3} & \num{0.15} & \num{7.3e+02}\\
ibm\_pittsburgh & LiH & trex & \num{1024} & \num{1.6} & \num{2.4464} & \num{0.65} & \num{0.71} & \num{2.5e+03}\\
ibm\_pittsburgh & LiH & zne & \num{1024} & \num{2.0} & \num{4.0704} & \num{0.49} & \num{0.8} & \num{4e+03}\\
ibm\_pittsburgh & O2 & raw & \num{1024} & \num{250} & \num{13.231} & \num{19} & \num{0.0028} & \num{2.9}\\
ibm\_pittsburgh & O2 & raw & \num{4096} & \num{246} & \num{6.5475} & \num{38} & \num{0.00071} & \num{2.9}\\
ibm\_pittsburgh & O2 & trex & \num{1024} & \num{51} & \num{17.138} & \num{3} & \num{0.1} & \num{1.2e+02}\\
ibm\_pittsburgh & O2 & zne & \num{1024} & \num{41.3} & \num{44.752} & \num{0.92} & \num{0.54} & \num{1.2e+03}\\
\bottomrule
\end{tabular}
}
\caption{All \IsolatedCells{} isolated treatment and shot cells: measured mean residual and
returned uncertainty, bias-dominance ratio $R$, maximum model MSE leverage $L$
and model-dependent nominal crossover $N^*$. Units are mHa except ratios and shot settings;
each cell has three windows, raw-\ShotsLow{} with six jobs and other supported cells three.
The crossover assumes a stationary model.}
\label{S-tab:crossover-all}\end{table}
\begin{table}[htbp]\centering\small
\begin{tabular}{llllll}
\toprule
System & Ansatz & Parameters & $V-A$ (\unit{\millihartree}) & CX depth & CX count\\
\midrule
H2 & UCCD & \num{1} & $<10^{-5}$ & \num{4} & \num{0}\\
H2 & UCCSD & \num{1} & $<10^{-5}$ & \num{4} & \num{0}\\
LiH & UCCD & \num{1} & \num{0.0156} & \num{11} & \num{2}\\
LiH & UCCSD & \num{3} & $<10^{-5}$ & \num{12} & \num{2}\\
O2 & UCCD & \num{1} & $<10^{-5}$ & \num{24} & \num{11}\\
O2 & UCCSD & \num{1} & $<10^{-5}$ & \num{24} & \num{11}\\
BeH2 & UCCD & \num{8} & \num{0.0007} & \num{274} & \num{137}\\
BeH2 & UCCSD & \num{12} & \num{0.0007} & \num{281} & \num{137}\\
H2O & UCCD & \num{8} & \num{0.1133} & \num{274} & \num{142}\\
H2O & UCCSD & \num{12} & \num{0.0003} & \num{290} & \num{148}\\
\bottomrule
\end{tabular}

\caption{CPU-only stationary UCCD and UCCSD comparison from frozen active integrals,
with three starts per ansatz. Gaps $V-A$ are mHa, depths and gate counts use a common logical CX basis. O$_2$ retains the same single parameter
after tapering. The selected O$_2$ sector is described in Sec.~\ref{S-sec:orbital-audit}.
}\label{S-tab:ansatz}
\end{table}

Adding singles lowers LiH's gap from \qty{\RoundThreeLiHGap}{\millihartree} to below $10^{-5}$~mHa and
H$_2$O's from \qty{\RoundThreeWaterGap}{\millihartree} to \qty{\RoundThreeWaterSinglesGap}{\millihartree} (four-decimal values in the table),
while common logical depths increase
from \AnsatzLiHUCCDlogicalcxdepth{} to \AnsatzLiHUCCSDlogicalcxdepth{} and \AnsatzHTwoOUCCDlogicalcxdepth{} to \AnsatzHTwoOUCCSDlogicalcxdepth{}. All stored $S^2$ expectations match the requested
sector. UCCD was selected to reuse validated circuits and isolate execution
effects with already small ansatz gaps.

The local ideal coverage comparison uses exactly the same \IdealRepeats{} sampled energies
per cell and replaces only the estimated within-group uncertainty by Eq.~(\ref{eq:variance})'s
known $\sigma$. LiH at nominal \ShotsLow{} includes $V$ \LiHEstimatedHits{}/\IdealRepeats{} times within $2s$
and \LiHExactHits{}/\IdealRepeats{} times within $2\sigma$. Other cells and one-standard-deviation
comparisons are archived. The estimated-uncertainty LiH inclusion has a 95\% Wilson interval of \LithiumCoverageLow--\LithiumCoverageHigh\%, below the nominal two-standard-deviation Gaussian inclusion. The exact-variance comparison isolates the effect of estimating within-group variances in the local sampler.

\begin{table}[htbp]\centering\begin{tabular}{llrrrr}
\toprule
Backend & System & Raw 1024 & Raw 4096 & TREX & ZNE\\
\midrule
Pittsburgh & LiH & 6/6 & 1/3 & 3/3 & 3/3\\
Pittsburgh & O2 & 0/6 & 0/3 & 0/3 & 3/3\\
Pittsburgh & BeH2 & 0/6 & 0/3 & 0/3 & --\\
Aachen & LiH & 1/6 & 0/3 & 1/3 & 3/3\\
Aachen & O2 & 0/6 & 0/3 & 0/3 & 1/3\\
Aachen & BeH2 & 0/6 & 0/3 & 0/3 & --\\
\bottomrule
\end{tabular}

\caption{Inclusion of $V$ within $2s$ in isolated molecular jobs, with raw-\ShotsLow{}
six jobs and other supported arms three per cell.  }\label{S-tab:coverage}
\end{table}
The data archive includes complete per-job listings and a data dictionary.
Empty metadata are null. The proposed follow-up experiments have not been performed.

\subsection{Decisive workload-response test}\label{S-sec:prospective}
The decisive test is randomized same-window one-PUB versus multi-PUB execution of identical circuits on fixed qubits with counterbalanced order. The measured TREX charges give an estimate of \qty{\RoundThreeControlCost}{\second} charge per processor for the TREX-only jobs at the archived mean \qty{\RoundThreeControlJobCost}{\second}/job, with a \qty{\RoundThreeControlCeiling}{\second} ceiling using the maximum observed job charge. No low-shot discount is assumed. 

\section{Accounting, timing and derived statistical diagnostics}
\begin{figure}[t]\centering
\includegraphics[width=.92\linewidth]{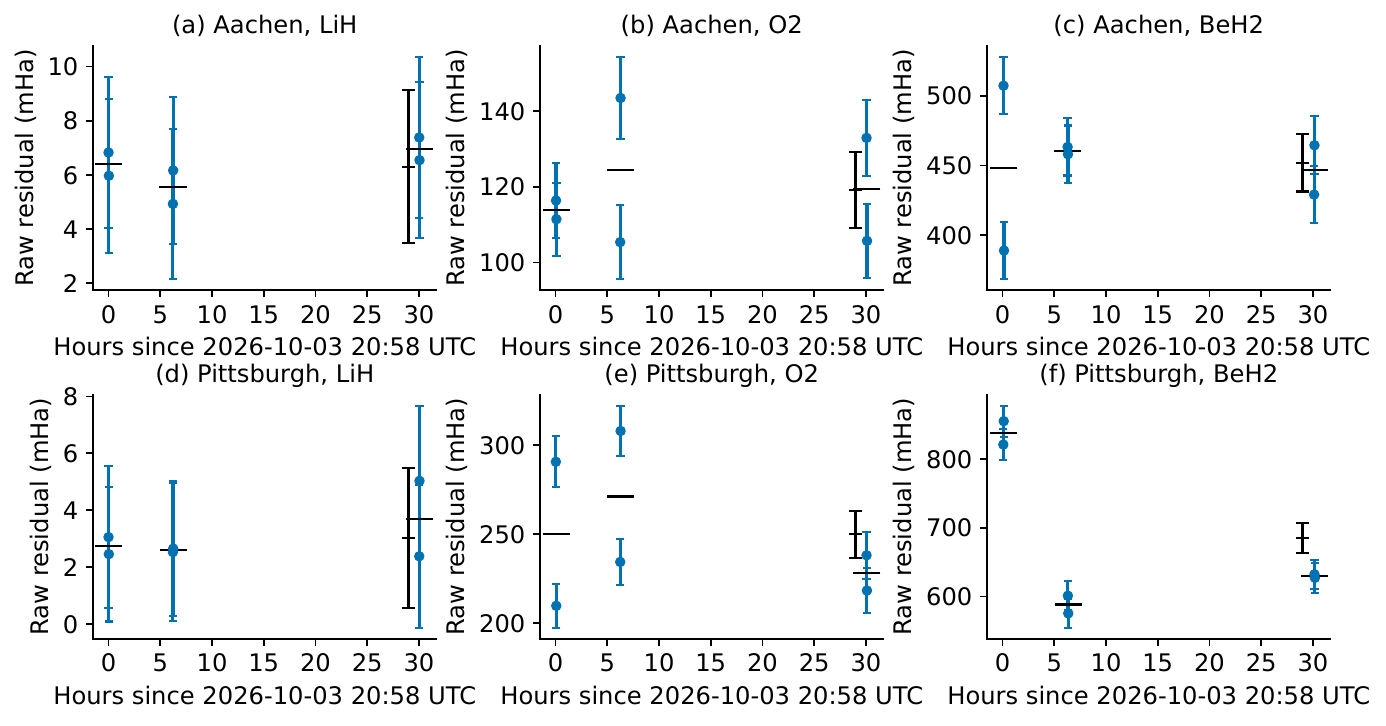}
\caption{Raw-\ShotsLow{} residuals from $V$ over provider execution times, six jobs
per cell. Panels (a--c) Aachen and (d--f) Pittsburgh have different vertical scales.
Bars are returned $s$, black marks are two-job window means, and the reference
bar shows mean returned $s$. All circuits and layouts are unchanged across windows.
}
\label{S-fig:supp-temporal}\end{figure}

\begin{table}[htbp]\centering\small\begin{tabular}{llll}
\toprule
Arm & Jobs & $|Y-V|\leq s$ & $|Y-V|\leq2s$\\
\midrule
Raw 1024 & \num{36} & \num{1} & \num{7}\\
Raw 4096 & \num{18} & \num{0} & \num{1}\\
TREX & \num{18} & \num{3} & \num{4}\\
ZNE & \num{12} & \num{7} & \num{10}\\
\bottomrule
\end{tabular}

\caption{Inclusion of the frozen ideal energy $V$ within one and two returned Estimator standard errors over all isolated jobs.}
\label{S-tab:fifth-inclusion}\end{table}
\begin{table}[htbp]\centering\small\begin{tabular}{lllll}
\toprule
Processor & System & $F(2,3)$ & $p$ & Adjusted $p$\\
\midrule
Aachen & BeH2 & \num{0.05} & \num{0.955530} & \num{1.000000}\\
Aachen & LiH & \num{2.06} & \num{0.273608} & \num{1.000000}\\
Aachen & O2 & \num{0.15} & \num{0.866823} & \num{1.000000}\\
Pittsburgh & BeH2 & \num{116.33} & \num{0.001436} & \num{0.008617}\\
Pittsburgh & LiH & \num{0.59} & \num{0.608664} & \num{1.000000}\\
Pittsburgh & O2 & \num{0.45} & \num{0.675640} & \num{1.000000}\\
\bottomrule
\end{tabular}

\caption{Exploratory balanced one-way F tests of between-window differences for the six Raw 1024 cells. The adjusted $p$ values use a six-cell Bonferroni correction.}
\label{S-tab:fifth-anova}\end{table}
\begin{table}[htbp]\centering\small\begin{tabular}{llllll}
\toprule
Processor & System & Mean & SD & Min & Max\\
\midrule
Aachen & BeH2 & \num{-29.3} & \num{79.6} & \num{-118.2} & \num{35.4}\\
Aachen & LiH & \num{-0.2} & \num{1.2} & \num{-0.9} & \num{1.2}\\
Aachen & O2 & \num{2.0} & \num{33.2} & \num{-27.2} & \num{38.1}\\
Pittsburgh & BeH2 & \num{1.1} & \num{30.3} & \num{-25.7} & \num{34.0}\\
Pittsburgh & LiH & \num{0.7} & \num{1.7} & \num{-0.6} & \num{2.6}\\
Pittsburgh & O2 & \num{-9.0} & \num{77.7} & \num{-80.9} & \num{73.5}\\
\bottomrule
\end{tabular}

\caption{Last minus first Raw 1024 residual in each window, summarized over three windows per device and system. Units are mHa; the entries quantify baseline movement across the fixed arm sequence.}
\label{S-tab:fourth-brackets}\end{table}
\begin{table}[htbp]\centering\small\begin{tabular}{lllll}
\toprule
Processor & System & $f_{\rm obs}$ & $f_{\rm cal}$ & Loss ratio\\
\midrule
Aachen & BeH2 & \num{0.619} & \num{0.828} & \num{2.22}\\
Aachen & LiH & \num{0.974} & \num{0.979} & \num{1.23}\\
Aachen & O2 & \num{0.930} & \num{0.961} & \num{1.80}\\
Pittsburgh & BeH2 & \num{0.422} & \num{0.719} & \num{2.06}\\
Pittsburgh & LiH & \num{0.988} & \num{0.984} & \num{0.75}\\
Pittsburgh & O2 & \num{0.854} & \num{0.853} & \num{0.99}\\
\bottomrule
\end{tabular}

\caption{Observed and calibration-product retained contrasts for the six isolated Raw 1024 cells. Loss ratio is $(1-f_{\rm obs})/(1-f_{\rm cal})$; both contrasts use the same tapered $E_{\rm mix}$.}
\label{S-tab:fourth-fidelity}\end{table}
Table~\ref{S-tab:supp-budget} gives acquisition charges; Fig.~\ref{S-fig:supp-temporal} shows the raw time series. Tables~\ref{S-tab:spectrum} and \ref{S-tab:variance-components} give spectral baselines and the balanced moment decomposition.
\begin{table}[t]\centering\small
\resizebox{\linewidth}{!}{\begin{tabular}{lllllll}
\toprule
Backend & System & $\bar r$ & $\bar s$ & $R$ & $N^*$ & $\mathrm{SD}_w$\\
\midrule
ibm\_aachen & BeH2 & \num{452} & \num{20.512} & \num{22} & \num{2.1} & \num{7.66}\\
ibm\_aachen & LiH & \num{6.3} & \num{2.8266} & \num{2.2} & \num{2.1e+02} & \num{0.712}\\
ibm\_aachen & O2 & \num{119} & \num{9.9959} & \num{12} & \num{7.2} & \num{5.26}\\
ibm\_pittsburgh & BeH2 & \num{685} & \num{21.637} & \num{32} & \num{1} & \num{134}\\
ibm\_pittsburgh & LiH & \num{3.0} & \num{2.4641} & \num{1.2} & \num{6.8e+02} & \num{0.602}\\
ibm\_pittsburgh & O2 & \num{250} & \num{13.231} & \num{19} & \num{2.9} & \num{21.5}\\
\bottomrule
\end{tabular}
}
\caption{Measured raw-\ShotsLow{} mean residual $\bar r$ and mean returned $\bar s$
(mHa), bias-dominance ratio $R$, model-dependent nominal crossover $N^*$,
and SD of the three window means (mHa). Each cell has six jobs in three windows.
Values below the tested shot range extrapolate the model. Other arms appear in
Table~\ref{S-tab:crossover-all}.}
\label{S-tab:supp-crossover}\end{table}
\begin{table}[t]\centering\small
\begin{tabular}{llll}
\toprule
Acquisition & Jobs/processor & Aachen charge (\unit{\second}) & Pittsburgh charge (\unit{\second})\\
\midrule
Pilot & \num{5} & \num{124} & \num{126}\\
Known-state & \num{8} & \num{125} & \num{125}\\
Isolated window 1 & \num{14} & \num{197} & \num{201}\\
Isolated window 2 & \num{14} & \num{197} & \num{201}\\
Isolated window 3 & \num{14} & \num{197} & \num{201}\\
Total & \num{55} & \num{840} & \num{854}\\
\bottomrule
\end{tabular}

\caption{Complete provider job charges: \JobCount{} jobs, \BenchmarkCount{} benchmark molecular energies,
\DiagnosticCount{} diagnostic scalar estimates (including \EnergyControlCount{} H$_2$ energy controls) and
\qty{\ChargeSeconds}{\second} (\qty{\Scalarbilledminutes}{\minute}). Charges use
\texttt{usage.qpu\_charge\_time\_seconds}, including the complete billed job;
learning and twirling components are not separately itemized.}
\label{S-tab:supp-budget}\end{table}
\begin{table}[htbp]\centering\scriptsize
\resizebox{\linewidth}{!}{\begin{tabular}{llllllll}
\toprule
Backend & Cohort & Jobs & Benchmark & H2 controls & Pauli & PUBs & QPU charge (\unit{\second})\\
\midrule
ibm\_aachen & pilot & \num{5} & \num{19} & \num{0} & \num{0} & \num{19} & \num{124}\\
ibm\_aachen & known\_state & \num{8} & \num{0} & \num{14} & \num{84} & \num{98} & \num{125}\\
ibm\_aachen & isolated & \num{42} & \num{42} & \num{0} & \num{0} & \num{42} & \num{591}\\
ibm\_pittsburgh & pilot & \num{5} & \num{19} & \num{0} & \num{0} & \num{19} & \num{126}\\
ibm\_pittsburgh & known\_state & \num{8} & \num{0} & \num{14} & \num{84} & \num{98} & \num{125}\\
ibm\_pittsburgh & isolated & \num{42} & \num{42} & \num{0} & \num{0} & \num{42} & \num{603}\\
\bottomrule
\end{tabular}
}
\caption{Jobs, benchmark energies, H$_2$ control energies, Pauli controls, PUBs
and billed seconds by cohort and backend. Counts deduplicate provider job identities.}\label{S-tab:cohorts}
\end{table}
\begin{table}[htbp]\centering\scriptsize
\resizebox{\linewidth}{!}{\begin{tabular}{lllll}
\toprule
Backend & Cohort/window & First start UTC & Last finish UTC & Elapsed (\unit{\minute})\\
\midrule
ibm\_aachen & pilot & 2026-10-03 19:28:30 & 2026-10-03 19:32:08 & \num{3.65}\\
ibm\_aachen & known-state & 2026-10-03 20:49:14 & 2026-10-03 20:56:30 & \num{7.26}\\
ibm\_aachen & validation-2026-10-03-eu & 2026-10-03 20:58:34 & 2026-10-03 21:07:48 & \num{9.24}\\
ibm\_aachen & validation-window2-2026-10-03-eu & 2026-10-04 03:11:49 & 2026-10-04 03:18:16 & \num{6.46}\\
ibm\_aachen & validation-window3-2026-10-05-eu & 2026-10-05 03:00:10 & 2026-10-05 03:07:41 & \num{7.51}\\
ibm\_pittsburgh & pilot & 2026-10-03 19:32:53 & 2026-10-03 19:36:10 & \num{3.29}\\
ibm\_pittsburgh & known-state & 2026-10-03 20:49:49 & 2026-10-03 20:56:42 & \num{6.87}\\
ibm\_pittsburgh & validation-2026-10-03-us & 2026-10-03 20:58:33 & 2026-10-03 21:05:54 & \num{7.35}\\
ibm\_pittsburgh & validation-window2-2026-10-03-us & 2026-10-04 03:11:50 & 2026-10-04 03:20:06 & \num{8.27}\\
ibm\_pittsburgh & validation-window3-2026-10-05-us & 2026-10-05 03:00:09 & 2026-10-05 03:09:56 & \num{9.79}\\
\bottomrule
\end{tabular}
}
\caption{First provider start and final finish per cohort or isolated window, in UTC.
Elapsed minutes are wall time, distinct from billed QPU seconds. The isolated
cohort spans \qty{\IsolatedSpanHours}{\hour}. Including pilot and known-state controls the campaign
spans \qty{\ScalarroundTwocampaignhours}{\hour}.}\label{S-tab:timing}
\end{table}
\begin{table}[htbp]\centering\scriptsize
\resizebox{\linewidth}{!}{\begin{tabular}{lllll}
\toprule
System & $\lambda_{\min}$ (\unit{\hartree}) & $\lambda_{\max}$ (\unit{\hartree}) & $E_{\rm mix}$ (\unit{\hartree}) & $E_{\rm HF}$ (\unit{\hartree})\\
\midrule
H2 & \num{-1.137253} & \num{0.476542} & \num{-0.330355} & \num{-1.116606}\\
LiH & \num{-7.861408} & \num{-7.167646} & \num{-7.614969} & \num{-7.861134}\\
O2 & \num{-147.633158} & \num{-144.601727} & \num{-145.919457} & \num{-147.632409}\\
BeH2 & \num{-15.566024} & \num{-11.648275} & \num{-14.379967} & \num{-15.560131}\\
H2O & \num{-74.972318} & \num{-68.826504} & \num{-72.918692} & \num{-74.964517}\\
\bottomrule
\end{tabular}
}
\caption{Exact spectral bounds, maximally mixed energy and frozen HF energy of
the tapered observables, including common total-energy offsets.}
\label{S-tab:spectrum}\end{table}
\begin{table}[htbp]\centering\scriptsize
\resizebox{\linewidth}{!}{\begin{tabular}{llllll}
\toprule
Backend & System & $\mathrm{SD}_{\rm within}$ & $\mathrm{SD}_{\rm epoch}$ & $\mathrm{SD}_{w}$ & $\bar s$\\
\midrule
ibm\_aachen & BeH2 & \num{50.40} & \num{0.00} & \num{7.66} & \num{20.512}\\
ibm\_aachen & LiH & \num{0.70} & \num{0.51} & \num{0.71} & \num{2.8266}\\
ibm\_aachen & O2 & \num{19.21} & \num{0.00} & \num{5.26} & \num{9.9959}\\
ibm\_pittsburgh & BeH2 & \num{17.54} & \num{133.19} & \num{133.76} & \num{21.637}\\
ibm\_pittsburgh & LiH & \num{1.11} & \num{0.00} & \num{0.60} & \num{2.4641}\\
ibm\_pittsburgh & O2 & \num{45.34} & \num{0.00} & \num{21.46} & \num{13.231}\\
\bottomrule
\end{tabular}
}
\caption{Raw-\ShotsLow{} balanced moment decomposition, in mHa: within-window SD,
between-window-component SD, SD of window means and mean returned $s$. Each cell has six
jobs, with within/between degrees of freedom three/two; negative signed between-window
variance estimates keep their signs in the archive and are clipped at zero only when converted to a displayed standard deviation.}
\label{S-tab:variance-components}\end{table}
\begin{table}[htbp]\centering\scriptsize
\resizebox{\linewidth}{!}{\begin{tabular}{llllllll}
\toprule
Backend & System & Arm & $N$ & Mean charge (\unit{\second}) & Min charge (\unit{\second}) & Max charge (\unit{\second}) & Ratio\\
\midrule
ibm\_aachen & BeH2 & raw & \num{1024} & \num{12.00} & \num{12} & \num{12} & \num{1.00}\\
ibm\_aachen & BeH2 & raw & \num{4096} & \num{43.00} & \num{43} & \num{43} & \num{3.58}\\
ibm\_aachen & BeH2 & trex & \num{1024} & \num{23.00} & \num{23} & \num{23} & \num{1.92}\\
ibm\_aachen & LiH & raw & \num{1024} & \num{3.00} & \num{3} & \num{3} & \num{1.00}\\
ibm\_aachen & LiH & raw & \num{4096} & \num{6.00} & \num{6} & \num{6} & \num{2.00}\\
ibm\_aachen & LiH & trex & \num{1024} & \num{12.00} & \num{12} & \num{12} & \num{4.00}\\
ibm\_aachen & LiH & zne & \num{1024} & \num{15.00} & \num{15} & \num{15} & \num{5.00}\\
ibm\_aachen & O2 & raw & \num{1024} & \num{6.00} & \num{6} & \num{6} & \num{1.00}\\
ibm\_aachen & O2 & raw & \num{4096} & \num{17.00} & \num{17} & \num{17} & \num{2.83}\\
ibm\_aachen & O2 & trex & \num{1024} & \num{15.00} & \num{15} & \num{15} & \num{2.50}\\
ibm\_aachen & O2 & zne & \num{1024} & \num{24.00} & \num{24} & \num{24} & \num{4.00}\\
ibm\_pittsburgh & BeH2 & raw & \num{1024} & \num{13.00} & \num{13} & \num{13} & \num{1.00}\\
ibm\_pittsburgh & BeH2 & raw & \num{4096} & \num{44.00} & \num{44} & \num{44} & \num{3.38}\\
ibm\_pittsburgh & BeH2 & trex & \num{1024} & \num{23.00} & \num{23} & \num{23} & \num{1.77}\\
ibm\_pittsburgh & LiH & raw & \num{1024} & \num{3.00} & \num{3} & \num{3} & \num{1.00}\\
ibm\_pittsburgh & LiH & raw & \num{4096} & \num{6.00} & \num{6} & \num{6} & \num{2.00}\\
ibm\_pittsburgh & LiH & trex & \num{1024} & \num{12.00} & \num{12} & \num{12} & \num{4.00}\\
ibm\_pittsburgh & LiH & zne & \num{1024} & \num{15.00} & \num{15} & \num{15} & \num{5.00}\\
ibm\_pittsburgh & O2 & raw & \num{1024} & \num{6.00} & \num{6} & \num{6} & \num{1.00}\\
ibm\_pittsburgh & O2 & raw & \num{4096} & \num{17.00} & \num{17} & \num{17} & \num{2.83}\\
ibm\_pittsburgh & O2 & trex & \num{1024} & \num{15.00} & \num{15} & \num{15} & \num{2.50}\\
ibm\_pittsburgh & O2 & zne & \num{1024} & \num{25.00} & \num{25} & \num{25} & \num{4.17}\\
\bottomrule
\end{tabular}
}
\caption{Per-system isolated job charges and ratios to the same system and backend's
raw-\ShotsLow{} charge. Values are billed seconds.}
\label{S-tab:system-costs}
\end{table}

The global-depolarization energy inversion is $f_{\rm obs}=(Y-E_{\rm mix})/(V-E_{\rm mix})$. $E_{\rm mix}$ is the maximally mixed energy of the tapered observable; because tapering removes symmetry sectors, device depolarization differs from depolarization within the reduced space, which is a further reason to read $f_{\rm obs}$ as a signed energy contrast.
It is signed and unclipped. Values outside $[0,1]$ reject that single-parameter
model, and values inside it do not measure state fidelity. All cells, HF baselines
and spectrum and correlation-energy normalized residuals are archived. Calibration products include
native gate occurrences and readout. Layer-fidelity metrics for the used layouts were not archived.

For independent and identically distributed (iid) Gaussian jobs, nominal chi-square SD intervals and Student-t signed-mean intervals
are archived for sensitivity analysis.
They are not claimed as valid hardware confidence intervals: a three-window sample
with substantial between-window variation violates the iid model. Likewise, the O$_2$
one-sided and BeH$_2$ two-sided sign tests give $p=\OxygenSignP$ and $p=\BerylliumSignP$ only under independent signs, with no multiplicity correction. The six O$_2$ contrasts share three acquisition windows, so their independence is unverified.
For the six-job Pittsburgh BeH$_2$ raw-\ShotsLow{} headline, the iid-reference
95\% dispersion-ratio interval is \RepeatabilityIIDLow--\RepeatabilityIIDHigh,
and the Student-t signed-bias-ratio interval is
\BiasRatioIIDLow--\BiasRatioIIDHigh.

The main text compares the independent assignment-removal model with the Pittsburgh O$_2$ TREX improvement (main Sec.~\ref{sec:hardware}).

The Aachen multi-PUB pilot makes the O$_2$ absolute residual worse
(\CellpilotibmaachenOTworawOneZeroTwoFourmeanmha{} to \qty{\CellpilotibmaachenOTwotrexOneZeroTwoFourAbsoluteMean}{\millihartree}), while isolated acquisitions improve it. Window and PUB
multiplicity change together. Hypothesis H2 is a workload-dependent mitigation
response, to be tested by randomized same-window one-PUB versus multi-PUB executions
of identical circuits. This is separate from the healthy-qubit response-sharing H1.

\begin{figure}[htbp]\centering
\includegraphics[width=.95\linewidth]{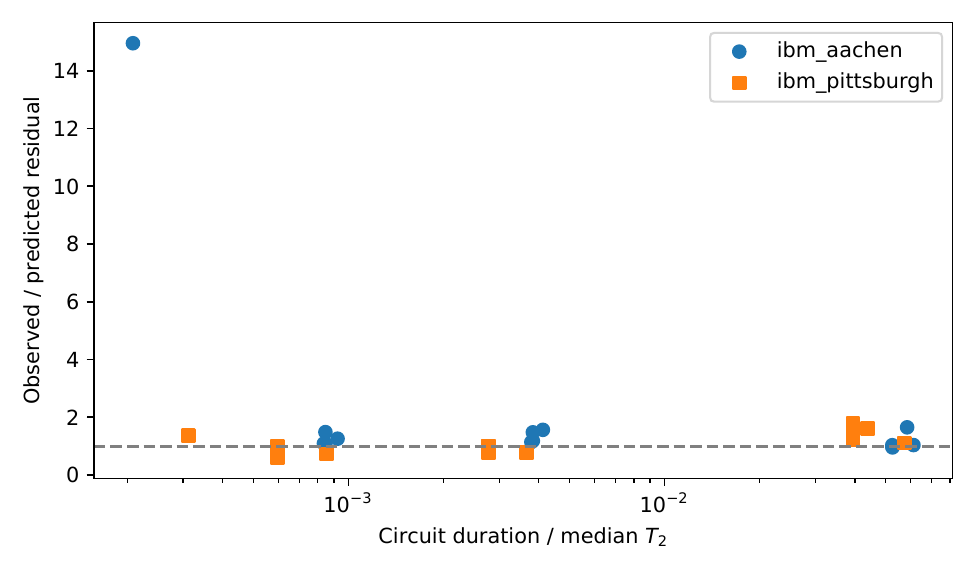}
\caption{Observed raw residuals from $V$ divided by the duration-adjusted calibration model, plotted against total scheduled state-preparation duration over median used-qubit $T_2$. Markers identify processors; pilot and isolated cohorts are shown. The attenuation uses total duration, including gate intervals, and is distinct from the channel simulation with noise confined to scheduled idle gaps. Archived gate lengths define an ASAP schedule excluding provider measurement rotations and alignment.}
\label{S-fig:idle-budget}\end{figure}

\section{Energy and Pauli physicality audit}\label{S-sec:physicality}
Tables~\ref{S-tab:fourth-physicality}, \ref{S-tab:physicality-arms} and \ref{S-tab:physicality-detail} give the campaign totals, arm-level split and nonzero energy violations, respectively.
\begin{table}[htbp]\centering\small\resizebox{\linewidth}{!}{\begin{tabular}{llllll}
\toprule
Arm & Processor & Qubit & $N$ & Outside & Max $|P|$\\
\midrule
RAW & Aachen & \num{27} & \num{18} & \num{0} & \num{0.996}\\
RAW & Aachen & \num{38} & \num{6} & \num{0} & \num{0.997}\\
RAW & Aachen & \num{124} & \num{12} & \num{0} & \num{0.753}\\
RAW & Pittsburgh & \num{61} & \num{12} & \num{0} & \num{0.996}\\
RAW & Pittsburgh & \num{77} & \num{12} & \num{0} & \num{1.000}\\
RAW & Pittsburgh & \num{112} & \num{6} & \num{0} & \num{0.999}\\
RAW & Pittsburgh & \num{114} & \num{6} & \num{0} & \num{1.000}\\
TREX & Aachen & \num{27} & \num{24} & \num{12} & \num{1.740}\\
TREX & Aachen & \num{38} & \num{6} & \num{4} & \num{1.008}\\
TREX & Aachen & \num{124} & \num{18} & \num{2} & \num{1.035}\\
TREX & Pittsburgh & \num{61} & \num{18} & \num{4} & \num{1.004}\\
TREX & Pittsburgh & \num{77} & \num{18} & \num{7} & \num{1.009}\\
TREX & Pittsburgh & \num{112} & \num{6} & \num{0} & \num{0.999}\\
TREX & Pittsburgh & \num{114} & \num{6} & \num{2} & \num{1.004}\\
\bottomrule
\end{tabular}
}
\caption{Individual Pauli-control spectral checks by arm, processor and physical qubit. Raw controls have no unit-magnitude violations.}
\label{S-tab:fifth-pauli}\end{table}
\begin{table}[htbp]\centering\small
\begin{tabular}{lllll}
\toprule
Cohort & Arm & $N$ & Outside & Max (\unit{\millihartree})\\
\midrule
benchmark & RAW & \num{80} & \num{0} & \num{0.0}\\
benchmark & TREX & \num{26} & \num{3} & \num{349.5}\\
benchmark & ZNE & \num{16} & \num{2} & \num{0.2}\\
H2 controls & RAW & \num{12} & \num{0} & \num{0.0}\\
H2 controls & TREX & \num{16} & \num{8} & \num{581.1}\\
\bottomrule
\end{tabular}

\caption{Energy estimates by cohort and arm. ``Outside'' counts estimates beyond the spectrum of their own mapped Hamiltonian, including the same energy offsets. The 150-energy denominator combines 122 benchmark and 28 molecular control estimates; the 168 Pauli controls have a separate unit-magnitude check.}
\label{S-tab:fourth-physicality}\end{table}
Control Hamiltonian outputs are archived as electronic energies. Before applying total-energy spectral bounds, the audit adds the same nuclear and inactive-orbital constants used for the benchmark energies. This conversion leaves residuals unchanged and prevents an offset mismatch from being mistaken for a physicality violation.
\begin{table}[htbp]\centering\small
\resizebox{\linewidth}{!}{\begin{tabular}{lllllll}
\toprule
Backend & Arm & Energies & Outside & Paulis & Outside Pauli & Max excess\\
\midrule
ibm\_aachen & raw & \num{46} & \num{0} & \num{36} & \num{0} & \num{0.000}\\
ibm\_aachen & trex & \num{21} & \num{4} & \num{48} & \num{18} & \num{0.740}\\
ibm\_aachen & zne & \num{8} & \num{1} & \num{0} & \num{0} & \num{0.000}\\
ibm\_pittsburgh & raw & \num{46} & \num{0} & \num{36} & \num{0} & \num{0.000}\\
ibm\_pittsburgh & trex & \num{21} & \num{7} & \num{48} & \num{13} & \num{0.009}\\
ibm\_pittsburgh & zne & \num{8} & \num{1} & \num{0} & \num{0} & \num{0.000}\\
\bottomrule
\end{tabular}
}
\caption{Energy and individual Pauli-control violations by backend and arm. No raw energy or raw Pauli control violates its physical range. Energy counts include benchmark and H$_2$ control estimates; Pauli counts concern individual known-state controls.}\label{S-tab:physicality-arms}
\end{table}
\begin{table}[htbp]\centering\small
\resizebox{\linewidth}{!}{\begin{tabular}{llllll}
\toprule
Cohort & Backend & System & Arm & Estimates & Outside\\
\midrule
benchmark & ibm\_aachen & LiH & zne & \num{4} & \num{1}\\
benchmark & ibm\_aachen & O2 & trex & \num{4} & \num{1}\\
benchmark & ibm\_pittsburgh & H2 & trex & \num{1} & \num{1}\\
benchmark & ibm\_pittsburgh & LiH & trex & \num{4} & \num{1}\\
benchmark & ibm\_pittsburgh & LiH & zne & \num{4} & \num{1}\\
h2\_control & ibm\_aachen & H2 & trex & \num{8} & \num{3}\\
h2\_control & ibm\_pittsburgh & H2 & trex & \num{8} & \num{5}\\
\bottomrule
\end{tabular}
}
\caption{Nonzero out-of-spectrum energy counts by cohort, system, backend and arm. Each energy is compared with the spectrum of its own mapped Hamiltonian, including identical offsets.}\label{S-tab:physicality-detail}
\end{table}

\section{Chemistry definitions and orbital audit}\label{S-sec:orbital-audit}
\begin{table}[htbp]\centering\small
\begin{tabular}{llllll}
\toprule
System & Atom & Element & $x$ (\unit{\angstrom}) & $y$ (\unit{\angstrom}) & $z$ (\unit{\angstrom})\\
\midrule
H2 & \num{0} & H & \num{0.0000000} & \num{0.0000000} & \num{0.0000000}\\
H2 & \num{1} & H & \num{0.0000000} & \num{0.0000000} & \num{0.7428000}\\
LiH & \num{0} & Li & \num{0.0000000} & \num{0.0000000} & \num{0.0000000}\\
LiH & \num{1} & H & \num{0.0000000} & \num{0.0000000} & \num{1.6204000}\\
O2 & \num{0} & O & \num{0.0000000} & \num{0.0000000} & \num{0.0000000}\\
O2 & \num{1} & O & \num{0.0000000} & \num{0.0000000} & \num{1.2160000}\\
BeH2 & \num{0} & H & \num{0.0000000} & \num{0.0000000} & \num{0.0000000}\\
BeH2 & \num{1} & Be & \num{0.0000000} & \num{0.0000000} & \num{1.3294454}\\
BeH2 & \num{2} & H & \num{0.0013027} & \num{0.0000000} & \num{2.6589464}\\
H2O & \num{0} & H & \num{0.0000000} & \num{0.0000000} & \num{0.0000000}\\
H2O & \num{1} & O & \num{0.0000000} & \num{0.0000000} & \num{0.9685435}\\
H2O & \num{2} & H & \num{0.9414237} & \num{0.0000000} & \num{1.1963137}\\
\bottomrule
\end{tabular}

\caption{Cartesian coordinates in angstroms, taken from the frozen molecular definitions. Geometry sources are inherited from the validated molecular specifications in the preceding studies~\cite{larrucea2026h2,larrucea2026selection}. All integrals and references used here were newly computed.}\label{S-tab:geometries}
\end{table}
The O$_2$ archive contains molecular-orbital integrals and active selection matrices. Atomic-orbital coefficients were not archived. We reconstructed the ROHF solution at the archived geometry and aligned the relative rotations of the two degenerate $\pi$ pairs and the orbital phases to both archived integral tensors. Agreement below the reference replay tolerance establishes a matching gauge. Transverse atomic-orbital weight and inversion parity identify orbital 6 as bonding $\pi_u$, 7 and 8 as antibonding $\pi_g^*$, and 9 as antibonding $\sigma_u^*$. The other bonding $\pi_u$ orbital is inactive. This is not a complete $\pi$ active space.
Projectors for particle number, spin projection and total spin in the tapered space select a sector of dimension \OxygenSectorDimension. Its Hamiltonian-connected block containing Hartree--Fock has the same dimension, and the one-parameter doubles ansatz spans that real block. Exact agreement with the active reference therefore tests a small symmetry sector. It does not establish broad ansatz expressiveness for oxygen chemistry.

\section{Bracket sensitivity and separated calibration products}
\begin{table}[!htbp]\centering\small
\begin{tabular}{llrrrr}
\toprule
Backend & System & $n$ & Mean (\unit{\millihartree}) & SD (\unit{\millihartree}) & $t(2)$\\
\midrule
Aachen & BeH2 & \num{3} & \num{-5.6} & \num{22.9} & \num{-0.43}\\
Aachen & LiH & \num{3} & \num{2.4} & \num{0.1} & \num{44}\\
Aachen & O2 & \num{3} & \num{0.0} & \num{14.2} & \num{0.00022}\\
Pittsburgh & BeH2 & \num{3} & \num{18.9} & \num{11.9} & \num{2.7}\\
Pittsburgh & LiH & \num{3} & \num{-0.2} & \num{0.8} & \num{-0.43}\\
Pittsburgh & O2 & \num{3} & \num{-3.8} & \num{10.3} & \num{-0.64}\\
\bottomrule
\end{tabular}

\caption{Within-window Raw 4096 minus the mean of two bracketing Raw 1024 residuals, paired across three windows per system and backend. The statistic $t(2)$ uses the SD of the three window differences; six exploratory cells were inspected without multiplicity adjustment.}
\label{S-tab:paired-all}\end{table}
\begin{table}[htbp]\centering\scriptsize
\resizebox{\linewidth}{!}{\begin{tabular}{lllllll}
\toprule
Backend & System & Arm & $N$ & Mean bracket & Mean linear & Positive linear\\
\midrule
ibm\_aachen & BeH2 & raw & \num{4096} & \num{5.63} & \num{0.22} & \num{2}\\
ibm\_aachen & BeH2 & trex & \num{1024} & \num{-5.53} & \num{0.71} & \num{2}\\
ibm\_aachen & LiH & raw & \num{4096} & \num{-2.40} & \num{-2.45} & \num{0}\\
ibm\_aachen & LiH & trex & \num{1024} & \num{-0.21} & \num{-0.14} & \num{1}\\
ibm\_aachen & LiH & zne & \num{1024} & \num{3.02} & \num{3.00} & \num{2}\\
ibm\_aachen & O2 & raw & \num{4096} & \num{-0.00} & \num{0.07} & \num{1}\\
ibm\_aachen & O2 & trex & \num{1024} & \num{32.41} & \num{31.20} & \num{3}\\
ibm\_aachen & O2 & zne & \num{1024} & \num{59.19} & \num{59.13} & \num{3}\\
ibm\_pittsburgh & BeH2 & raw & \num{4096} & \num{-18.89} & \num{-18.77} & \num{0}\\
ibm\_pittsburgh & BeH2 & trex & \num{1024} & \num{-6.93} & \num{-7.24} & \num{2}\\
ibm\_pittsburgh & LiH & raw & \num{4096} & \num{0.20} & \num{0.44} & \num{2}\\
ibm\_pittsburgh & LiH & trex & \num{1024} & \num{1.45} & \num{1.22} & \num{3}\\
ibm\_pittsburgh & LiH & zne & \num{1024} & \num{0.93} & \num{0.93} & \num{2}\\
ibm\_pittsburgh & O2 & raw & \num{4096} & \num{3.80} & \num{0.69} & \num{1}\\
ibm\_pittsburgh & O2 & trex & \num{1024} & \num{199.03} & \num{199.72} & \num{3}\\
ibm\_pittsburgh & O2 & zne & \num{1024} & \num{208.56} & \num{207.67} & \num{3}\\
\bottomrule
\end{tabular}
}
\caption{Mean absolute-residual improvements relative to the arithmetic raw bracket and a raw bracket interpolated linearly at each treatment's provider start time, in mHa. Positive-linear counts are out of three windows. The interpolation is a sensitivity model for fixed treatment order.}\label{S-tab:brackets}
\end{table}
\begin{table}[htbp]\centering\scriptsize
\resizebox{\linewidth}{!}{\begin{tabular}{lllllll}
\toprule
Backend & System & Pilot n & Pilot mean & Isolated n & Isolated mean & Difference\\
\midrule
ibm\_aachen & BeH2 & \num{3} & \num{424.71} & \num{6} & \num{451.86} & \num{27.15}\\
ibm\_aachen & LiH & \num{3} & \num{8.79} & \num{6} & \num{6.30} & \num{-2.49}\\
ibm\_aachen & O2 & \num{3} & \num{142.33} & \num{6} & \num{119.19} & \num{-23.13}\\
ibm\_pittsburgh & BeH2 & \num{3} & \num{798.11} & \num{6} & \num{685.28} & \num{-112.84}\\
ibm\_pittsburgh & LiH & \num{3} & \num{3.67} & \num{6} & \num{3.02} & \num{-0.65}\\
ibm\_pittsburgh & O2 & \num{3} & \num{238.59} & \num{6} & \num{249.86} & \num{11.27}\\
\bottomrule
\end{tabular}
}
\caption{Pilot and isolated raw-\ShotsLow{} means from $V$, in mHa. Workload multiplicity changes with acquisition time, so these are cohort contrasts.}\label{S-tab:cohort-comparison}
\end{table}
\begin{table}[htbp]\centering\scriptsize
\resizebox{\linewidth}{!}{\begin{tabular}{llllllll}
\toprule
Backend & System & Window & $f_{1q}$ & $f_{2q}$ & $f_{gate}$ & $f_{ro}$ & $f_{prod}$\\
\midrule
ibm\_aachen & BeH2 & pilot & \num{0.967} & \num{0.896} & \num{0.866} & \num{0.965} & \num{0.836}\\
ibm\_aachen & BeH2 & validation-2026-10-03-eu & \num{0.967} & \num{0.896} & \num{0.866} & \num{0.965} & \num{0.836}\\
ibm\_aachen & BeH2 & validation-window2-2026-10-03-eu & \num{0.961} & \num{0.896} & \num{0.860} & \num{0.976} & \num{0.840}\\
ibm\_aachen & BeH2 & validation-window3-2026-10-05-eu & \num{0.961} & \num{0.873} & \num{0.839} & \num{0.965} & \num{0.810}\\
ibm\_aachen & H2 & pilot & \num{1.000} & \num{1.000} & \num{1.000} & \num{0.959} & \num{0.959}\\
ibm\_aachen & H2O & pilot & \num{0.957} & \num{0.869} & \num{0.831} & \num{0.965} & \num{0.802}\\
ibm\_aachen & LiH & pilot & \num{0.999} & \num{0.999} & \num{0.998} & \num{0.980} & \num{0.978}\\
ibm\_aachen & LiH & validation-2026-10-03-eu & \num{0.999} & \num{0.999} & \num{0.998} & \num{0.980} & \num{0.978}\\
ibm\_aachen & LiH & validation-window2-2026-10-03-eu & \num{0.999} & \num{0.999} & \num{0.998} & \num{0.986} & \num{0.984}\\
ibm\_aachen & LiH & validation-window3-2026-10-05-eu & \num{0.999} & \num{0.998} & \num{0.997} & \num{0.979} & \num{0.976}\\
ibm\_aachen & O2 & pilot & \num{0.997} & \num{0.992} & \num{0.990} & \num{0.968} & \num{0.959}\\
ibm\_aachen & O2 & validation-2026-10-03-eu & \num{0.997} & \num{0.992} & \num{0.990} & \num{0.968} & \num{0.959}\\
ibm\_aachen & O2 & validation-window2-2026-10-03-eu & \num{0.997} & \num{0.992} & \num{0.990} & \num{0.980} & \num{0.970}\\
ibm\_aachen & O2 & validation-window3-2026-10-05-eu & \num{0.997} & \num{0.990} & \num{0.988} & \num{0.968} & \num{0.955}\\
ibm\_pittsburgh & BeH2 & pilot & \num{0.965} & \num{0.872} & \num{0.841} & \num{0.835} & \num{0.703}\\
ibm\_pittsburgh & BeH2 & validation-2026-10-03-us & \num{0.965} & \num{0.872} & \num{0.841} & \num{0.864} & \num{0.727}\\
ibm\_pittsburgh & BeH2 & validation-window2-2026-10-03-us & \num{0.965} & \num{0.872} & \num{0.841} & \num{0.856} & \num{0.720}\\
ibm\_pittsburgh & BeH2 & validation-window3-2026-10-05-us & \num{0.958} & \num{0.873} & \num{0.836} & \num{0.849} & \num{0.709}\\
ibm\_pittsburgh & H2 & pilot & \num{1.000} & \num{1.000} & \num{1.000} & \num{0.990} & \num{0.990}\\
ibm\_pittsburgh & H2O & pilot & \num{0.956} & \num{0.842} & \num{0.805} & \num{0.835} & \num{0.673}\\
ibm\_pittsburgh & LiH & pilot & \num{0.999} & \num{0.998} & \num{0.998} & \num{0.988} & \num{0.986}\\
ibm\_pittsburgh & LiH & validation-2026-10-03-us & \num{0.999} & \num{0.998} & \num{0.998} & \num{0.989} & \num{0.986}\\
ibm\_pittsburgh & LiH & validation-window2-2026-10-03-us & \num{0.999} & \num{0.998} & \num{0.998} & \num{0.986} & \num{0.983}\\
ibm\_pittsburgh & LiH & validation-window3-2026-10-05-us & \num{0.999} & \num{0.998} & \num{0.997} & \num{0.984} & \num{0.981}\\
ibm\_pittsburgh & O2 & pilot & \num{0.998} & \num{0.990} & \num{0.988} & \num{0.839} & \num{0.829}\\
ibm\_pittsburgh & O2 & validation-2026-10-03-us & \num{0.998} & \num{0.990} & \num{0.988} & \num{0.873} & \num{0.862}\\
ibm\_pittsburgh & O2 & validation-window2-2026-10-03-us & \num{0.998} & \num{0.990} & \num{0.988} & \num{0.862} & \num{0.851}\\
ibm\_pittsburgh & O2 & validation-window3-2026-10-05-us & \num{0.997} & \num{0.991} & \num{0.988} & \num{0.854} & \num{0.844}\\
\bottomrule
\end{tabular}
}
\caption{Native one-qubit, two-qubit, combined gate and readout products on each frozen layout, using the calibration snapshot matching its acquisition window. Virtual gates contribute their archived zero error.}\label{S-tab:products}
\end{table}
\begin{figure}[htbp]\centering
\includegraphics[width=.92\linewidth]{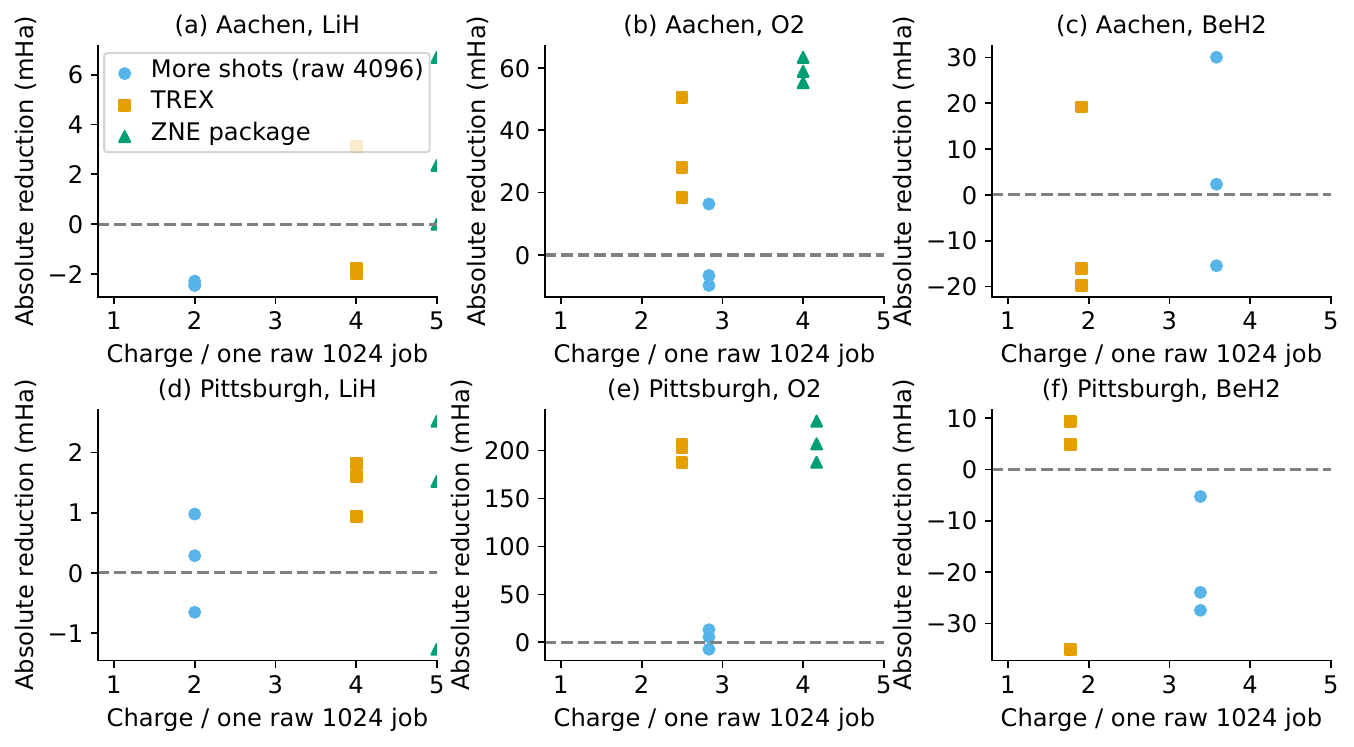}
\caption{Within-window absolute-residual reductions from $V$ relative to the two raw-\ShotsLow{} bracketing jobs. Panels (a--c) show Aachen and (d--f) Pittsburgh. Each treatment has three windows and the cost axis is normalized within system and backend. Colors and markers distinguish more shots, TREX and the ZNE package; BeH$_2$ ZNE was not run.}\label{S-fig:absolute-benefit}
\end{figure}

\section{Detailed execution comparisons}\label{S-sec:execution-details}
\subsection{Worked oxygen cancellation}\label{S-sec:oxygen-cancellation}
For the larger problems, truncation dominates the stationary gap. For example,
O$_2$ has $A-F=\RefOTwoactivetruncationmha{}$~mHa and negligible $V-A$, but Aachen pilot TREX has
$Y-V=\CellpilotibmaachenOTwotrexOneZeroTwoFourmeanmha{}$~mHa, yielding $Y-F=\PilotOxygenibmaachenTREXFull{}$~mHa.
The pilot raw mean has $Y-F=\ScalarAachenpilotOTworawmeanYF{}$~mHa: TREX appears \qty{\PilotOxygenibmaachenCancellation}{\millihartree} closer to $F$
while its absolute execution residual increases from \CellpilotibmaachenOTworawOneZeroTwoFourmeanmha{} to \qty{\CellpilotibmaachenOTwotrexOneZeroTwoFourAbsoluteMean}{\millihartree}.
The returned TREX energy lies below the minimum eigenvalue of the tapered
Hamiltonian. Its lower value therefore does not demonstrate a more accurate
prepared state. The reported estimate violates the mapped variational bound.

\subsection{Shot scaling and window-level ranges}
As a consistency check of the provider-reported scaling, across \ShotContrasts{} shot contrasts, $s_{\ShotsHigh{}}/\bar s_{\ShotsLow{}}=\RoundThreeShotRatioMin{}$--$\RoundThreeShotRatioMax{}$, close to the expected square-root scaling of the returned uncertainty. Raw O$_2$ mean residuals remain \CellisolatedibmaachenOTworawOneZeroTwoFourmeanmha{} and
\qty{\CellisolatedibmpittsburghOTworawOneZeroTwoFourmeanmha}{\millihartree} at low shots (window ranges \CellisolatedibmaachenOTworawOneZeroTwoFourwindowmeanminmha{}--\CellisolatedibmaachenOTworawOneZeroTwoFourwindowmeanmaxmha{} and \CellisolatedibmpittsburghOTworawOneZeroTwoFourwindowmeanminmha{}--\qty{\CellisolatedibmpittsburghOTworawOneZeroTwoFourwindowmeanmaxmha}{\millihartree}),
versus \CellisolatedibmaachenOTworawFourZeroNineSixmeanmha{} and \qty{\CellisolatedibmpittsburghOTworawFourZeroNineSixmeanmha}{\millihartree} at high shots (single-job ranges \CellisolatedibmaachenOTworawFourZeroNineSixminmha{}--\CellisolatedibmaachenOTworawFourZeroNineSixmaxmha{}
and \CellisolatedibmpittsburghOTworawFourZeroNineSixminmha{}--\qty{\CellisolatedibmpittsburghOTworawFourZeroNineSixmaxmha}{\millihartree}), respectively. The low-shot ranges summarize two-job
window means, whereas each high-shot job is its window mean.
Pittsburgh BeH$_2$ window means are \CellisolatedibmpittsburghBeHTworawOneZeroTwoFourwindowmeanmaxmha{}, \CellisolatedibmpittsburghBeHTworawOneZeroTwoFourwindowmeanminmha{} and \qty{\ModelibmpittsburghBeHTwovalidationwindowThreeTwoZeroTwoSixOneZeroZeroFiveusobservedrawmeanresidualmha}{\millihartree} (range
\qty{\BerylliumWindowRange}{\millihartree}). Its six-job SD is \qty{\CellisolatedibmpittsburghBeHTworawOneZeroTwoFoursamplesdmha}{\millihartree}, whereas mean returned $s$ is
\qty{\CellisolatedibmpittsburghBeHTworawOneZeroTwoFourmeanreturnedsmha}{\millihartree}. The SD of the three two-job window means is \qty{\VaribmpittsburghBeHTwowindowmeansdmha}{\millihartree} (\WindowDispersionibmpittsburghBeHTwo{} times the unrounded $\bar s$). Under an iid Gaussian reference model, the six-job SD has a 95\% interval of \HeadlineSDLow--\qty{\HeadlineSDHigh}{\millihartree}. Cross-window changes violate that reference model, so this interval describes its small-sample sensitivity under its assumptions only.

\subsection{Duration and calibration-model comparisons}\label{S-sec:duration-models}
The calibration-product model counts gate and readout errors. To examine its sensitivity to circuit duration, we also apply a phenomenological attenuation factor. Scheduling each instruction as soon as its qubits are free gives \qty{\ScalarroundTwodurationBeHTwo}{\micro\second} for Pittsburgh BeH$_2$, using archived gate durations. The mean of the window-specific median $T_2$ values over its five mapped qubits is \qty{\RoundThreeWireTtwo}{\micro\second}. Initially that median is \qty{\RoundThreeWireInitialTtwo}{\micro\second}, compared with the device-wide \qty{\RoundThreeDeviceTtwo}{\micro\second} in Table~\ref{S-tab:device-calibration}. LiH and O$_2$ execute in \DurationLiH{} and \qty{\DurationOTwo}{\micro\second}, respectively. Here $\tau$ is total state-preparation duration, including gate intervals as well as idle gaps. Multiplying the calibration product by a phenomenological per-qubit factor $\exp(-\tau/T_2)$ moves its mean contrast from \Contrastfprod{} to \Contrastfprodidle{}, toward the measured contrast \ContrastObserved{}. Across isolated BeH$_2$ cells, ratios of observed to predicted residuals fall from \ModelibmpittsburghBeHTwovalidationwindowTwoTwoZeroTwoSixOneZeroZeroThreeusobservedovernaiveprediction{}--\BudgetRatioBeforeMax{} to \BudgetRatioAfterMin{}--\BudgetRatioAfterMax{} (Fig.~\ref{S-fig:idle-budget}).

 The factor can therefore overlap decoherence already represented by calibrated gate errors; closer agreement does not isolate an idle-noise contribution. A separate channel simulation applies thermal relaxation only to scheduled idle gaps and predicts \RoundThreeThermalIdleMin--\qty{\RoundThreeThermalIdleMax}{\millihartree} for Pittsburgh BeH$_2$, still below its observed window means (Table~\ref{S-tab:calibration-model}).

\subsection{Mitigation uncertainty and extrapolation diagnostics}
At the same nominal \ShotsLow{} shots, returned raw uncertainties exceed the exact ideal QWC standard deviations of Table~\ref{tab:sampling} by factors \ScalarroundTworatiomin--\ScalarroundTworatiomax. Different measurement policies and total work prevent attributing this excess solely to state variance or sampling overhead.

O$_2$ TREX reduces absolute residuals in every window on both processors;
Pittsburgh improvements span \OxygenTREXibmpittsburghMin{}--\qty{\OxygenTREXibmpittsburghMax}{\millihartree} at \OxygenTREXCostibmpittsburgh{} times raw job cost.
ZNE-package mean residuals are \qty{\CellisolatedibmaachenOTwozneOneZeroTwoFourmeanmha}{\millihartree} on Aachen (window range \CellisolatedibmaachenOTwozneOneZeroTwoFourminmha{}--\CellisolatedibmaachenOTwozneOneZeroTwoFourmaxmha{})
and \qty{\CellisolatedibmpittsburghOTwozneOneZeroTwoFourmeanmha}{\millihartree} on Pittsburgh (\CellisolatedibmpittsburghOTwozneOneZeroTwoFourminmha{}--\CellisolatedibmpittsburghOTwozneOneZeroTwoFourmaxmha{}), at about four times raw cost.
For O$_2$, the ZNE package's returned-$s$ overhead relative to bracketing raw jobs spans \RoundThreeZNEAachenMin--\RoundThreeZNEAachenMax~on Aachen and \RoundThreeZNEPittsburghMin--\RoundThreeZNEPittsburghMax~on Pittsburgh. An independent, equal-variance linear intercept at factors
$(1,3,5)$ has weights $(13/12,1/3,-5/12)$ and SD amplification $\sqrt{35/24}=\LinearZNEAmplification{}$;
heterogeneous fits, unequal factor variances and twirling limit transfer of this comparison model to the package.

O$_2$ ZNE sample SD divided by mean returned $s$ is only
\AachenOxygenZNEDispersion~on Aachen and \PittsburghOxygenZNEDispersion~on
Pittsburgh. Returned uncertainty
can thus exceed the spread of three ZNE executions even while raw-arm cross-job scatter exceeds returned uncertainty. The largest isolated dispersion ratio is
\LargestDispersion~for Pittsburgh BeH$_2$ raw \ShotsHigh{}, with three jobs.

\section{Additional figure views and complete statistics}
\begin{figure}[htbp]\centering
\includegraphics[width=\linewidth]{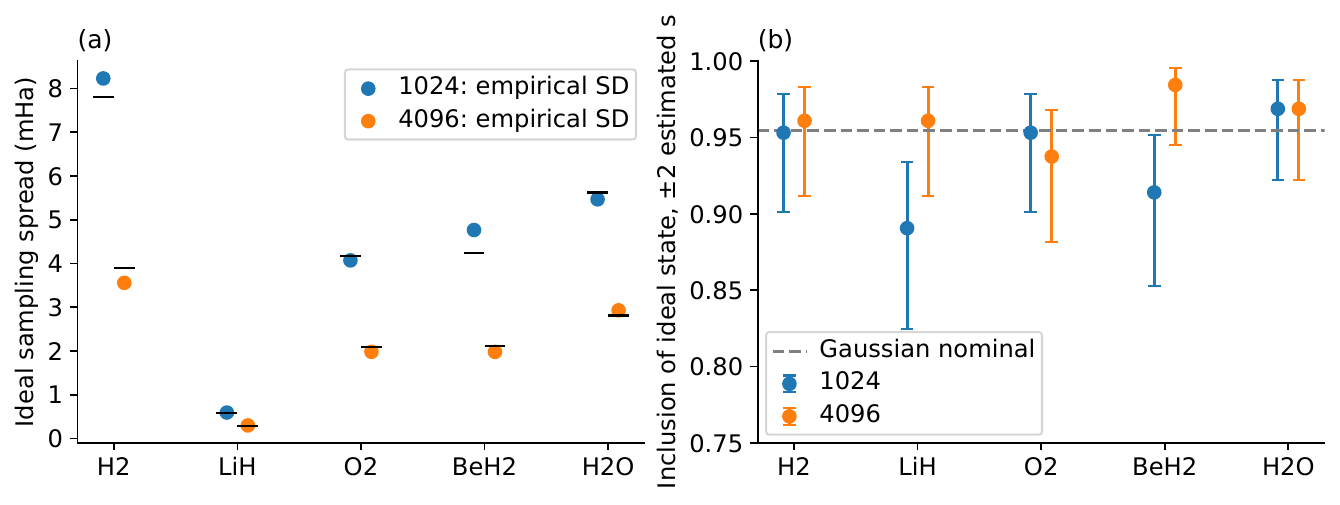}
\caption{Ideal QWC validation at frozen $V$: (a) empirical SD and exact ideal $\sigma$; (b) inclusion within estimated $\pm2s$, with Wilson intervals over \IdealRepeats{} independent repetitions per molecule and shot setting. This checks the local sampling and uncertainty implementation.}\label{S-fig:sampling}
\end{figure}
\begin{figure}[htbp]\centering
\includegraphics[width=\linewidth]{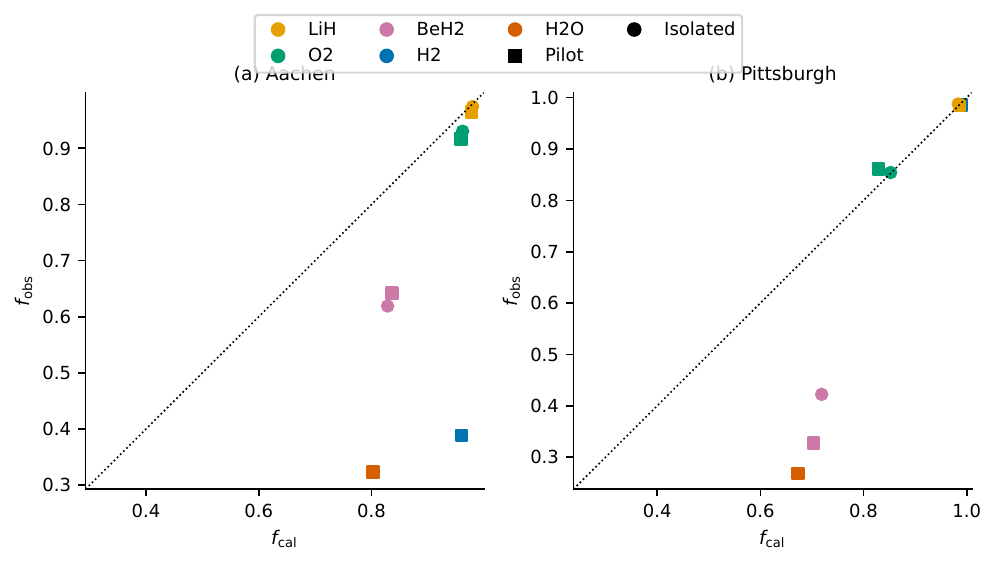}
\caption{Retained energy contrast $f_{\rm obs}=(E_{\rm mix}-Y)/(E_{\rm mix}-V)$ versus calibration product $f_{\rm cal}=f_{\rm gate}f_{\rm ro}$ for raw \ShotsLow{} estimates on Aachen (left) and Pittsburgh (right). Colors identify systems, squares pilot means and circles isolated means. The dotted line is equality. Pilot means use three jobs except H$_2$O (one), and isolated means use six.}\label{S-fig:contrast-scatter}
\end{figure}
\begin{figure}[htbp]\centering
\includegraphics[width=\linewidth]{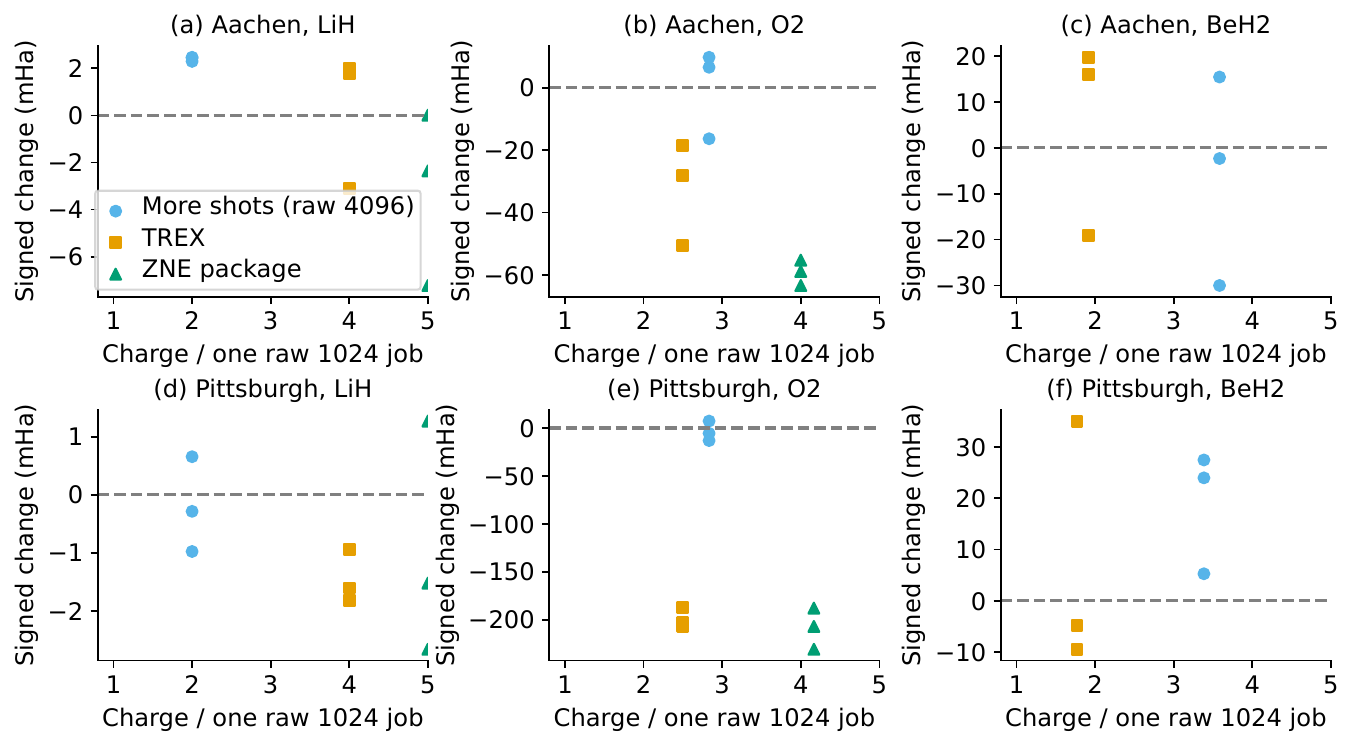}
\caption{Per-system signed residual changes from $V$ against the raw-\ShotsLow{} bracket, versus charge relative to one raw job of the same system and backend. Panels (a--c) show Aachen and (d--f) Pittsburgh. Each supported treatment has three window points. Colors and markers distinguish more shots, TREX and the ZNE package.}\label{S-fig:cost-panels}
\end{figure}
\begin{table}[htbp]\centering\scriptsize
\resizebox{\linewidth}{!}{\begin{tabular}{lllrrrrrrrr}
\toprule
System & Arm & $N$ & $n$ & Mean & SE$_w$ & SD & Min & Max & $\bar s$ & SD/$\bar s$\\
\midrule
\multicolumn{11}{l}{\textbf{Aachen}}\\
BeH2 & raw & \num{1024} & \num{6} & \num{452} & \num{4.42} & \num{39.64} & \num{389} & \num{507} & \num{20.512} & \num{1.9}\\
BeH2 & raw & \num{4096} & \num{3} & \num{446} & \num{9.55} & \num{16.53} & \num{431} & \num{464} & \num{10.305} & \num{1.6}\\
BeH2 & trex & \num{1024} & \num{3} & \num{457} & \num{8.1} & \num{14.00} & \num{441} & \num{468} & \num{20.544} & \num{0.68}\\
LiH & raw & \num{1024} & \num{6} & \num{6.3} & \num{0.41} & \num{0.84} & \num{4.9} & \num{7.4} & \num{2.8266} & \num{0.3}\\
LiH & raw & \num{4096} & \num{3} & \num{8.7} & \num{0.46} & \num{0.80} & \num{7.8} & \num{9.4} & \num{1.4874} & \num{0.54}\\
LiH & trex & \num{1024} & \num{3} & \num{6.5} & \num{1.68} & \num{2.91} & \num{3.3} & \num{8.9} & \num{3.1042} & \num{0.94}\\
LiH & zne & \num{1024} & \num{3} & \num{3.1} & \num{1.9} & \num{3.33} & \num{-0.3} & \num{6.4} & \num{6.7559} & \num{0.49}\\
O2 & raw & \num{1024} & \num{6} & \num{119} & \num{3.0} & \num{15.61} & \num{105} & \num{143} & \num{9.9959} & \num{1.6}\\
O2 & raw & \num{4096} & \num{3} & \num{119} & \num{6.1} & \num{10.57} & \num{108} & \num{129} & \num{4.9564} & \num{2.1}\\
O2 & trex & \num{1024} & \num{3} & \num{87} & \num{7.8} & \num{13.56} & \num{74} & \num{101} & \num{10.707} & \num{1.3}\\
O2 & zne & \num{1024} & \num{3} & \num{60.0} & \num{0.721} & \num{1.25} & \num{58.6} & \num{61.0} & \num{22.886} & \num{0.055}\\
\multicolumn{11}{l}{\textbf{Pittsburgh}}\\
BeH2 & raw & \num{1024} & \num{6} & \num{685} & \num{77.2} & \num{120.41} & \num{575} & \num{855} & \num{21.637} & \num{5.6}\\
BeH2 & raw & \num{4096} & \num{3} & \num{704} & \num{80.9} & \num{140.18} & \num{594} & \num{862} & \num{10.793} & \num{13}\\
BeH2 & trex & \num{1024} & \num{3} & \num{692} & \num{91.0} & \num{157.56} & \num{583} & \num{873} & \num{24.415} & \num{6.5}\\
LiH & raw & \num{1024} & \num{6} & \num{3.0} & \num{0.35} & \num{1.01} & \num{2.4} & \num{5.0} & \num{2.4641} & \num{0.41}\\
LiH & raw & \num{4096} & \num{3} & \num{2.8} & \num{0.6} & \num{1.04} & \num{1.6} & \num{3.4} & \num{1.1918} & \num{0.87}\\
LiH & trex & \num{1024} & \num{3} & \num{1.6} & \num{0.34} & \num{0.59} & \num{0.9} & \num{2.1} & \num{2.4464} & \num{0.24}\\
LiH & zne & \num{1024} & \num{3} & \num{2.0} & \num{1.5} & \num{2.62} & \num{-0.1} & \num{5.0} & \num{4.0704} & \num{0.64}\\
O2 & raw & \num{1024} & \num{6} & \num{250} & \num{12.4} & \num{40.02} & \num{210} & \num{308} & \num{13.231} & \num{3}\\
O2 & raw & \num{4096} & \num{3} & \num{246} & \num{9.8} & \num{16.92} & \num{236} & \num{266} & \num{6.5475} & \num{2.6}\\
O2 & trex & \num{1024} & \num{3} & \num{51} & \num{8.5} & \num{14.78} & \num{41} & \num{68} & \num{17.138} & \num{0.86}\\
O2 & zne & \num{1024} & \num{3} & \num{41.3} & \num{0.94} & \num{1.63} & \num{40.3} & \num{43.2} & \num{44.752} & \num{0.036}\\
\bottomrule
\end{tabular}
}
\caption{Complete isolated descriptive statistics, including minimum and maximum residuals from $V$, in mHa. Six raw-\ShotsLow{} or three other jobs contribute per system, device and arm; SE$_w$ is the standard error of three window means.}\label{S-tab:full-descriptive}
\end{table}
\begin{table}[htbp]\centering\small
\begin{tabular}{lllll}
\toprule
System & $F$ (\unit{\hartree}) & $A$ (\unit{\hartree}) & $V$ (\unit{\hartree}) & $E_{nuc}$ (\unit{\hartree})\\
\midrule
H2 & \num{-1.137253} & \num{-1.137253} & \num{-1.137253} & \num{0.712409}\\
LiH & \num{-7.881936} & \num{-7.861408} & \num{-7.861392} & \num{0.979716}\\
O2 & \num{-147.746483} & \num{-147.633158} & \num{-147.633158} & \num{27.851432}\\
BeH2 & \num{-15.595127} & \num{-15.566024} & \num{-15.566024} & \num{3.383299}\\
H2O & \num{-75.015607} & \num{-74.972318} & \num{-74.972205} & \num{9.089246}\\
\bottomrule
\end{tabular}
\par\smallskip
\begin{tabular}{llllll}
\toprule
System & Bonds (\unit{\angstrom}) & Active orbitals & Paulis & $\|H_q\|_1$ (\unit{\hartree}) & Inactive offset (\unit{\hartree})\\
\midrule
H2 & \num{0.7428} & 0;1 & \num{3} & \num{2.010} & \num{0.000000}\\
LiH & \num{1.6204} & 1;2 & \num{9} & \num{1.368} & \num{-7.787901}\\
O2 & \num{1.2160} & 6;7;8;9 & \num{50} & \num{6.077} & \num{-170.121940}\\
BeH2 & 1.3294, 1.3295 & 1;2;3;4 & \num{119} & \num{6.309} & \num{-15.031574}\\
H2O & 0.9685, 0.9686 & 3;4;5;6 & \num{155} & \num{10.563} & \num{-77.915751}\\
\bottomrule
\end{tabular}

\caption{Total references, nuclear repulsion, inactive-orbital offsets, active orbitals and electronic coefficient norms. All Hamiltonians use STO-3G with an active-space transformer and no separate frozen-core transformation.}\label{S-tab:energy-offsets}
\end{table}
\begin{table}[htbp]\centering\small
\resizebox{\linewidth}{!}{\begin{tabular}{lllllll}
\toprule
Qubit & Basis & $p(-|+)$ & $p(+|-)$ & Mean error & Calibration & Factor\\
\midrule
\num{124} & Z & \num{0.3242} & \num{0.1313} & \num{0.2278} & \num{0.0405} & \num{5.6}\\
\num{124} & X & \num{0.3022} & \num{0.1338} & \num{0.2180} & \num{0.0405} & \num{5.4}\\
\num{27} & Z & \num{0.0020} & \num{0.0034} & \num{0.0027} & \num{0.0016} & \num{1.7}\\
\num{27} & X & \num{0.0049} & \num{0.0049} & \num{0.0049} & \num{0.0016} & \num{3.1}\\
\bottomrule
\end{tabular}
}
\caption{Effective assignment interpretation of raw eigenstate controls in the mixed Aachen job. In Z, plus and minus mean $|0\rangle$ and $|1\rangle$. In X they mean $|+\rangle$ and $|-\rangle$. The factors compare basis-average control error with calibrated mean readout error; preparation and basis rotation also contribute.}\label{S-tab:effective-assignment}
\end{table}

\end{document}